\documentclass[11pt,a4paper,dvips]{article}
\usepackage{a4p}
\usepackage{graphicx}
\usepackage{cite,./mcite}
\usepackage{amsmath}
\usepackage{xspace}
\newcommand{\W}{\mbox{$W$}~}

\newcommand{\pt}{\mbox{$p_{T}$}~}
\newcommand{\ptx}{\mbox{$p_{T}$}}
\newcommand{\TeV}{\mbox{\rm ~TeV}~}

\newcommand{\GeV}{\mbox{\rm ~GeV}~}
\newcommand{\GeVx}{\mbox{\rm ~GeV}}
\newcommand{\MeV}{\mbox{\rm ~MeV}~}

\newcommand{\as}{\mbox{$\alpha_s$}\xspace}
\newcommand{\vud}{V_{ud}^2}
\newcommand{\vus}{V_{us}^2}
\newcommand{\vcd}{V_{cd}^2}
\newcommand{\vcs}{V_{cs}^2}

\newcommand{\ROOT}{\texttt{ROOT}~}
\newcommand{\MCFM}{\texttt{MCFM}~}
\newcommand{\NLOJET}{\texttt{NLOJET++}~}

\newcommand{\MCFMx}{\texttt{MCFM}}
\newcommand{\NLOJETx}{\texttt{NLOJET++}}
\newcommand{\nsub}{n_\mathrm{sub}}

\begin{document}
\title{A posteriori inclusion of parton density functions in NLO QCD final-state calculations at hadron colliders: \\ The APPLGRID Project
}
\author{Tancredi Carli$^{1}$,
        Dan Clements$^{2}$,
        Amanda Cooper-Sarkar$^{3}$,
        Claire Gwenlan$^{3}$,\\
        Gavin P. Salam$^{4}$, 
        Frank Siegert$^{5}$, 
        Pavel  Starovoitov$^{1,6}$,
        Mark Sutton$^{7}$ \\ \\
%}
%\institute{
%\begin{center}
$^{1}$ CERN, Department of Physics, Geneva (Switzerland), \\
$^{2}$ University of Glasgow, Glasgow (UK), \\
$^{3}$ University of Oxford, Oxford (UK), \\
$^{4}$ LPTHE, UPMC Univ. Paris 6 and CNRS UMR 7589, Paris 05, France, \\
$^{5}$ IPPP, Durham University, Durham (UK), \\
$^{6}$ Nat. Sci. Educ. Center of Part.and HEP, Minsk (Belorussia), \\
% $^{7}$ University College London, London (UK). \\
$^{7}$ University of Sheffield, Sheffield (UK). \\
%}
%\end{center}
}
\maketitle
% Give date: \today for drafts, a fixed date for final papers
\def\thedate{\today}

\begin{abstract}
A method to facilitate the consistent inclusion of
cross-section measurements based on complex
final-states from HERA, TEVATRON and the LHC in proton parton 
density function (PDF) fits has been developed.
This can be used to increase the sensitivity of LHC data to deviations from
Standard Model predictions.
The method stores perturbative coefficients of NLO QCD calculations of
final-state observables measured in hadron colliders in look-up tables.
This allows the {\em a posteriori} inclusion of parton density functions (PDFs),
and of the strong coupling, as well as the  {\em a posteriori} variation of the
renormalisation and factorisation scales in cross-section calculations.
The main novelties in comparison to 
%GPS: previous work -> 
original work on the subject
are the use
of higher-order interpolation, which substantially improves the
trade-off between accuracy and memory use, and a CPU and computer 
memory optimised way to construct and store the look-up table
using modern software tools.
It is demonstrated that a sufficient accuracy on the cross-section
calculation can be achieved with reasonably small look-up table size by using 
the examples of jet production and electro-weak boson ($Z$, $W$) production 
in proton-proton collisions at a center-of-mass energy of $14$ TeV
at the LHC. 
The use of this technique in PDF fitting is demonstrated in a PDF-fit to 
HERA data and
simulated LHC jet cross-sections
as well as in a study of the jet cross-section uncertainties
at various centre-of-mass energies.
\end{abstract}
\newpage
\section{Introduction}
The Large Hadron Collider (LHC) at CERN
will collide protons at a centre-of-mass energy of up to $14000$~GeV. 
The combination of its high collision rate and centre-of-mass energy will make it
possible to probe new interactions at very short distances. Such interactions might
be revealed in the production of cross-sections of particles
at very high transverse momentum (\ptx) as a deviation from the Standard Model theory.

The sensitivity to new physics depends on experimental uncertainties in the measurements
and on theoretical uncertainties in the Standard Model predictions. 
It is therefore important to work out
a strategy to minimise both the experimental and theoretical uncertainties from LHC data.
Residual renormalisation and factorisation scale uncertainties
in next-to-leading order (NLO) QCD calculations for single inclusive
jet cross-sections are typically about $5-10\%$ and should hopefully be reduced
as NNLO calculations become available. 
% mandy proposal
%The impact of PDF uncertainties on the other
%hand can be substantially larger in some regions, especially at large \ptx. 
%For example, at $\ptx = 2000$~{\rm GeV} PDF uncertainties dominate the overall
%theoretical uncertainty of $20\%$.
However, in some kinematic regimes, PDF uncertainties can be substantially larger
than the uncertainties from higher-order corrections, for example
at large \ptx.
One strategy to reduce such uncertainty is to use 
single inclusive jet  or Drell-Yan
cross-sections at lower \pt to constrain the proton parton density function
(PDF) uncertainties at high \ptx.

In order to further constrain PDF uncertainties, it would be useful to be able 
to include final state data such as \pt and rapidity 
distributions for $W/Z$-boson and jet production in global NLO 
QCD PDF fits, without recourse to inexact methods like the use of
simple factor correcting of LO cross-sections ($k-$factors). 
We propose here a method for a consistent
inclusion of final-state observables in global QCD analyses.

For inclusive data, like the proton structure function $F_2$ in deep-inelastic
scattering (DIS) the perturbative coefficients
are known analytically. During the fit the cross-section
can therefore be quickly calculated from the strong coupling ($\as$) and the PDFs
and then be compared to the measurements.
However, final state observables, where detector acceptances
or jet algorithms are involved in the definition of the perturbative coefficients
(called ``weights'' in the following),
have to be calculated using NLO QCD Monte Carlo programs. 
Typically such programs
need about one day of CPU time to accurately calculate the cross-section.
It is therefore necessary to find less time consuming methods.
%a way to calculate the perturbative
%coefficients with high precision in a long run and to include
% $\as$ and the PDFs  ``a posteriori''.

Any NLO QCD calculation of a final-state observable involves Monte
Carlo integration over a large number of events. For deep-inelastic
scattering and at hadron colliders 
this must usually be repeated for each new PDF set, making
it impractical to consider many `error' PDF sets, or carry out PDF
fits.
Here, the ``{\em a posteriori}'' inclusion of PDFs is discussed, whereby the
Monte Carlo run calculates a look-up table (in momentum fraction, $x$, 
and momentum transfer, $Q$) of cross-section
weights that can subsequently be combined with an arbitrary PDF.
The procedure is numerically equivalent to using an interpolated
form of the PDF.

 Many methods have been proposed to solve this problem in the past
\cite{Graudenz:1995sk,Kosower:1997vj,Stratmann:2001pb,wobisch,zeusjets}.  In principle the
highest efficiencies can be obtained by taking moments with respect to
Bjorken-$x$ \cite{Graudenz:1995sk,Kosower:1997vj}, because this
converts convolutions into multiplications. This can have notable
advantages with respect to memory consumption, especially in cases
with two incoming hadrons.  On the other hand, there are complications
such as the need for PDFs in moment space and the
associated inverse Mellin transforms.

Methods in $x$-space have traditionally been somewhat less efficient,
both in terms of speed 
%(in the `a posteriori' steps --- not a major issue here)
 and in terms
of memory consumption. They are, however, somewhat more transparent
since they provide direct information on the $x$ values of relevance.
Furthermore they can be used with any PDF.
The use of $x$-space methods can be further improved by using
methods developed originally for PDF evolution \cite{Ratcliffe:2000kp,Dasgupta:2001eq,Salam:2008qg}.

Our method \cite{Carli:2005ji} bears a number of similarities to that of the \texttt{fastNLO}
project \cite{fastnlo} and the two approaches were to some extent developed in parallel. 
Relative to \texttt{fastNLO}, we take better advantage of the
sparse nature of the $x$-dependent weights, 
allow for more flexibility in the scale choice by keeping 
explicitly the scale dependence as an additional dimension in the weighting table
and provide a means to
evaluate renormalisation and factorisation scale-dependence {\em a posteriori}. 
We also provide a broader range of processes, since in addition to di-jet production, 
we include W- and Z-boson production.  
%The
%main advantage of the \texttt{fastNLO} package compared to our approach is
%perhaps the ready availability of a large number of weight files for
%practically all inclusive jet $p_T$ spectra and di-jet mass spectra.
In order to make easy use of the large number of weight files for
practically all inclusive jet $p_T$ spectra and di-jet mass spectra
made available by the \texttt{fastNLO} project,
we provide a software interface to make use of these weight tables
within the \texttt{APPLGRID} framework.

\section{PDF-independent representation of cross-sections}
\subsection{Representing the PDF on a grid}
We make the assumption that PDFs can be accurately represented by storing their values on a
two-dimensional grid of points and using $n^{\mathrm{th}}$-order interpolations between those points.
Instead of using the parton momentum fraction $x$ and the factorisation scale $Q^2$, we use a
variable transformation that provides good coverage of the full $x$
and $Q^2$ range
%to be able to work (for simplicity)
with uniformly spaced grid points:%
\begin{equation}
\label{eq:ytau}
y(x) = \ln \frac{1}{x} + a (1 - x ) \; \; \; {\rm and} \; \; \;
\tau(Q^2) = \ln \ln \frac{Q^2}{\Lambda^2}.
\end{equation}
%Here for simplicity we assume the renormalisation and factorisation scales to coincide and call them $Q^2$,
%while the Bjorken $x$ as usual represents the momentum fraction carried by the parton involved in the scattering
The parameter $\Lambda$ should be chosen of the order of $\Lambda_{\mathrm{QCD}}$, but need not necessarily be identical.
The parameter $a$ serves to increase the density of points in the
large $x$ region%
\footnote{For a fixed total number of bins, as the bins at large $x$
  get finer, the low-$x$ ones become wider.} %
and can
be chosen according to the needs of the concrete application.\footnote{In case of $a=0$ the function
is analytically invertible, for $a \neq 0$ a numerical inversion has to be applied.}

The PDF $f(x,Q^2)$ is then represented by its values $q_{i_y,i_\tau}$ at the 2-dimensional
grid point $(i_y \, \delta y, i_\tau \, \delta \tau)$, where $\delta
y$ and $\delta \tau$ denote the grid spacings,
and is obtained elsewhere by interpolation:
\begin{equation}
\label{eq:interp}
f(x,Q^2) = \sum_{i=0}^n \sum_{\iota=0}^{n'} f_{k+i,\kappa+\iota} \,\,
I_i^{(n)}\left(  \frac{y(x)}{\delta y}  - k \right)\,
I_\iota^{(n')}\left(  \frac{\tau(Q^2)}{\delta\tau}-\kappa  \right),
\end{equation}
where $n$, $n'$ are the interpolation orders.
The interpolation function $I_i^{(n)}(u)$ is 1 for $u=i$, and otherwise is given by:
\begin{equation}
\label{eq:Ii}
I_i^{(n)}(u) = \frac{(-1)^{n-i}}{i!(n-i)!} \frac{u (u-1) \ldots (u-n)}{u-i}\,.
\end{equation}
%There is some freedom in the choice of $k$ and $\kappa$.
Defining $\mathrm{int} (u)$ to be the largest integer such that $\mathrm{int}(u) \le u$,
$k$ and $\kappa$ are defined as:
\begin{eqnarray}
\label{eq:kchoice}
%k(x) =& 1 +  \mathrm{int} \left( y(x) / \delta y \right),
%&\kappa(x) = \mathrm{int} \left( \tau(Q^2) / \delta \tau \right) %\\
k(x) =& \mathrm{int} \left( \frac{y(x)}{\delta y} - \frac{n-1}{2} \right), &
\kappa(Q^2) = \mathrm{int} \left( \frac{\tau(Q^2)}{\delta \tau} - \frac{n'-1}{2} \right).
\end{eqnarray}
%The former reduces the dependence of an observable on `kinematically inaccessible' points of the PDF
%(i.e.\ $x$ values smaller than the lowest accessible values), while the latter should lead to slightly
%higher accuracy, at least for the intermediate regions.
%Note that for large $x$ values the sum
%in eq. (\ref{eq:interp}) will run over negative values of $k-i$ and $\kappa-\iota$ where $q_{k-i,\kappa+\iota}$
%is by definition zero.\footnote{This corresponds to extending the PDF beyond $x=1$ with a uniformly zero value
%there. One can avoid this by instead requiring $k\ge n$. Numerically, the rapid vanishing of the PDFs as
%$x\to1$ tends to ensure that both options are fairly similar.}
%
Given finite grids whose vertex indices range from, $0\ldots N_y-1$, for
the $y$ grid and, $0\ldots N_\tau-1$, for the $\tau$ grid, one should
additionally require that eq.~(\ref{eq:interp}) only uses available
grid points. This can be achieved by remapping, $k \to
\max(0,\min(N_y-1-n,k))$, and, $\kappa \to
\max(0,\min(N_\tau-1-n',\kappa))$.

\subsection{Representing the final state cross-section weights on a grid (DIS case)}
To illustrate the method we take the case of a single flavour in 
deep-inelastic scattering (DIS).

Suppose that we have an NLO Monte Carlo program that produces events, $m=1\dots N$.
Each event $m$ has an $x$ value, $x_m$, a $Q^2$ value, $Q^2_m$, as well as a weight, $w_m$.
We define $p_m$ as the number of powers in the strong coupling
$\as$ in event $m$.
Normally one would obtain the final result $W$ of the Monte Carlo integration 
for one sub-process from:\footnote{Here, and in the following,
renormalisation and factorisation scales have been set equal for simplicity.} 
\begin{equation}
\label{eq:normalint}
 W = \sum_{m=1}^N \,w_m \, \left( \frac{\alpha_s(Q_m^2)} {2\pi}\right)^{p_m}  \, f(x_m,Q^2_m),
\end{equation}
where $f(x, Q^2)$ is the PDF of the flavour under consideration.

Instead one introduces a weight grid $W_{i_y,i_\tau}^{(p)}$ and then for each 
event one updates
a portion of the grid with:\\
$i = 0\dots n,\; \iota = 0\dots n':$
\begin{eqnarray}
\label{eq:weight2evolve}
W_{k+i,\kappa + \iota}^{(p_m)} \to W_{k+i,\kappa + \iota}^{(p_m)} + w_m\,
  I_i^{(n)} \left(\frac{y(x_m)}{\delta y} - k\right)
  I_{\iota}^{(n')}\left(\frac{\tau(Q^2_m)}{\delta \tau} - \kappa \right), \\
 \;\;\; {\rm where} \;\;\;
  k \equiv k(x_m),\; \kappa \equiv \kappa(Q^2_m). \nonumber
  \end{eqnarray}
The final result for $W$, for an arbitrary PDF and an arbitray \as, can then be obtained \emph{subsequent}
to the Monte Carlo run:
\begin{equation}
\label{eq:WfinalxQ}
W = \sum_p \sum_{i_y} \sum_{i_\tau}
W_{i_y,i_\tau}^{(p)} \, \left( \frac{\alpha_s\left({Q^2}^{(i_\tau)}\right)}{2\pi}\right)^{p} f \!\left(x^{(i_y)}, {Q^2}^{(i_\tau)} \right)\,,
\end{equation}
where the sums with indices $i_y$ and $i_\tau$ run over the number of  grid points and
we have explicitly introduced $x^{(i_y)}$ and ${Q^2}^{(i_\tau)}$ such that:
\begin{equation}
\label{eq:xQdefs}
 y(x^{(i_y)}) = i_y \, \delta y \quad {\rm and} \quad
\tau\left({Q^2}^{(i_\tau)}\right) =  i_\tau \, \delta \tau.
\end{equation}

\subsection{Including renormalisation and factorisation scale dependence }
%Regardless of the representation used for the cross-sections (on or
%more grids in $x$-space, or even Mellin moments),
If one has the weight matrix $W_{i_y,i_\tau}^{(p)}$ determined separately order by order in
$\as$, it is straightforward to vary the renormalisation $\mu_R$  and
factorisation $\mu_F$ scales {\em a posteriori} (we assume that they were set equal
in the original calculation).

It is helpful to introduce some notation related to the DGLAP evolution %\cite{dglap}
equation:
\begin{equation}
  \label{eq:DGLAP}
  \frac{d f(x,Q^2)}{d \ln Q^2} = \frac{\alpha_s(Q^2)}{2\pi} (P_0 \otimes f)(x,Q^2)
                + \left(\frac{\alpha_s(Q^2)}{2\pi}\right)^2 (P_1
                \otimes f)(x,Q^2) + \ldots,
\end{equation}
where the $P_0$ and $P_1$ are the LO and NLO matrices of DGLAP
splitting functions
that operate on vectors (in flavour space) $f$ of the
PDFs.
% For any group that is already performing PDF fits the DGLAP evolution program
%should be able to provide $(P_0 \otimes q)(x,Q^2)$ relatively straightforwardly.
Let us now restrict our attention to the NLO case where we have just
two values of $p$ in eq.~\ref{eq:WfinalxQ}.
For example, in jet production in DIS, $p_{\mathrm{LO}}=1$ and
$p_{\mathrm{NLO}}=2$.
Introducing $\xi_R$ and $\xi_F$ corresponding to the factors by which
one varies $\mu_R$ and $\mu_F$ respectively, 
for arbitrary $\xi_R$ and $\xi_F$ we may then write:
\begin{eqnarray}
  \label{eq:Wfinalxi}
  W(\xi_R, \xi_F) = \sum_{i_y}  \sum_{i_\tau}
  \left\{
    \left(\frac{\alpha_s\left(\xi_R^2 {Q^2}^{(i_\tau)}\right)\,
               }{2\pi}
     \right)^{p_{\mathrm{LO}}}
  W_{i_y,i_\tau}^{(p_{\mathrm{LO}})}
   f \!\left(x^{(i_y)}, \xi_F^2  {Q^2}^{(i_\tau)} \right) +
   \right.
   \nonumber \\
  \left(\frac{\alpha_s\left(\xi_R^2 {Q^2}^{(i_\tau)} \right)\,
             }{2\pi}
   \right)^{p_{\mathrm{NLO}}}
  \left[
    \left( W_{i_y,i_\tau}^{(p_{\mathrm{NLO}})} + 2\pi  \beta_0 p_{\mathrm{LO}} \ln \xi_R^2
         \,W_{i_y,i_\tau}^{(p_{\mathrm{LO}})}
    \right)  f \!\left(x^{(i_y)}, \xi_F^2 {Q^2}^{(i_\tau)} \right)
    \right. \\
    \left.
     - \ln \xi_F^2 \,W_{i_y,i_\tau}^{(p_{\mathrm{LO}})}
     (P_0\otimes f) \!\left(x^{(i_y)}, \xi_F^2 {Q^2}^{(i_\tau)}
     \right) 
  \right] \Bigg\} \,, \nonumber
\end{eqnarray}
where $\beta_0 = (11 N_c - 2n_f)/(12\pi)$ and $N_c$ ($n_f$) is the number of colours (flavours).
Though this formula is
given for an $x$-space based approach, a similar formula applies for
moment-space approaches. Furthermore it is straightforward to extend
it to higher perturbative orders.

To obtain the full DIS cross-section a summation 
of the weights and the parton densities
over the contributing sub-processes is required.

 \subsection{The case of two incoming hadrons}
\label{sec:twohadrons}
In hadron-hadron scattering one can use analogous procedures but with one more dimension.
Besides $Q^2$, the weight grid depends on the momentum fractions of the first ($x_1$) and
second ($x_2$) hadrons.

%\subsubsection{Representing the weights in the case of two incoming hadrons}
The analogue of eq.~\ref{eq:WfinalxQ} is given by:
\begin{equation}
\label{eq:WfinalxQ_twohadrons}
W = \sum_p \sum_{l=0}^{\nsub} \sum_{i_{y_1}} \sum_{i_{y_2}} \sum_{i_\tau}
W_{i_{y_1},i_{y_2},i_\tau}^{(p)(l)} \, \left( \frac{\alpha_s\left({Q^2}^{(i_\tau)}\right)}{2\pi}\right)^{p}
F^{(l)}\left(x_1^{(i_{y_1})}, x_2^{(i_{y_1})},  {Q^2}^{(i_\tau)}\right),
\end{equation}
where 
%$p$ is the number of orders of $\as$ in the leading order process, 
$\nsub$ is the number of sub-processes
and the initial state parton combinations $F$ are specified in 
eqs.~\ref{eq:jetgenpdf}, \ref{Wprocesses} and \ref{Zprocesses}.

The combinations of the incoming parton densities 
(defining the number of sub-processes)
often can be simplified
by making use of the symmetries in the weights.
In the case of jet production only seven sub-processes are needed (see section~\ref{sec:jetsubprocesses}).
The case of $W$-boson and $Z$-boson production is treated in the Appendix A.
The case of $b$-quark production is discussed in ref.~\cite{Banfi:2007gu}.
%section \ref{sec:wzsubprocesses}.

An automated way to find the sub-processes is discussed in Appendix B.

% GPS: removed "Break-up" from title (too colloquial)
\subsubsection{Sub-processes for jet production in hadron-hadron collisions}
\label{sec:jetsubprocesses}
In the case of jet production in proton-proton collisions
the weights generated by the Monte Carlo program 
%as well as the PDFs
can be organised in seven possible initial-state combinations of partons:
\begin{eqnarray}
\label{eq:jetgenpdf}
\mathrm{gg}: \;\; F^{(0)}(x_{1}, x_{2}; Q^{2}) &=& G_{1}(x_{1})G_{2}(x_{2})\notag\\
\mathrm{qg}: \;\; F^{(1)}(x_{1}, x_{2}; Q^{2}) &=& \left(Q_{1}(x_{1})+
                  \overline Q_{1}(x_{1})\right) G_{2}(x_{2})\notag\\
\mathrm{gq}: \;\; F^{(2)}(x_{1}, x_{2}; Q^{2}) &=&  G_{1}(x_{1})\left(Q_{2}(x_{2})+
                  \overline Q_{2}(x_{2})\right)\notag\\
\mathrm{qr}: \;\; F^{(3)}(x_{1}, x_{2}; Q^{2}) &=&  Q_{1}(x_{1}) Q_{2}(x_{2})
                                        + \overline Q_{1}(x_{1}) \overline Q_{2}(x_{2}) -D(x_{1}, x_{2}) \notag\\
\mathrm{qq}: \;\; F^{(4)}(x_{1}, x_{2}; Q^{2}) &=& D(x_{1}, x_{2}) \notag\\
\mathrm{q\bar q}: \;\; F^{(5)}(x_{1}, x_{2}; Q^{2}) &=& \overline D(x_{1}, x_{2})\notag\\
\mathrm{q\bar r}: \;\; F^{(6)}(x_{1}, x_{2}; Q^{2}) &=& Q_{1}(x_{1}) \overline Q_{2}(x_{2})
                   + \overline Q_{1}(x_{1}) Q_{2}(x_{2})       -\overline D(x_{1}, x_{2}),
\end{eqnarray}
where $g$ denotes gluons, $q$, quarks and $r$, quarks of different flavour, $q \neq r$
and we have used the generalised PDFs defined as:
\begin{eqnarray}
 G_{H}(x) = f_{0/H}(x,Q^{2}), &&
 Q_{H}(x) = \sum_{i = 1}^{6} f_{i/H}(x,Q^{2}), \;\;
 \overline Q_{H}(x) = \sum_{i = -6}^{-1} f_{i/H}(x,Q^{2}), \nonumber \\
 D(x_{1}, x_{2}) &=& \mathop{\sum_{i = -6}^{6}}_{i\neq0} f_{i/H_1}(x_{1},Q^2) f_{i/H_2}(x_{2},Q^{2}), \\
 \overline D(x_{1}, x_{2}) &=&
  \mathop{\sum_{i = -6}^{6}}_{i\neq0} f_{i/H_1}(x_{1},Q^{2}) f_{-i/H_2}(x_{2},Q^{2}), \nonumber \;\;
\end{eqnarray}
where $f_{i/H}$ is the PDF of flavour $i=-6 \dots 6$ for hadron $H$
and $H_1$ ($H_2$) denotes the first or second hadron.\footnote{
In the above equation and in the following we follow the standard PDG Monte Carlo numbering
scheme \cite{Eidelman:2004wy}, where gluons
are denoted as $0$, quarks have values from $1$-$6$ and anti-quarks have the corresponding
negative values.}

 \subsection{Including scale dependence in the case of two incoming hadrons}
It is again possible to choose arbitrary renormalisation and
factorisation scales. Specifically for NLO accuracy:
\begin{eqnarray}
  \label{eq:Wfinalxi_twohadrons}
  W(\xi_R, \xi_F) = \sum_{l=0}^{\nsub-1} \sum_{i_{y_1}} \sum_{i_{y_2}} \sum_{i_\tau}
  \left\{
  \left(\frac{\alpha_s\left(\xi_R^2 {Q^2}^{(i_\tau)}\right)\,
               }{2\pi}
     \right)^{p_{\mathrm{LO}}}
     W_{i_{y_1},i_{y_2},i_\tau}^{(p_{\mathrm{LO}})(l)}
  F^{(l)}\left(x_1^{(i_{y_1})}, x_2^{(i_{y_1})}, \xi_F^2{Q^2}^{(i_\tau)}\right)
    +
    \right.
   \nonumber \\
  \left(\frac{\alpha_s\left(\xi_R^2 {Q^2}^{(i_\tau)} \right)\,
             }{2\pi}
   \right)^{p_{\mathrm{NLO}}}
  \left[
    \left(
      W_{i_{y_1},i_{y_2},i_\tau}^{(p_{\mathrm{NLO}})(l)}
      + 2\pi  \beta_0 p_{\mathrm{LO}} \ln \xi_R^2
         \,
         W_{i_{y_1},i_{y_2},i_\tau}^{(p_{\mathrm{LO}})(l)}
    \right)
    F^{(l)}\left(x_1^{(i_{y_1})}, x_2^{(i_{y_1})}, \xi_F^2{Q^2}^{(i_\tau)}\right)
    \right. \\\left.
     - \ln \xi_F^2 \,
     W_{i_{y_1},i_{y_2},i_\tau}^{(p_{\mathrm{LO}})(l)}
     \left(
     F^{(l)}_{q_1 \to P_0\otimes q_1}\left(x_1^{(i_{y_1})}, x_2^{(i_{y_1})},
       \xi_F^2{Q^2}^{(i_\tau)}\right) +
     F^{(l)}_{q_2 \to P_0\otimes q_2}\left(x_1^{(i_{y_1})}, x_2^{(i_{y_1})},
       \xi_F^2{Q^2}^{(i_\tau)}\right)
     \right)
  \right] \Bigg\} \,, \nonumber
\end{eqnarray}
where $F^{(l)}_{q_1 \to P_0\otimes q_1}$ is calculated as $F^{(l)}$,
but with $q_1$ replaced with $P_0 \otimes q_1$, and analogously for
$F^{(l)}_{q_2 \to P_0\otimes q_2}$.

%%%%%%%%%%%%%%%%%%%%%%%%%%%%%%%%%%%%%%%%%%
\subsection{Reweighting to a different center-of-mass energy}
\label{sec:cmsreweighting}
From a weight grid $W$ calculated at a particular 
centre-of-mass energy 
$\sqrt{s}$ it is also possible to calculate a cross-section at
a different centre-of-mass energy $\sqrt{s^\prime}$ by using
transformed parton momentum fractions $x^\prime_{1/2}$
and adding a flux factor in the cross-section convolution
as given by eq.~\ref{eq:WfinalxQ_twohadrons} or eq.~\ref{eq:Wfinalxi_twohadrons}:
\begin{eqnarray}
\label{eq:cmsreweightW}
W^\prime( \xi_R, \xi_F) = \frac{s}{s^\prime} \; W( \xi_R, \xi_F),
\end{eqnarray}
and the momentum fractions $x_{1/2}$ in the generalised parton densities
$F(x_1, x_2, Q^2)$ are replaced by:
\begin{eqnarray}
\label{eq:cmsreweightX}
x^\prime_{1,2} = \frac{\sqrt{s}}{\sqrt{s^\prime}} \; x_{1,2}. 
\end{eqnarray}
When $\sqrt{s'} < \sqrt{s}$ it can occur that $x^\prime_{1}>1$ or
$x^\prime_{2}>1$, in which case the parton densities should be set to
zero.
One should be aware that a jet transverse momentum that corresponds to
moderate $x$ values with centre-of-mass energy $\sqrt{s}$ (and
correspondingly low density of grid points in $x$) may correspond to
large $x$ when using a smaller $\sqrt{s'}$. In such cases, it can
happen that the low density of grid points in $x$ is no longer
sufficient, given that PDFs vary more rapidly at large $x$ than at
moderate $x$.

Special care is also needed when taking $\sqrt{s'} > \sqrt{s}$ insofar
as there will be kinematic regions accessible with the larger
$\sqrt{s'}$ values that were not probed at all in the original NLO
calculation at centre-of-mass energy $\sqrt{s}$.
As a concrete example, with $\sqrt{s'}=14\,\mathrm{TeV}$, there can be
events with three jets having respectively $p_T = 6,4,2
\,\mathrm{TeV}$. Such events contribute to the inclusive jet spectrum
at $p_T = 4\TeV$. However, taking a grid calculated with
$\sqrt{s}=10\,\mathrm{TeV}$ (where such events are kinematically
disallowed) and using it to determine the inclusive jet spectrum with
$\sqrt{s'}=14\,\mathrm{TeV}$, this kind of contribution will be left
out.

%\newpage
\section{Technical implementation}
\label{sec:technical}
To test the scheme discussed above, the NLO QCD Monte Carlo programs
\NLOJET\cite{Nagy:2003tz,*Nagy:2001fj,*Nagy:2001xb} for jet production
and \MCFM\cite{MCFM1,MCFM2} for the production of $W$- and $Z$-boson are used.
To illustrate the performance of the method jet and  $W$- and $Z$-boson production
are used as examples. However, it is worth noting that these
these two programs give access to many of the NLO QCD calculations presently
available.

The weight grid $W_{i_{y_1},i_{y_2},i_\tau}^{(p)(l)}$ 
of eq.~\ref{eq:WfinalxQ_twohadrons}
is filled (for each cross-section bin)
in the user module of the NLO program, where one has access to the event
weights and the partons' momenta. This object is called ``grid'' in the following.
At this point the cross-section definition is specified
and the physical observables that are being studied 
are defined (e.g. using a jet algorithm).

The weight grid for each value of the observable in question is represented
as a multidimensional object with one dimension each for $x_1$, $x_2$ and $Q^2$,
one for the sub-process in question, and one for the order in $\alpha_s$. 
The task is to store the weight grid in such a way that as little memory as possible
is used and the information can be extracted in a fast way.
In the following several options to reduce the necessary memory are discussed:

The simplest structure for a software implementation of the weight grid
is a multidimensional array (for $x_1$, $x_2$ and $Q^2$), like the 
\texttt{TH3D}-class available in the \ROOT analysis framework.

The overhead of storing empty bins can be  largely reduced by calculating the
$x_1$, $x_2$ and $Q^2$  boundaries of the weight grid using the NLO QCD program in a special
run before the actual filling step. At the beginning of the
filling step the adjusted boundaries of the weight grids are then read-in
and an optimised weight grid is constructed.

Since the rectilinear region bounded by limits in
$x_{1}$, $x_2$ and $Q^2$  may contain many phase-space points that are unoccupied,
additional memory can be saved by using methods to avoid storing elements in the weight grid
that are not filled. Since the occupied regions are continuous, but irregular,
grid formats for truly sparse matrices
%(such as the Harwell-Boeing format~\cite{harwelboeing}) %have been rejected in favour
(such as the Harwell-Boeing format) %have been rejected in favour
 are not used. Instead
a custom format is favoured where the grid, lower- and upper-limits in each dimension are stored
along with all the elements in between. 

\begin{figure}[htp]
\begin{minipage}[b]{7cm}
\includegraphics[width=7cm]{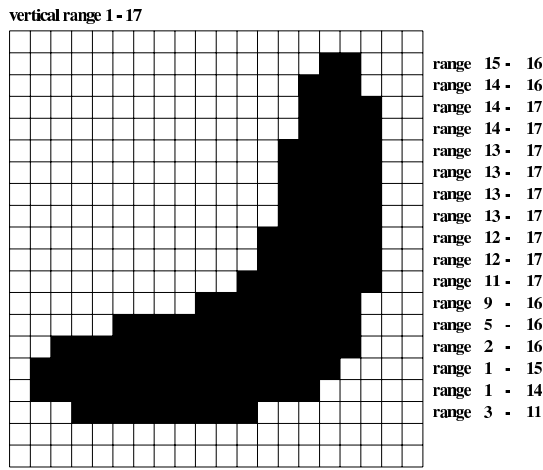}
\end{minipage}
\hspace{1cm}\begin{minipage}[b]{7cm}
\caption{An example of the custom two-dimensional sparse structure. Rows and columns are numbered from 0
 to 20
from the top left. The elements with data members are shown filled,
only rows 1 to 17 have data members, for each row the columns that have data members are shown on the right.
A total of 117 elements, from the maximum of 400 elements are stored, along with the single pair of row-
range
delimiters, and the 17 pairs of column range delimiters for each of the individual rows.}
\label{sparse}
\end{minipage}
\end{figure}

This is illustrated in Fig.~\ref{sparse} for a
simple two-dimensional grid. For the three-dimensional structure, each of the row-column
elements would itself be a column with its own lower
and upper range delimiters. The resulting saving of memory is usually around a factor of
four, even after taking into account the additional storage for the range delimiters.\footnote{
If additional savings are required in the future, packing the range delimiters for
each sparse one dimensional structure into a single integer will halve this additional
overhead, but will slightly increase the access time due to the unpacking.}

%%%%%%%%%%%%%%%%%%%%%%%%%%%%%%%%%%%%%%%%%%%%%%%%%%%%%%%%%%%%%%%%%%%%%
%%%%%%%%%%%%%%%%%%%%%%%%%%%%%%%%%%%%%%%%%%%%%%%%%%%%%%%%%%%%%%%%%%%%%
Since the grid itself knows the index of the first and last filled element in each row,
column etc.,
%the grid can immediately
%say which are the indices of the first filled row, or the first filled column for a given row,
%etc.
it is possible to only iterate over those elements of the grid that contain data. Similarly
when interrogating the grid for the value of an element, it is possible to ascertain
whether the element is in the occupied, or unoccupied region of the grid and return the
value of the filled element if filled, or 0 otherwise. This makes accessing the unfilled
members of the grid much faster than otherwise.

The actual implementation for the grid\footnote{ 
The complete code including the interfaces to \NLOJET and \MCFM 
is available from http://svn.hepforge.org/applgrid.}
involves a number of related
classes written in \texttt{C++}\footnote{A \texttt{FORTRAN} interface
is also available so that the basic functionality can be accessed from within user \texttt{FORTRAN}
code.}.
The grid for a given cross-section is represented by a concrete instance of a master class,
{\tt appl::grid}. This class has a number of constructors that allow the cross-section it will
calculate to be defined in terms of a fixed number of regular or variable width bins in the
cross-section observable.

For each bin in the observable, the master class has a number of instances of an internal
class -- one for each order of $\alpha_s$ -- so that for a cross-section with $10$ bins, with
contributions at leading order and next-to-leading order, the master class would contain
a total of $20$ instances of the internal class.

This internal class, {\tt appl:igrid},
encodes all the information required to create the cross-section, at one particular order,
for that bin. The class contains the $x$-to-$y$, $y$-to-$x$ and $Q^2$-to-$\tau$, $\tau$-to-$Q^2$
transform pairs, and a subclass that encodes information on how to generate the $N$
generalised internal sub-processes for the particular interaction from the basic
parton distribution functions. It also contains instances of the sparse grid
class in $x_1$, $x_2$ and $Q^2$ described above, for each of the $N$ sub-process.

When requested to perform the convolution, the master class calls the convolute
method of the subclass for each order of the cross-section in each bin.
The convolute method of the subclass performs the convolution over $x_1$, $x_2$ and $Q^2$
for each of the sub-processes.

For each bin in the observable, the master class takes the cross-sections from the subclasses
for each order from each bin and adds them to arrive at the final cross section for that bin.

The subclass for the generalised internal sub-processes are very basic classes which
encode the number of sub-processes, i.e. seven in the case of jet production, and twelve in the
case of $Z$-boson production, and simply take the 13 parton distribution values for
each incoming hadron at a given scale, and generate the $N$ internal processes from these.

%When writing the grid to the file, 
When the grid is saved to a \ROOT file\footnote{Technically,
the grid is transformed to \texttt{TH3D}-histograms that are stored in the output \ROOT file.}
the master class encodes the complete status of
the internal grids, which transform pair, and which sub-process is required etc., so that
once reading from the file, everything required to calculate the cross section
(e.g. sub-process definition, CKM matrix elements etc.) is available. 
In this way all information to perform the cross-section calculation is available
from the output file from a single function call by the user
and the only additional information required is an input function
for generating the PDFs and another one for calculating $\alpha_s$.
We use the \texttt{HOPPET} program \cite{Salam:2008qg} to calculate the
DGLAP splitting functions needed for the cross-section convolution when 
the renormalisation and factorisation is varied (see eq.~\ref{eq:Wfinalxi}).

%\footnote{
%This is the form required for use in a fit of the parton distributions where
%the PDF and $\alpha_s$ will be changing for  each evaluation of
%the cross-section and it wraps the complete generation of the cross-section
%into a single function call to greatly simplify the user code structure and
%interaction with the grid class.}.
%
%
%
%Many additional choices and parameters are possible, and these too are encoded in the grid,
%e.g. the choice of the coordinate transform function (eq.~\ref{eq:ytau}), where the user can choose
%between any of those defined or add their own if required.
%
% already said:
%
%After a call to initialise the grid from the file,
%the cross-section can be calculated via a method that % reads all necessary information
%performs the convolution from the structure in memory.
%This method reads the perturbative coefficients for each sub-process
%and observable bin from the structure in memory and  multiplies them with the arbitrary
%$\as$ and PDF from the input functions according to eq.~\ref{eq:WfinalxQ_twohadrons}.
%This convolution execution is very fast (see section~\ref{sec:cpuperformance})
%and so it is appropriate for use in a PDF fit.

All the various choices in the weight grid architecture
and other information needed to calculate the cross-section
are encoded in the output file. 
They are described in the following:
\begin{itemize}
\item The centre-of-mass energy at which the weight grid has been produced.
\item 
The choice of the coordinate transform function. By default the form of 
eq.~\ref{eq:ytau} is used. However, any other function can be provided by the user.
\item
The interpolation order as given by eq.~\ref{eq:interp}.
\item 
The number of grid points to be used for each dimension $x_1$, $x_2$ and $Q^2$.
\item 
The definition of the sub-processes via a $13$ x $13$ matrix.
\item 
The CKM matrix elements or other constants needed to calculate the cross-sections.
\item 
The required number of the points on the grid can be optionally
reduced with the aid of 
reweighting factor in the filling step. This
flattens out the PDF in the region where it is steeply falling.
%A tag for the PDF weight function is stored together with the weight grid so that
%it can be taken into account during the cross-section convolution.

By default the following functional form is used for the reweighting\footnote{Such a PDF
reweighting was first introduced in ref.\cite{fastnlo}.}:
\begin{equation}
            w(x) = x^{a_1} \; ( 1 - 0.99 \; x )^{a_2}.
\label{eq:pdfweight}
\end{equation}
The parameter $a_1$ can be adjusted to flatten out the change of the PDF at low-$x$
while the parameter $a_2$ can be optimised for the high-$x$ region. The factor $0.99$
prevents the weight from being zero for $x = 1$.

Reasonable values for the parameters $a_1$ and $a_2$
have been determined by fitting the sum of the up, down and gluon PDFs.
For the CTEQ6 PDFs \cite{Pumplin:2002vw},
values of $a_1 = -1.5$ to $-1.6$ and $a_2= 3.0$ to $3.4$ have been
found for the range $5 < Q < 5000$ \GeVx.
The variation comes from a slight dependence of the $a_1$ and $a_2$ parameters
on $Q^2$. 
For other PDFs, the results of the fit can be slightly different.
%, for example $a_1 = -1.7$ and $a_2 = 2.7$ for H12000ms \cite{h12000ms}.
%

The user can change the parameters or provide another functional form.

\end{itemize}
%\newpage
\section{Accuracy of the weight grids}
\label{sec:results}
The choice of the weight grid architecture depends on the required accuracy, on the exact cross-section definition
and on the available computer resources. 
For each possible application the weight grid architecture has to be carefully chosen
in order to achieve the required accuracy with the available computer memory
and computing time. 
For instance, for observables where the PDFs are steeply
falling, e.g. the inclusive jet cross-section at high transverse momentum in the forward region, a fine
grid in $x$ is needed. 
The memory usage of weight grids for one cross-section should be kept small
since, e.g. in global PDF fits, it might be necessary to read in a large number
of weight grids. In addition, the convolution time depends on the number of
grid nodes, and so keeping memory requirements 
as small as possible is in any case desirable. 
The number of points needed in the weight grid
is kept modest by using the higher-order interpolation functions of
eq.~\ref{eq:interp},  and optionally also by introducing
a PDF weight, as in eq.~\ref{eq:pdfweight} during the filling step,
or by using a sparse structure.

In the following, the influence of the grid architecture 
on the achievable accuracy in the cross-section calculation
is discussed. The computer memory use and execution speed are also investigated.
The production of jets and of $W$- and $Z$-bosons at LHC are used as examples. 

In our test runs, to be independent from statistical fluctuations
(which can be large, in particular in the NLO case),
in addition to the weight grid, reference histograms are filled 
using the NLO QCD calculation without weights in the standard way.
The result obtained from the weight grid is then compared to these reference histogram.

%%%%%%%%%%%%%%%%%%%%%%%%%%%%%%%%%%%%%%%%%%%%%%%%%%%%%%%%%%%%%%%%%%5
\begin{figure}[htbp]
\centering
\includegraphics[width=0.49\textwidth]{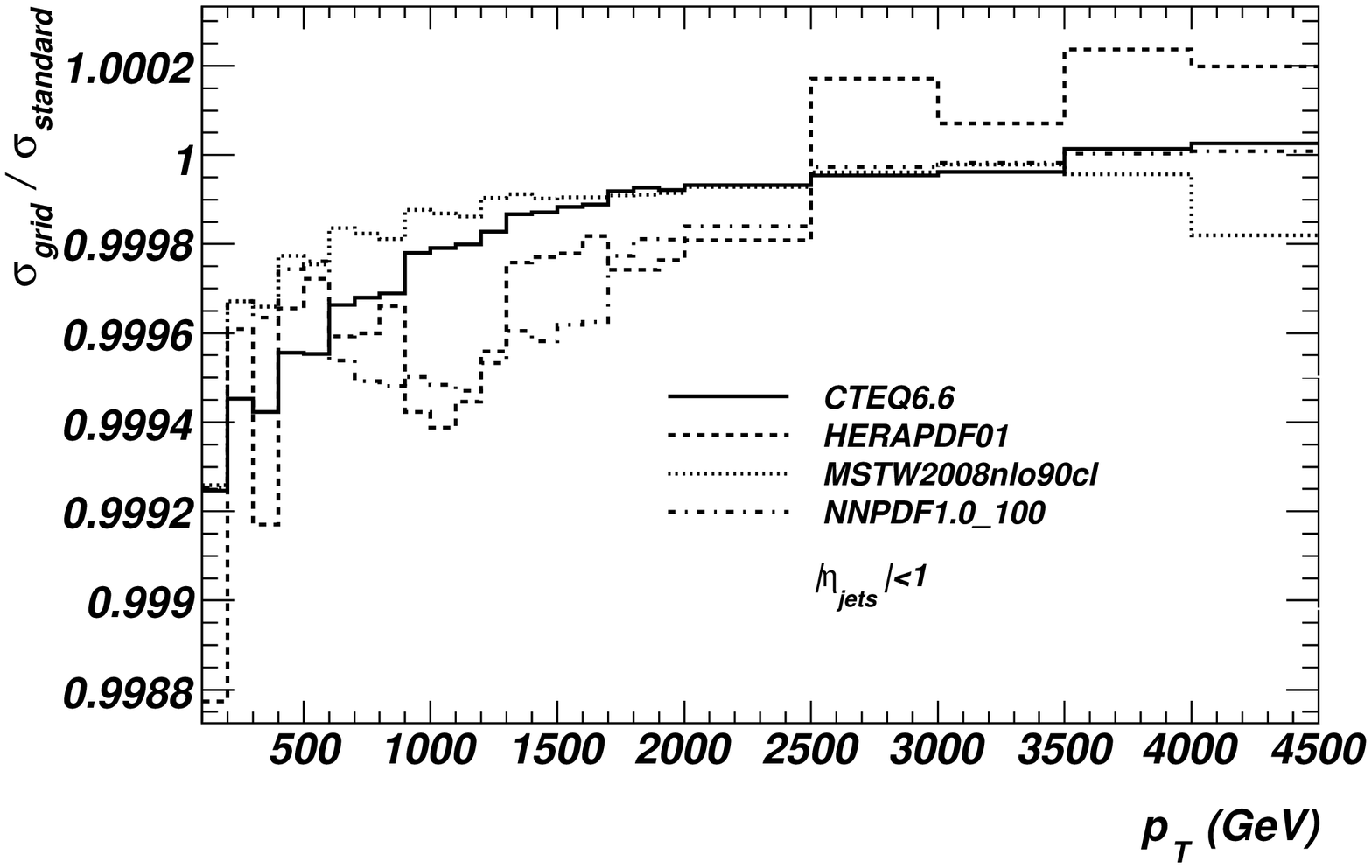}
\includegraphics[width=0.49\textwidth]{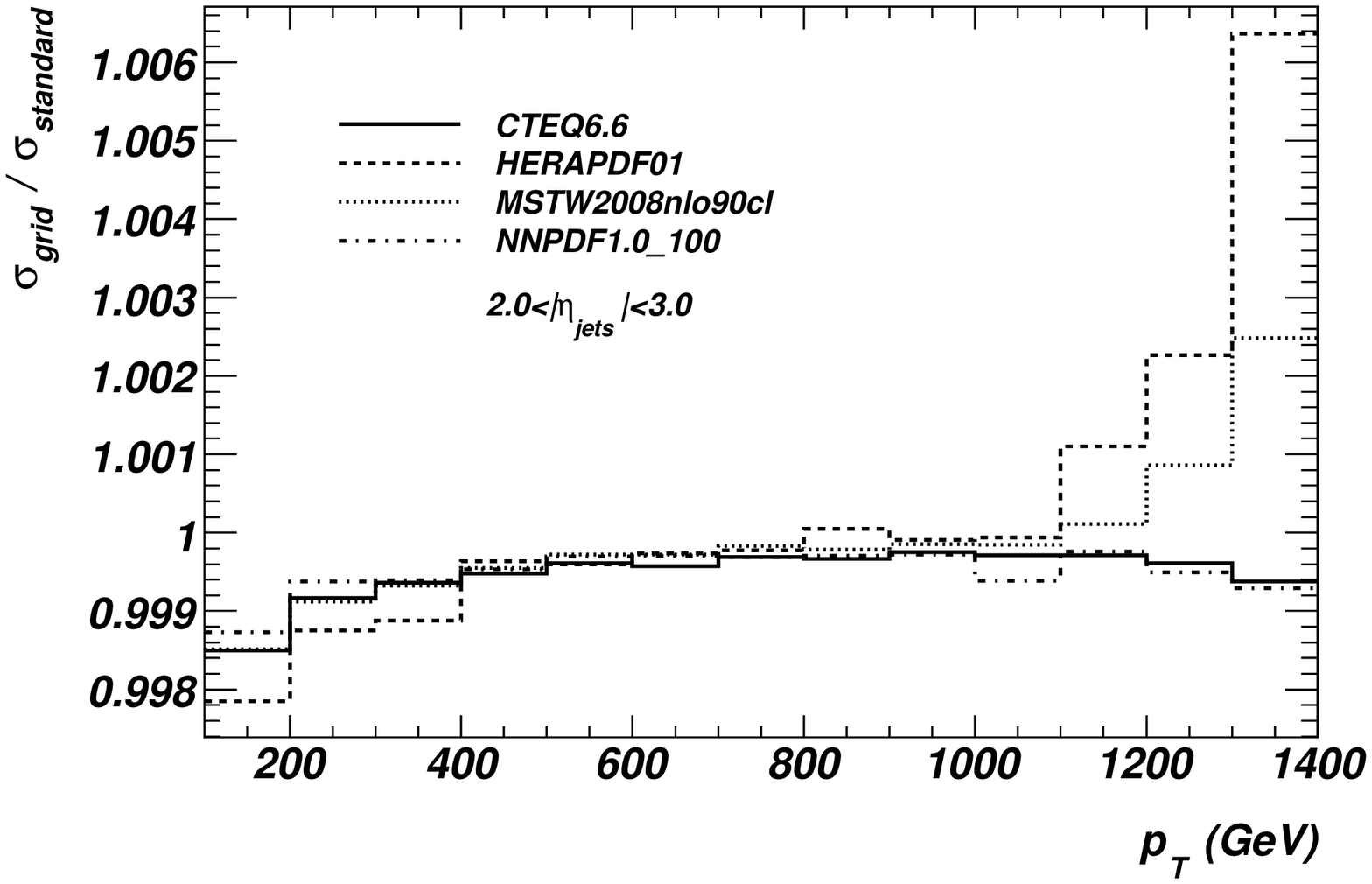}
\begin{picture}(0,0)
\put( -450,0){a)}
\put( -210,0){b)}
\end{picture}
\caption{
  Ratio of grid and standard calculations of the
  single inclusive jet \pt spectrum for $0 < y < 1$ (a) and for $2 < y
  < 3$ (b), for a variety of PDFs.
  The results are shown for the default weight-grid settings, 
  i.e.\  $30$ bins in $x$, $10$ bins in $Q^2$,
  a coordinate transform parameter $a=5$ and fifth order interpolation.
}
\label{fig:jetprodvarPDf}
\end{figure}
%%%%%%%%%%%%%%%%%%%%%%%%%%%%%%%%%%%%%%%%%%%%%%%%%%%%%%%%%%%%%%%%%%5
%%%%%%%%%%%%%%%%%%%%%%%%%%%%%%%%%%%%%%%%%%%%%%%%%%%%%%%%%%%%%%%%%%5
\begin{figure}[htbp]
\centering
\includegraphics[width=0.49\textwidth]{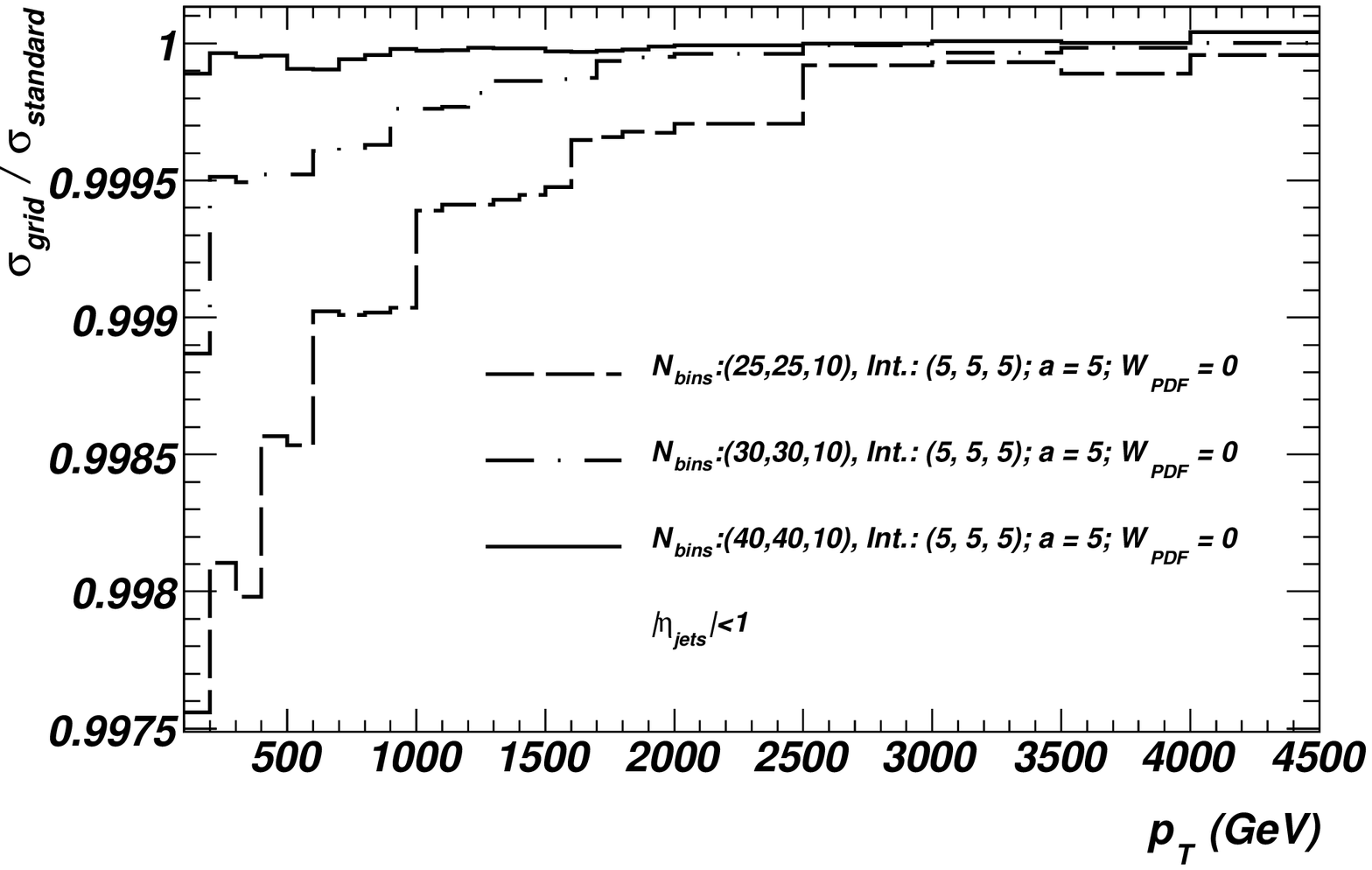}
\includegraphics[width=0.49\textwidth]{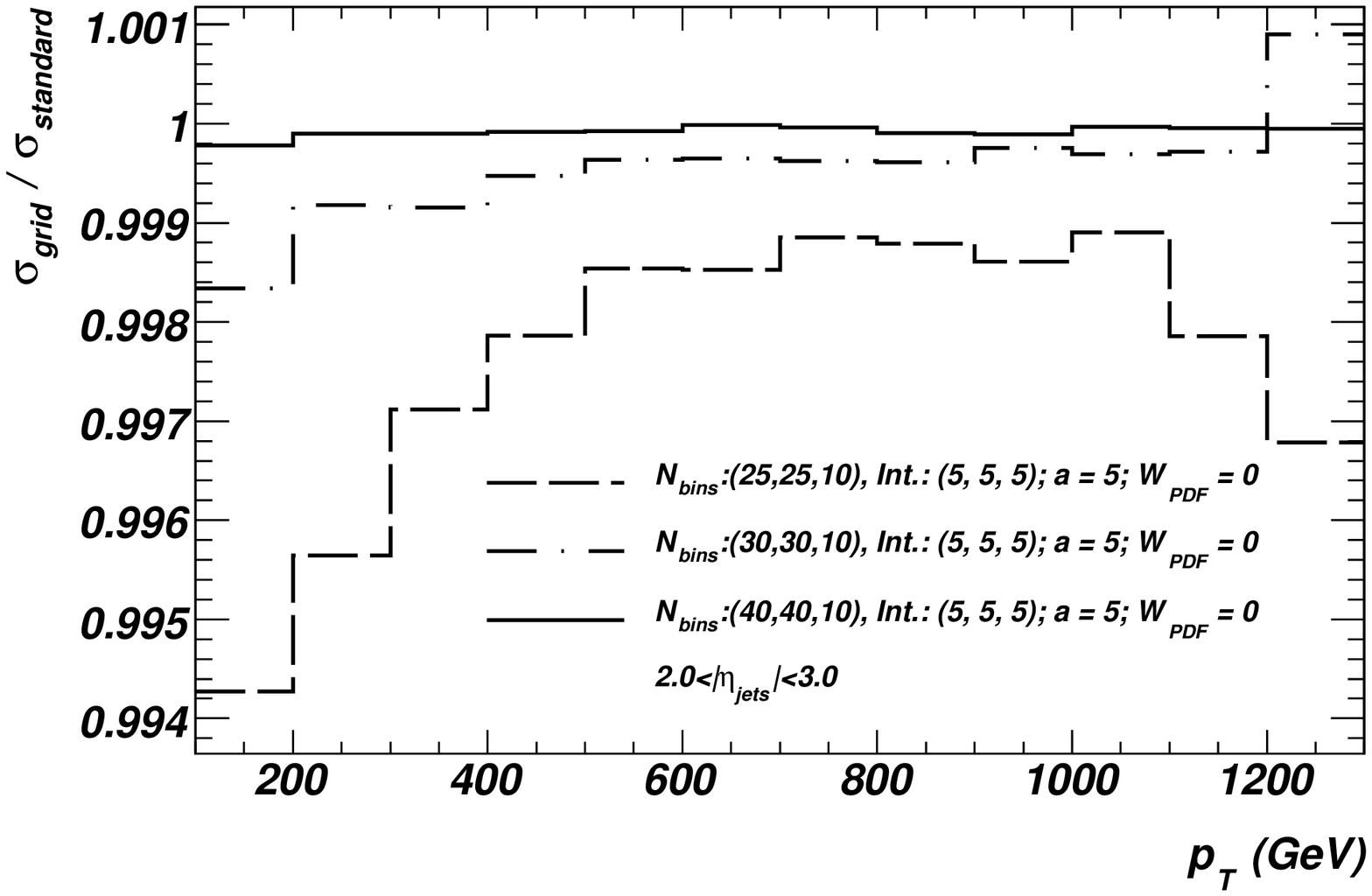}
\begin{picture}(0,0)
\put( -450,0){a)}
\put( -210,0){b)}
\end{picture}
\caption{
  Ratios of grid and standard calculations of the
  single inclusive jet \pt spectrum for $0 < y < 1$ (a) and for $2 < y
  < 3$ (b), illustrating the impact of varying the number of $x$-bins in
  the grid.
  All weight grids have $10$ bins in $Q^2$, a coordinate transform
  parameter $a=5$ and fifth order interpolation.  
  The PDF set is CTEQ6mE.  }
\label{fig:jetprodvarx}
\end{figure}
%%%%%%%%%%%%%%%%%%%%%%%%%%%%%%%%%%%%%%%%%%%%%%%%%%%%%%%%%%%%%%%%%%5

%%%%%%%%%%%%%%%%%%%%%%%%%%%%%%%%%%%%%%%%%%%%%%%%%%%%%%%%%%%%%%%%%%5
\begin{figure}[htbp]
\centering
\includegraphics[width=0.49\textwidth]{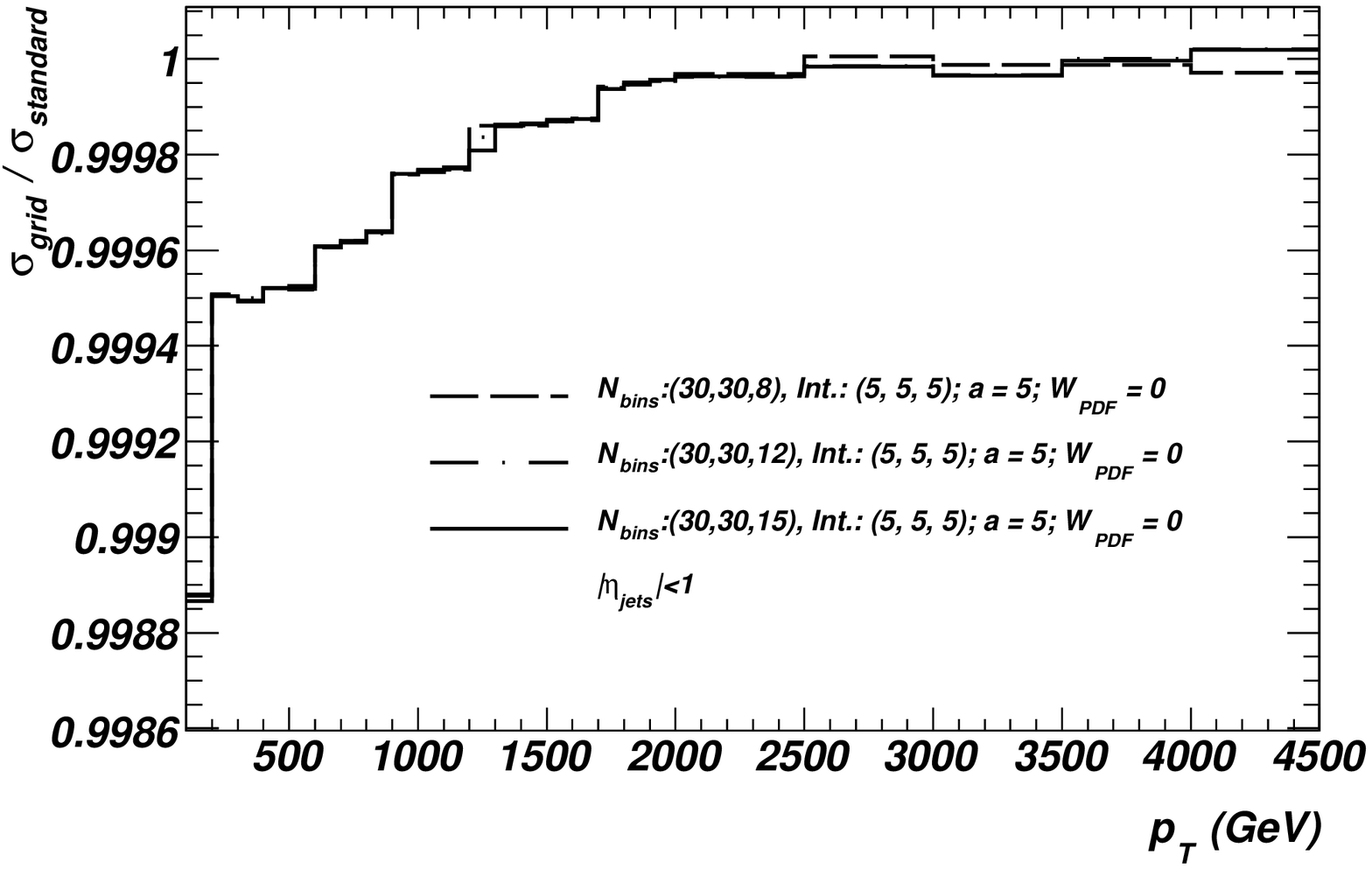}
\includegraphics[width=0.49\textwidth]{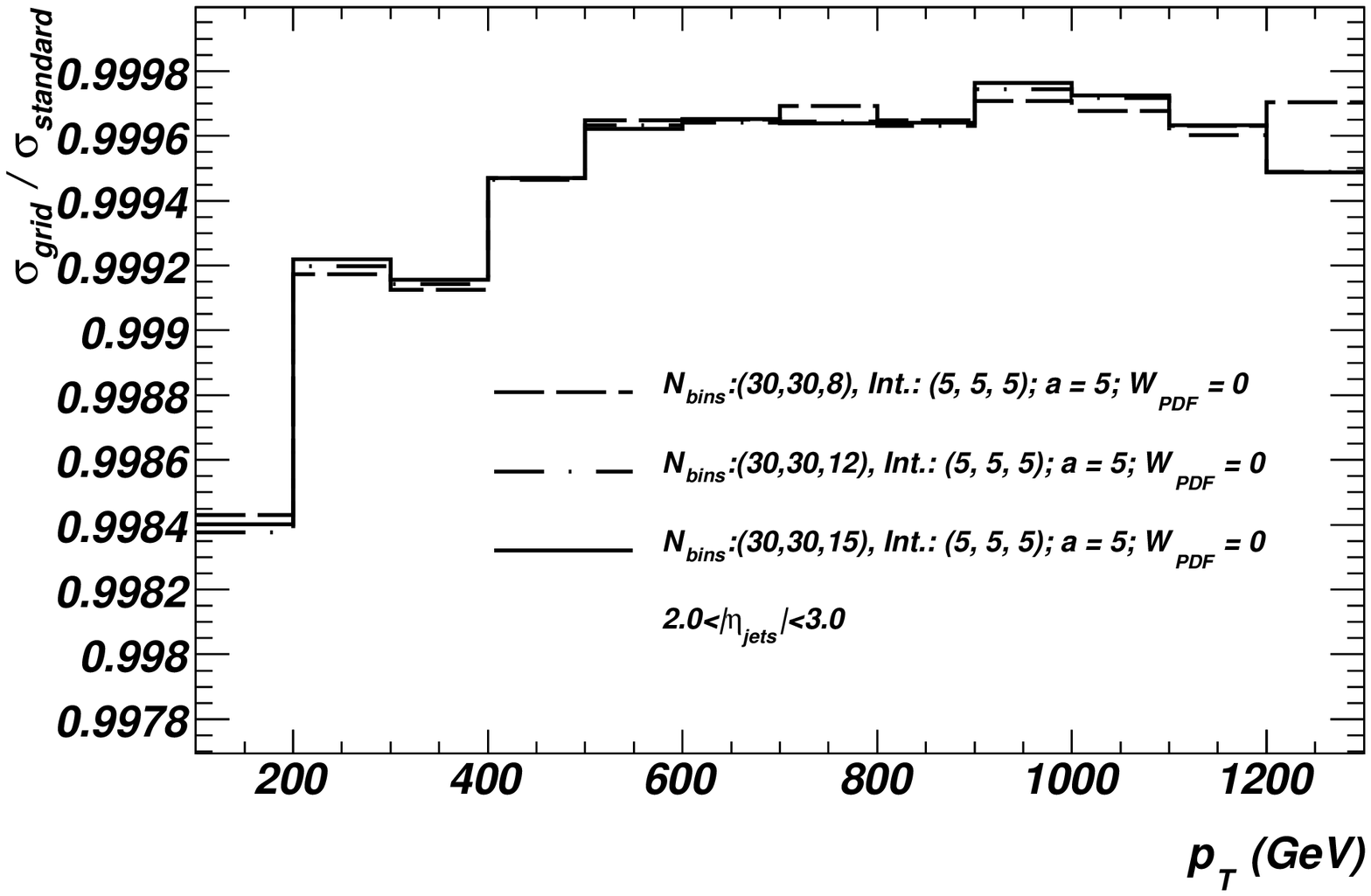}
\begin{picture}(0,0)
\put( -450,0){a)}
\put( -210,0){b)}
\end{picture}
\caption{
  Ratios of grid and standard calculations of the
  single inclusive jet \pt spectrum for $0 < y < 1$ (a) and for $2 < y
  < 3$ (b), illustrating the impact of varying the number of $Q^2$-bins in
  the grid.
  All weight grids have $30$ bins in $x$, a coordinate transform
  parameter $a=5$ and fifth order interpolation. The PDF set is
  CTEQ6mE.  }
\label{fig:jetprodvarQ2}
\end{figure}
%%%%%%%%%%%%%%%%%%%%%%%%%%%%%%%%%%%%%%%%%%%%%%%%%%%%%%%%%%%%%%%%%%5

%%%%%%%%%%%%%%%%%%%%%%%%%%%%%%%%%%%%%%%%%%%%%%%%%%%%%%%%%%%%%%%%%%5
\begin{figure}[htbp]
\centering
\includegraphics[width=0.49\textwidth]{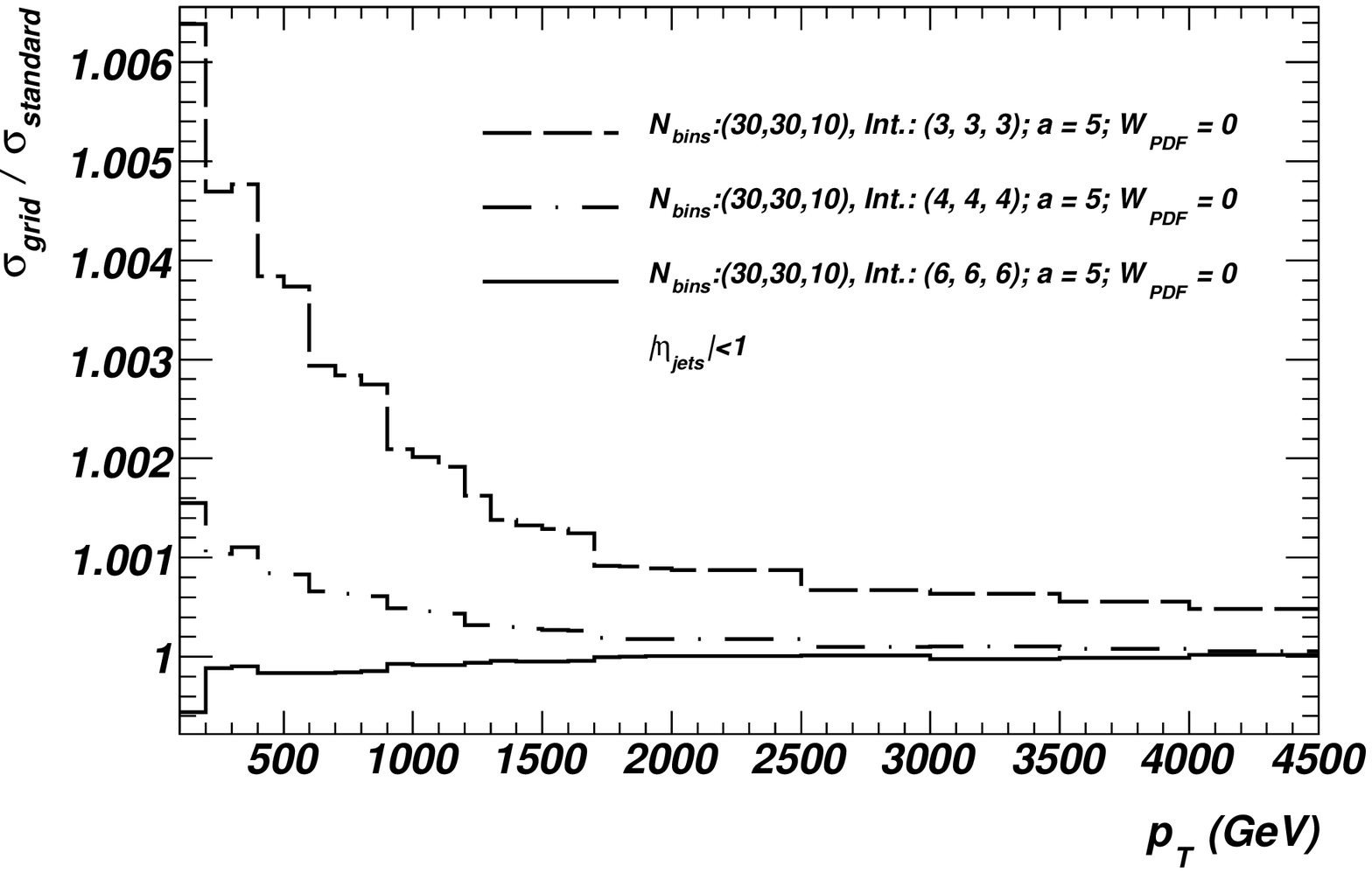}
\includegraphics[width=0.49\textwidth]{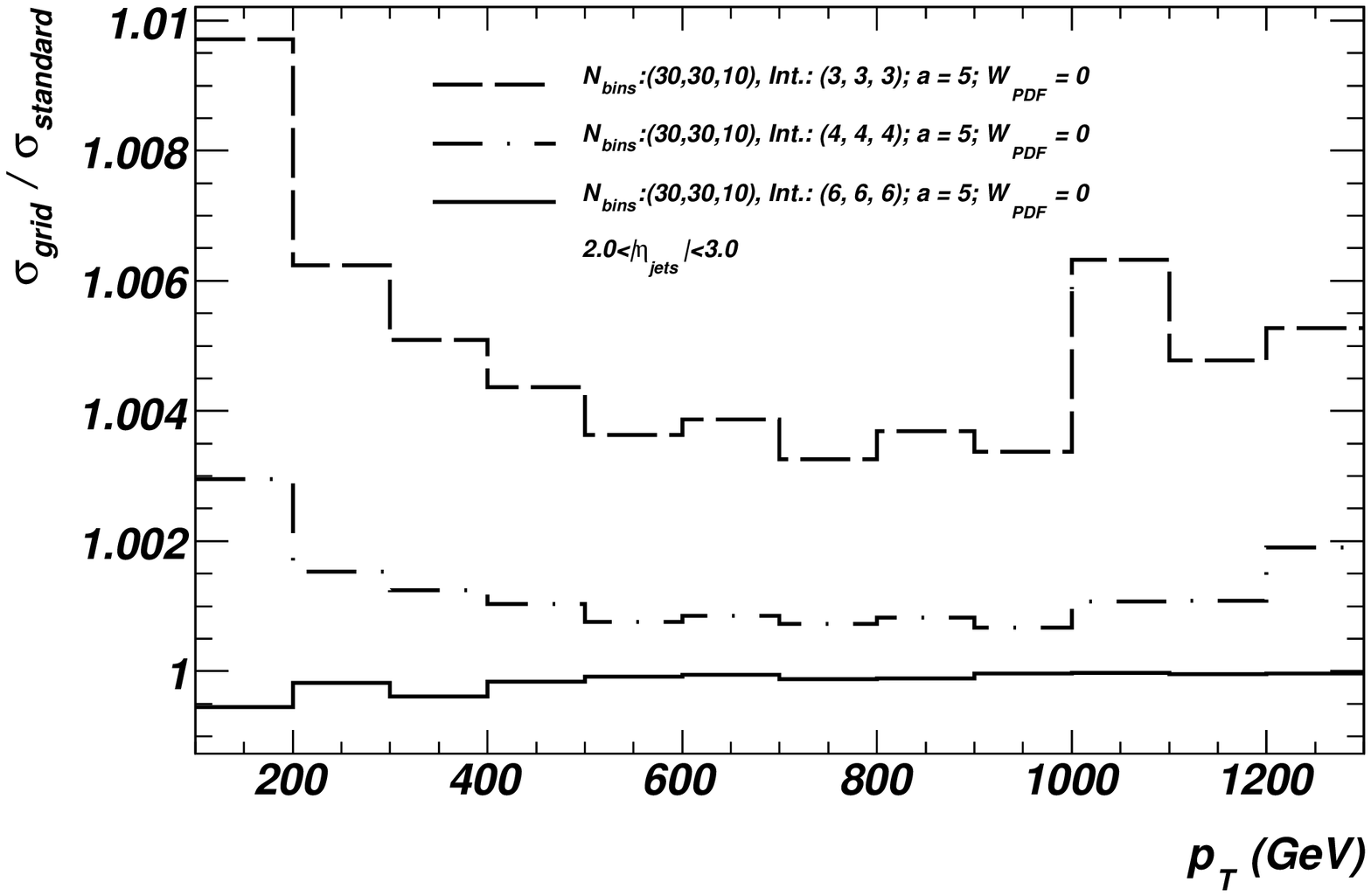}
\begin{picture}(0,0)
\put( -450,0){a)}
\put( -210,0){b)}
\end{picture}
\caption{
  Ratios of grid and standard calculations of the
  single inclusive jet \pt spectrum for $0 < y < 1$ (a) and for $2 < y
  < 3$ (b), illustrating the impact of varying the grid interpolation
  order. 
  All weight grids have $30$ bins in $x$, $10$ bins in $Q^2$ and a
  coordinate transform parameter $a=5$. The PDF set is CTEQ6mE.  
}
\label{fig:jetprodvarint}
\end{figure}
%%%%%%%%%%%%%%%%%%%%%%%%%%%%%%%%%%%%%%%%%%%%%%%%%%%%%%%%%%%%%%%%%%5

%%%%%%%%%%%%%%%%%%%%%%%%%%%%%%%%%%%%%%%%%%%%%%%%%%%%%%%%%%%%%%%%%%5
\begin{figure}[htbp]
\centering
\includegraphics[width=0.49\textwidth]{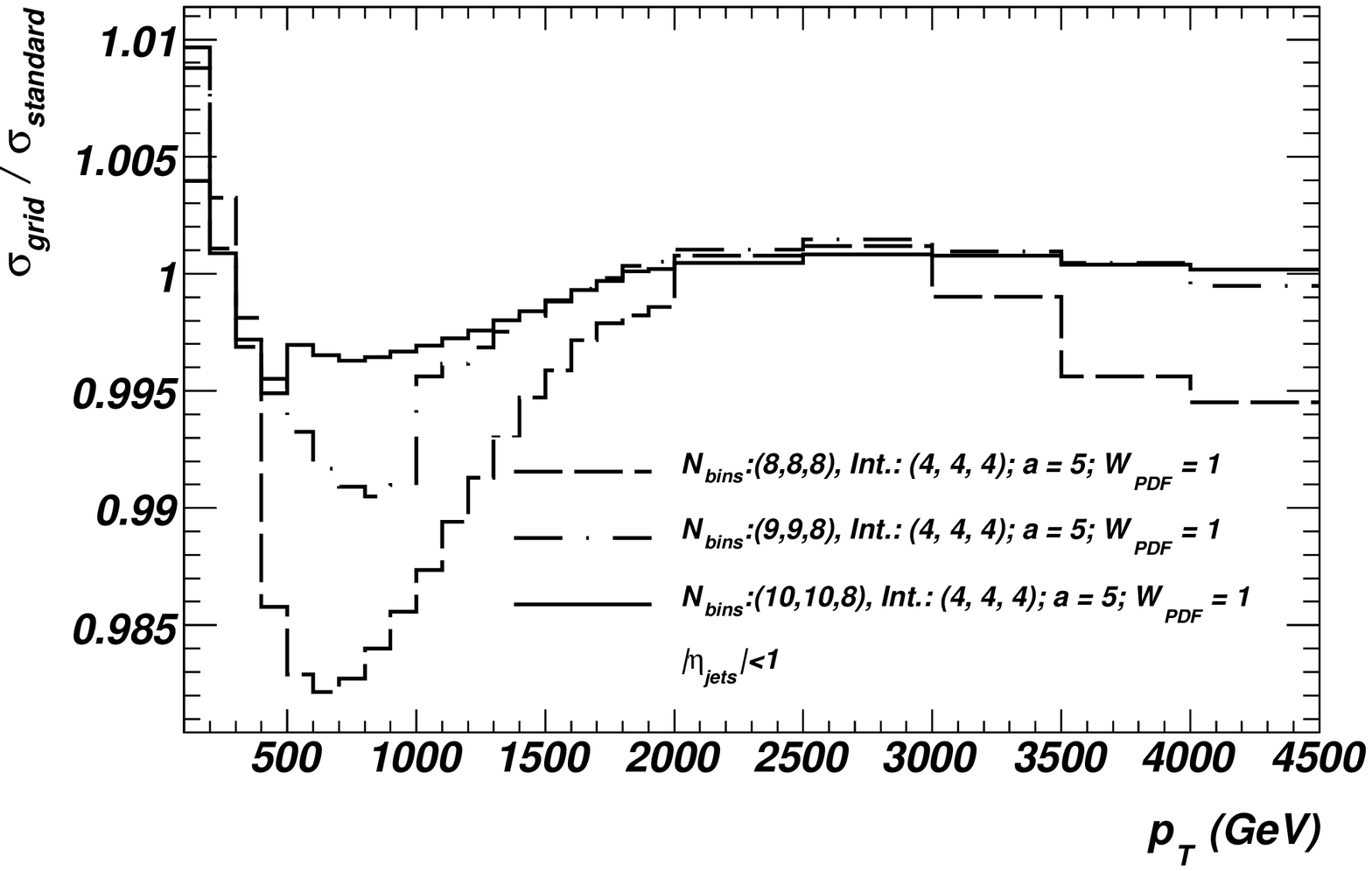}
\includegraphics[width=0.49\textwidth]{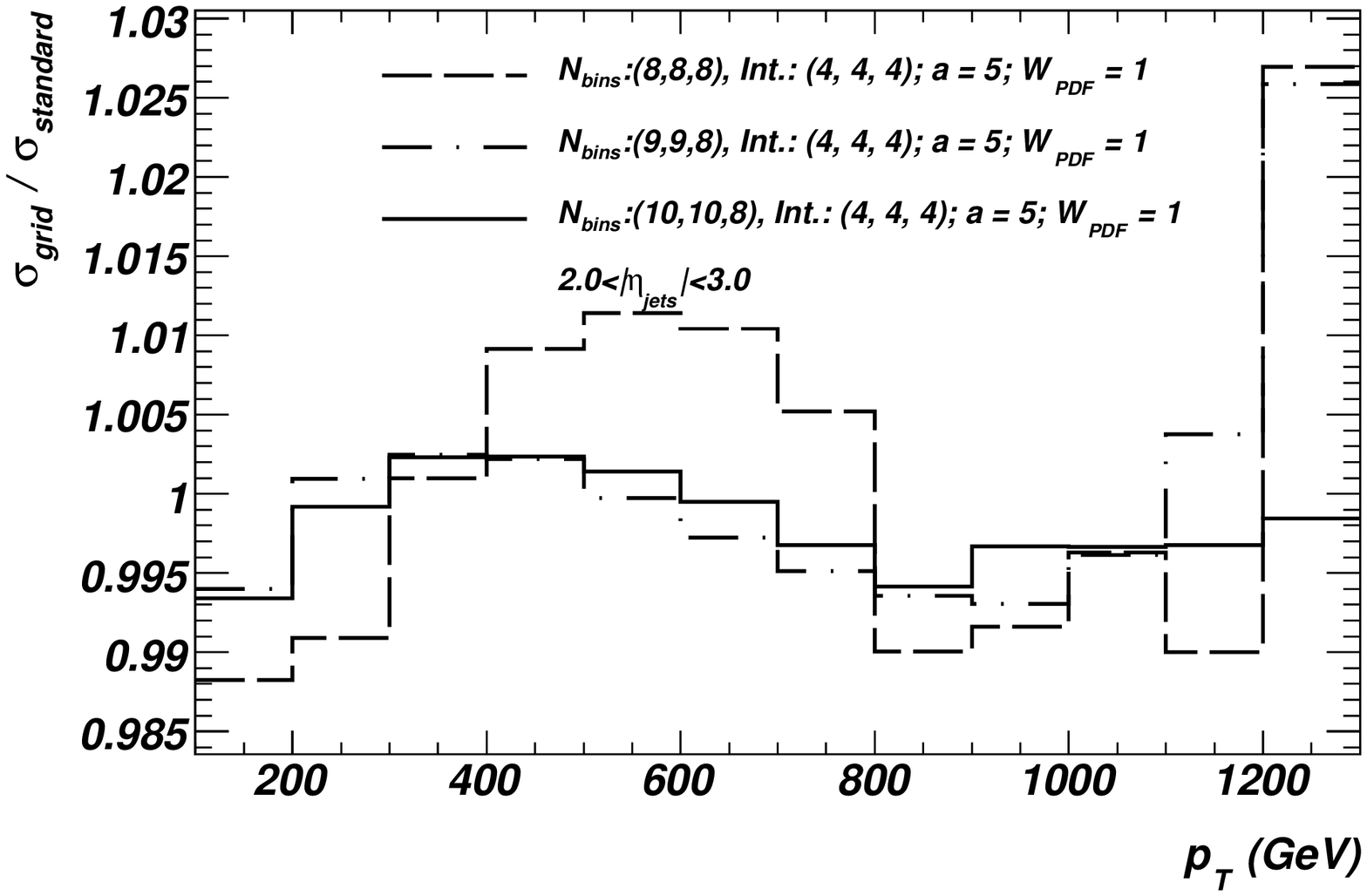}
\begin{picture}(0,0)
\put( -450,0){a)}
\put( -210,0){b)}
\end{picture}
\caption{
  Ratios of grid and standard calculations of the
  single inclusive jet \pt spectrum for $0 < y < 1$ (a) and for $2 < y
  < 3$ (b), illustrating the use of small grids with PDF
  reweighting. 
  The weight grids 
  have a low number of $x$-bins ($8$, $9$, $10$), $8$ bins in $Q^2$, 
  a coordinate transform parameter $a=5$, fourth order
  interpolation and  PDF reweighting with eq.~\ref{eq:pdfweight}.
  The PDF set is CTEQ6mE.  }
\label{fig:jetprodPDFweighting}
\end{figure}
%%%%%%%%%%%%%%%%%%%%%%%%%%%%%%%%%%%%%%%%%%%%%%%%%%%%%%%%%%%%%%%%%%5

%%%%%%%%%%%%%%%%%%%%%%%%%%%%%%%%%%%%%%%%%%%%%%%%%%%%%%%%%%%%%%%%%%5
\begin{figure}[htbp]
\centering
\includegraphics[width=0.49\textwidth]{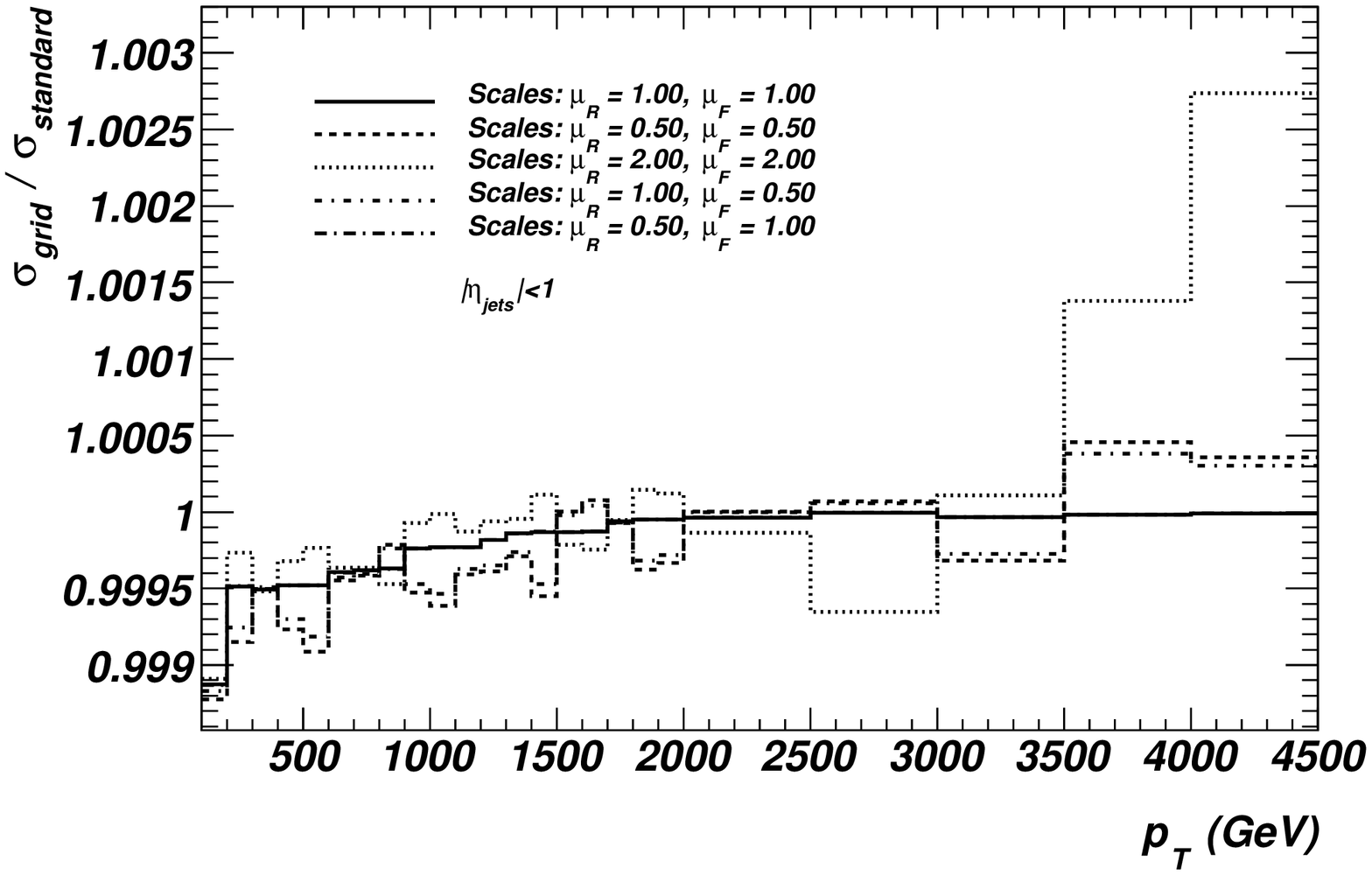}
\includegraphics[width=0.49\textwidth]{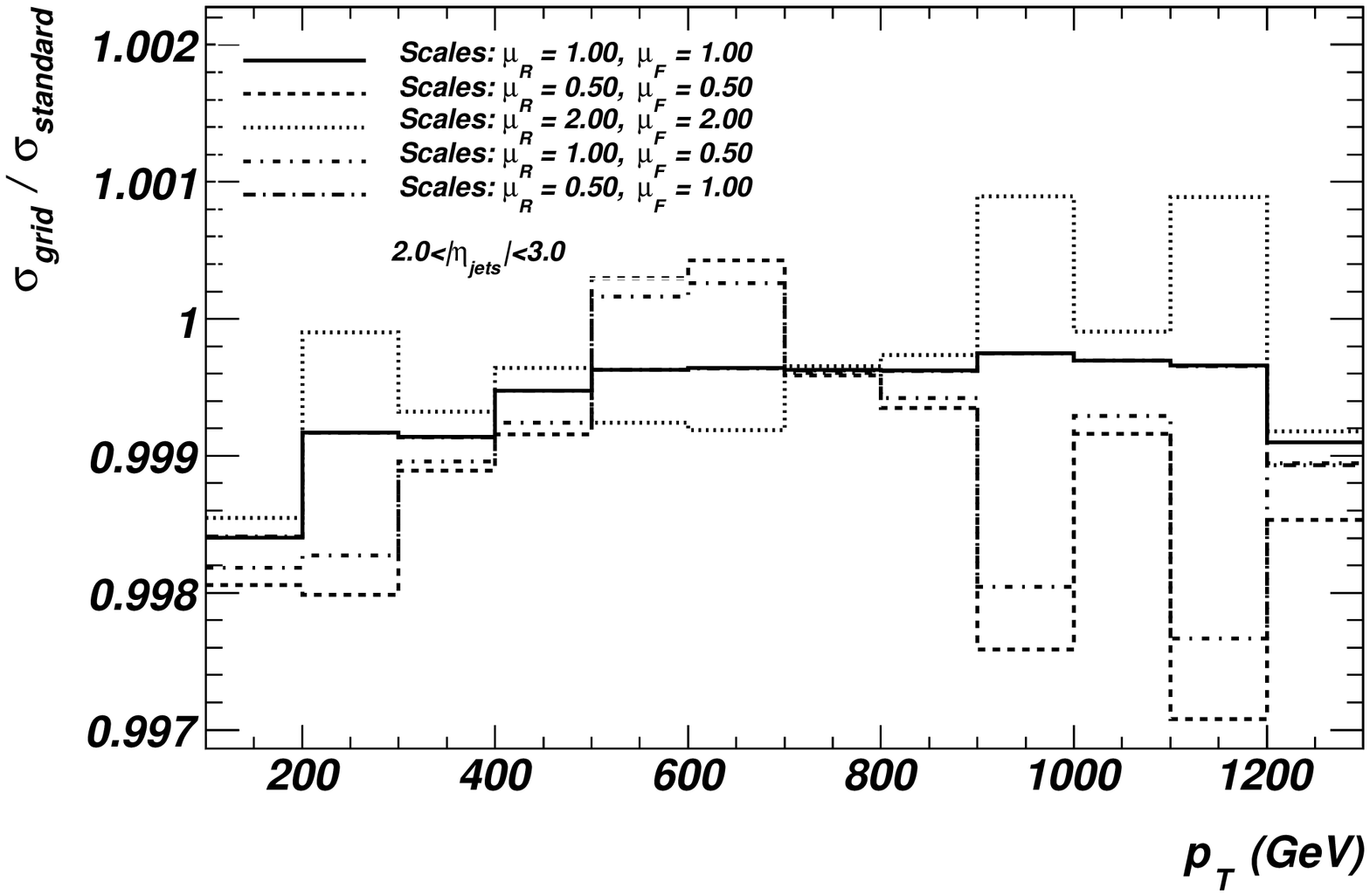}
\begin{picture}(0,0)
\put( -450,0){a)}
\put( -210,0){b)}
\end{picture}
\caption{
  Ratios of grid and standard calculations of the
  single inclusive jet \pt spectrum for $0 < y < 1$ (a) and for $2 < y
  < 3$ (b), with scale variation. 
  The default weight grid is used, with $30$ bins in $x$, $10$ bins in $Q^2$, a
  coordinate transform parameter $a=5$ and fifth order interpolation. 
  The grid results are based on {\em a posteriori}  variation of the
  renormalisation and 
  factorisation scales, using eq.~\ref{eq:Wfinalxi}, while the standard
  results have been obtained separately for each choice of
  renormalisation and factorisation scale.
  The PDF set is CTEQ6mE.  }
\label{fig:jetprodvarScale}
\end{figure}
%%%%%%%%%%%%%%%%%%%%%%%%%%%%%%%%%%%%%%%%%%%%%%%%%%%%%%%%%%%%%%%%%%5

\subsection{Jet production at hadron colliders}
\label{sec:jetsresults}
The single inclusive jet cross-section as a function of the jet transverse
momentum (\ptx) is calculated for jets in the central rapidity ($y$) region 
of $0< y <1$ and in the
forward rapidity region  of $2< y <3$.
%
%Jets are defined via the seedless cone jet algorithm as implemented in NLOJET++.
%Partons are combined to jets using the four-vector recombination scheme 
%and midpoints of the centroids between proto-jets are also used during the jet clustering. 
Jets are defined via the seedless cone jet algorithm as implemented
in \NLOJETx, which corresponds to the seedless algorithm of ref.~\cite{Blazey:2000qt} 
(or SISCone \cite{Salam:2007xv}), except for
small differences in the split--merge procedure which are irrelevant at this order.
The cone radius has been set to $R=0.7$, the overlap fraction
to $f=0.5$.\footnote{
  These choices are related to the fact that some of the \NLOJET runs were
  performed some time in the past. A modern cone-algorithm (in the
  class of those with a split--merge procedure) would be
  SISCone~\cite{Salam:2007xv}, and a value of $f=0.75$ would be
  recommended~\cite{Cacciari:2008gn}.
}
 The renormalisation and factorisation scales are set to $Q^2=p_{T,max}^2$,
where $p_{T,max}$ is the $p_T$ of the highest \pt jet in the required rapidity region\footnote{
Note that beyond LO the $p_{T,max}$ will in general differ from the \pt
of the other jets, so when binning an inclusive jet cross-section, the
\pt of a given jet may not correspond to the renormalisation scale
chosen for the event as a whole. For this reason separate
grid dimensions for the jet \pt and for the renormalisation scale are used. 
This requirement has been efficiently circumvented
in some moment-space approaches \cite{Kosower:1997vj}.}.

To discuss the dependence of the weight grid performance on the grid architecture,
a default weight grid is defined from which variations in a single parameter
are studied systematically. 
The default weight grid consists of $30$ bins in $x$ and $10$ bins in $Q^2$.
The points are distributed according to eq.~\ref{eq:ytau} with $a=5$ and  $5$th order
interpolation is used. No PDF reweighting (see eq.~\ref{eq:pdfweight}) is used.

The ratio of the cross-section calculated with the default weight grid to the
reference cross-section calculation is shown in Fig.~\ref{fig:jetprodvarPDf}
for the jet cross section in the central rapidity region ($0< y < 1$)
(a) and the forward rapidity region ($2< y < 3$) (b).
The weight grid is produced in a run where the CTEQ6mE PDF \cite{Pumplin:2002vw}
has been used to calculate the jet cross-section.
This PDF is used as standard in the following.
To show the independence of the weight grid performance on the used PDF,
%as examples, 
Fig.~\ref{fig:jetprodvarPDf} also includes more recent PDFs
based on the analyses of a large variety of data (global analysis) like
CTEQ6.6 \cite{CTEQ66} and MSTW2008 \cite{Martin:2009iq} or
only using inclusive DIS data based on combined H1 and ZEUS data
(HERAPDF01) \cite{herapdf01}. In addition, we include a PDF
that does not use a parameterised input distribution NNPDF  \cite{Ball:2008by}.
%
%
%measured at HERA. 
%CTEQ6.1 \cite{Pumplin:2002vw}, CTEQ4m \cite{CTEQ4m} , MRST2004nlo  \cite{MRS:2004}, 
%the H12000ms PDFs using inclusive DIS data from HERA and BCDMS \cite{h12000ms}
%
%The more recent PDF lead to jet cross-section that are $1-2\%$ higher
%than the one calculated with  CTEQ6.1 for 
%$ \pt < 1000$ GeV. For higher \pt  the predicted
%cross-section is smaller by up to $15 \%$ for $\pt = 4000$ GeV.
Further comparisons of the jet cross-sections calculated with these PDFs
can be found in section~\ref{sec:jetuncertainties}.

In the central region the cross-section calculated with the weight
grid reaches 
an accuracy of about $0.1$\% for all tranverse
jet momenta and all PDFs. 
In the forward region a similar performance 
is achieved for transverse jet momenta up to $1000$\GeVx. For  
transverse jet momenta above that value the performance
degrades to $0.6 \%$ and a variation with the PDF is observed.

%%%%%%%%%%%%%%%%%%%%%%%%%%%%%%%%%%%%%%%%%%%%%%%%%%%%%%%%%%%%%%%%%%%%%%%%%%%%%%%%%%%%%%%
%\begin{figure}[ht]
%\vspace{-2cm}
%\centering
%\includegraphics[width=0.49\textwidth]{large-x-h12000ms.eps}
%\caption{The up-quark distribution (rescaled by $(1-x)^{-4}$) for the
%  H12000ms and CTEQ61 PDFs. It illustrates the strong non-physical structure
%  that is present in the H12000ms set, which is the probable cause of the
%  limited accuracy of the grid in certain circumstances with this set.
%\label{fig:h1200ms}
%}
%\end{figure}
%%%%%%%%%%%%%%%%%%%%%%%%%%%%%%%%%%%%%%%%%%%%%%%%%%%%%%%%%%%%%%%%%%%%%%%%%%%%%%%%%%%%%%%%%%

The dependence of the accuracy on the number of $x$-bins is illustrated 
in Fig.~\ref{fig:jetprodvarx}. If only $25$ $x$-bins 
are used, the accuracy is $0.3\%$ in the central and $0.6\%$
in the forward rapidity region. The accuracy decreases towards low jet
transverse momenta. More accuracy is achieved by a larger number of $x$-bins.
For $30$ bins the accuracy is $0.1\%$. For $40$ $x$-bins the improvement
is small, but visible.
A very sensitive
kinematic region is the forward region with very high transverse momenta.
In this region at least $30$  $x$-bins are needed to get an accuracy
of $0.1\%$.

Fig.~\ref{fig:jetprodvarQ2} shows the dependence of the accuracy on the number of
$Q^2$-bins. This dependence is rather small. 
When a large enough number of $x$-bins is chosen, no change is observed
for $8$ to $15$ bins in $Q^2$.

The dependence on the interpolation order (as defined in eq.~\ref{eq:Ii}) is shown
in Fig.~\ref{fig:jetprodvarint}. While 
varying the default interpolation order $n=5$ to $n=4$ and $n=6$
gives similar results within $0.1\%$, 
the interpolation order $n=3$ leads to an accuracy loss
of $0.5\%$ at low transverse jet momenta in the central regions, 
and by $0.4 - 1 \%$ 
at low and high transverse jet momenta in the forward region. 

In conclusion, the results in Figs.~\ref{fig:jetprodvarPDf}-\ref{fig:jetprodvarint}
demonstrate that an accuracy of $0.1$ \% can be reached with a reasonable weight
grid size. The most critical parameter is the number of $x$-bins, which must be large enough
to accommodate strong PDF variations in certain phase space regions. 
In comparison, the dependence on the number
of $Q^2$ bins is rather weak. The interpolation between the grid points is 
sufficiently accurate to allow the 
grid technique to be used  and fifth order interpolation produces reasonable results. The achieved accuracy is 
probably sufficient for all practical applications.

In applications where a very small weight grid is needed, one can also introduce a 
PDF-weight to flatten out the $x$-dependence of the PDFs
(see eq.~\ref{eq:pdfweight}). 
The PDF weight is calculated using $a_1=-1.5$ and $a_2=3$.
This is illustrated in Fig.~\ref{fig:jetprodPDFweighting},
 where grids with very low number of $x$-bins ($8$, $9$, $10$)
and eight $Q^2$ bins are used, the interpolation is lowered to $n=4$.
Even with the smallest weight grid an accuracy of $1\%$ is achieved
using the PDF-weight. For a somewhat larger weight grid with $10$ $x$-bins
the accuracy is $0.5\%$ in all phase space regions.

 One of the important theoretical uncertainties in NLO QCD calculations is the
 variation of the results with the choice of the factorisation and renormalisation 
 scale. Eq.~\ref{eq:Wfinalxi_twohadrons} allows the calculation of the cross-section
 for any scale choice {\em a posteriori} from one weight grid produced at a fixed scale choice. 
 The results from scale variations
 by a factor of $2$ up and down is shown in Fig.~\ref{fig:jetprodvarScale}. 
 The renormalisation and factorisation scales are either varied together
 or varied individually.
 The weight grid result has been calculated with a single weight grid 
 and the reference cross-sections have been calculated by repeating the 
 standard NLO QCD calculation for each of the scale variations.
%three times.
%
 The cross-section calculated with the weight grid reproduces the standard results to
 within about $0.1$\% in the central region and $0.1 - 0.2$\% in the forward region.

\subsection{Reweighting jet cross-sections to a different centre-of-mass energy}
\label{sec:cmsreweightingperformance}
As outlined in section~\ref{sec:cmsreweighting} a weight grid produced at
a given centre-of-mass energy can also be used to calculate the
cross-section at a lower or higher centre-of-mass energy.
This procedure works if the coverage in $x$ in the weight grid is
large enough. For instance,
when lowering the centre-of-mass energy to calculate the jet
cross-section at a fixed transverse jet momentum, it might
happen that the required large $x$ values are not present in the weight
grid produced at a higher centre-of-mass energy. 
The variation of the centre-of-mass energy has therefore to be done
with care by the user.

%%%%%%%%%%%%%%%%%%%%%%%%%%%%%%%%%%%%%%%%%%%%%%%%%%%%%%%%%%%%%%%%%%
\begin{figure}[htp!]
\centering
\includegraphics[width=0.48\textwidth]{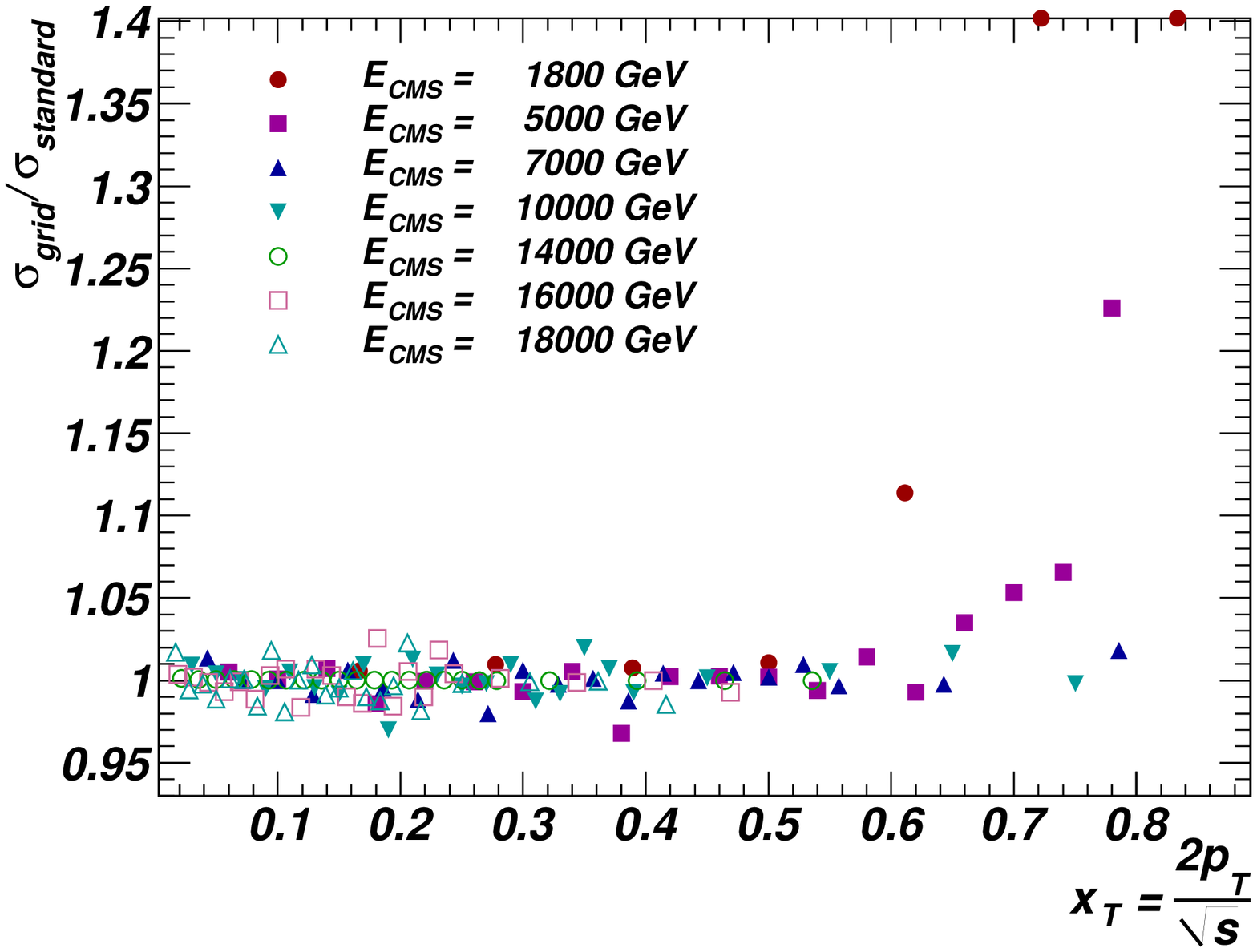}
\includegraphics[width=0.51\textwidth]{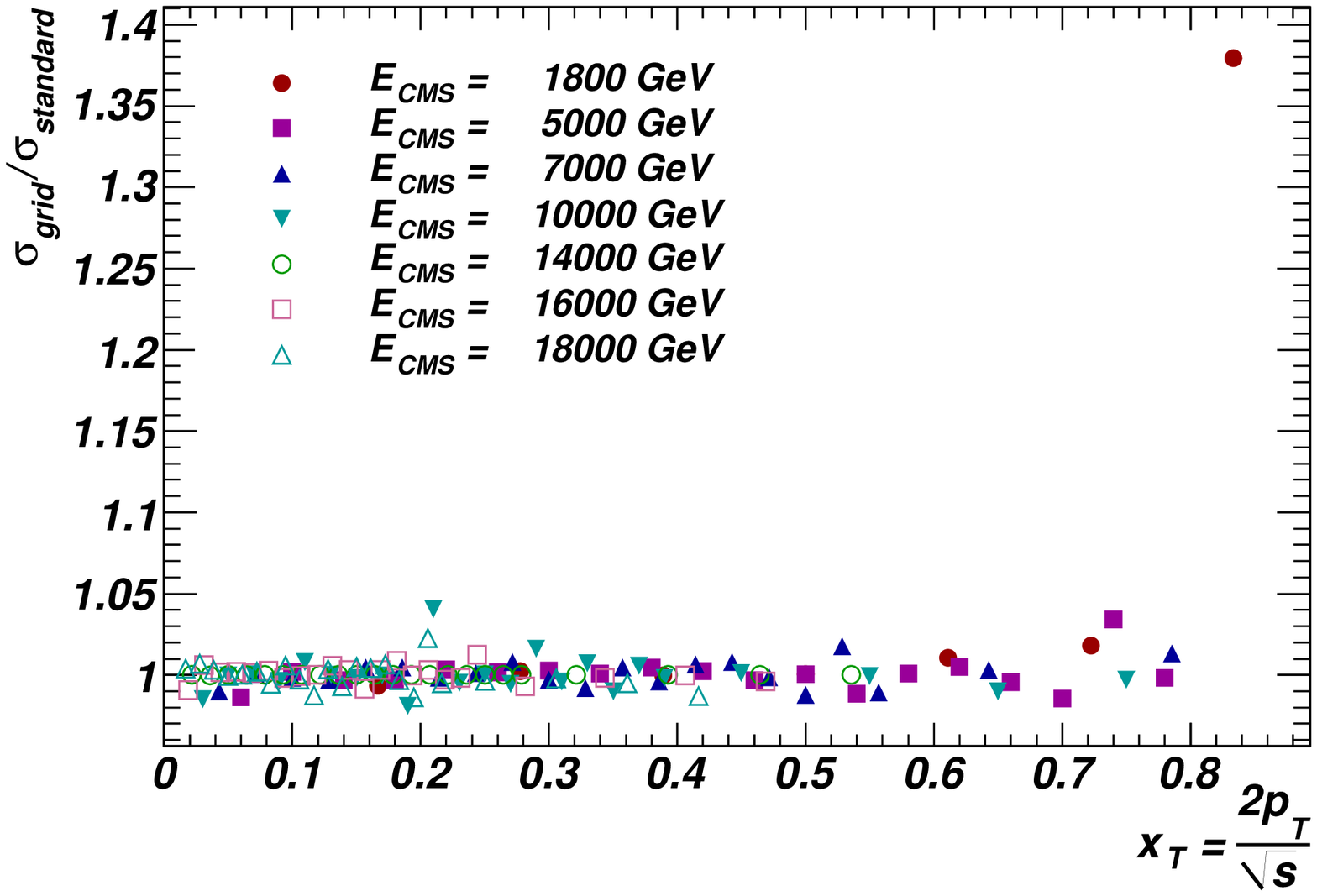}
\begin{picture}(0,0)
\put( -450,0){a)}
\put( -210,0){b)}
\end{picture}
\caption{
  Ratios of grid and standard calculations of the
  single inclusive jet $d\sigma/d x_T$ spectrum, with $x_T= 2
  p_T/\sqrt{s}$, for various centre-of-mass energies. The standard
  calculation has been performed separately for each
  centre-of-mass energy, while the grid results are all based on a common
  $\sqrt{s}=14000$~GeV grid.
  The PDF set is CTEQ6mE.
  In a) the default grid parameters are used ($30$ bins in $x$ and $10$ bins in $Q^2$).
  The last two points for $\sqrt{s} = 1800$~GeV are drawn at $1.3$ for
  better visibility, but their true values are very large.  
  In b) a larger grid with $50$ bins in $x$ and $30$ bins in $Q^2$ is used.
}
\label{cmsreweightingvalidation}
\end{figure}
%%%%%%%%%%%%%%%%%%%%%%%%%%%%%%%%%%%%%%%%%%%%%%%%%%%%%%%%%%%%%%%%%%

As an example, the accuracy of the jet cross-section calculation
using the default weight grid at a fixed centre-of-mass energy 
of $\sqrt{s} = 14000$~GeV is investigated. 
Reference cross-sections are calculated at
various centre-of-mass energies, i.e. 
$1800$, $5000$, $7000$, $10000$, $14000$, $16000$ and $18000$ GeV.
Since the calculations at the various centre-of-mass energies
are statistically independent, each reference cross-section as well
as the default weight grid at $\sqrt{s} = 14000$~GeV needs to be calculated with large
event samples. Each of the calculations is done with $50\, 000\, 000$ events
produced with \NLOJET.

In order to make the comparisons more meaningful the jet transverse
momentum \pt is transformed to $x_T = 2 \pt /\sqrt{s}$.
For central jets the variable $x_T$ gives approximately the momentum fraction of the 
incoming parton with respect to the proton.
Fig.~\ref{cmsreweightingvalidation}a) shows the ratio of the cross-section
calculated with the standard weight grid produced at $\sqrt{s} = 14000$~GeV to the cross-section
calculated at various centre-of-mass energies in the standard way
as a function of $x_T$. For most points, the calculations agree within $2\%$.
The observed fluctuations are statistical.

For large changes in centre-of-mass energy and large $x_T$ values
the approximation of the standard grid becomes inaccurate.
For instance, for $\sqrt{s} = 1800$~GeV and $x_T = 0.6$  
the weight grid calculation gives a result that is $10\%$ higher
than the standard calculation. This discrepancy increases further
for large $x_T$ values. The ratio of the last two $x_T$ values
becomes very large.\footnote{In Fig.~\ref{cmsreweightingvalidation}a) 
they are drawn at $1.3$ for better visibility of the rest of the points.}

Fig.~\ref{cmsreweightingvalidation}b) shows the result for a larger
grid using $50$ bins in $x$ and $20$ bins in $Q^2$. With such a grid
the deviations are mostly reduced to statistical fluctuations.
Only the largest $x_T$ value for the lowest centre-of-mass energy
exhibts a deviation by $30$~\% from the standard calculation. 

A small grid with a PDF weighting leads to large discrepancies
to the standard calculation and cannot be used. 

In conclusion, the grid technique gives a good accuracy to compute
the jet cross-section at various centre-of-mass energies.
For very high transverse momenta and extreme centre-of-mass
variations a large grid might be required.

% GPS we don't have Z-boson in there, so removed it from title!
\subsection{W-boson production at hadron colliders}
\label{sec:wzresults}
To further demonstrate the performance of the weight grid method, the production of
$W$-bosons at LHC energies is taken as example. 
The observable that will be examined is the transverse-momentum
distribution of the positron from W$^+$-boson decays, when the positron
is either central $|\eta|\le 0.5$, or very forward, $|\eta|\ge 3.0$.

% The differential cross-section
% in terms of the positron's (final state) transverse momentum in the central pseudo-rapidity
% region $|\eta|\le 0.5$ and in the very forward  pseudo-rapidity region $|\eta|\ge 3.0$ is used. 
% The positron is the decay product of the $W$-boson. 

%%%%%%%%%%%%%%%%%%%%%%%%%%%%%%%%%%%%%%%%%%%%%%%%%%%%%%%%%%%%%%%%%%%%%%%%%%%%%%%%%%%%%%%
\begin{figure}[t]
%\vspace{-2cm}
\centering
\includegraphics[width=0.49\textwidth]{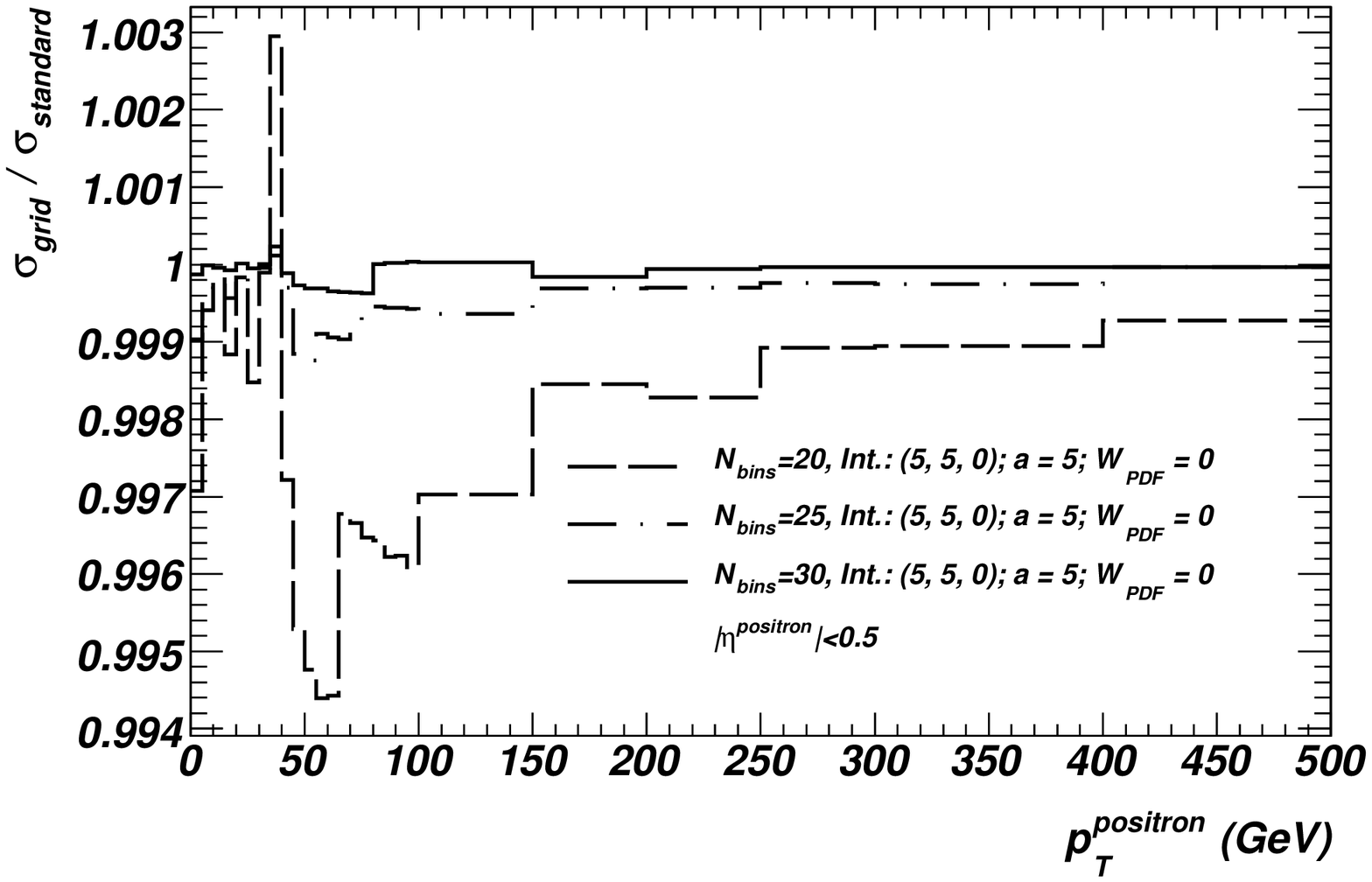}
\includegraphics[width=0.49\textwidth]{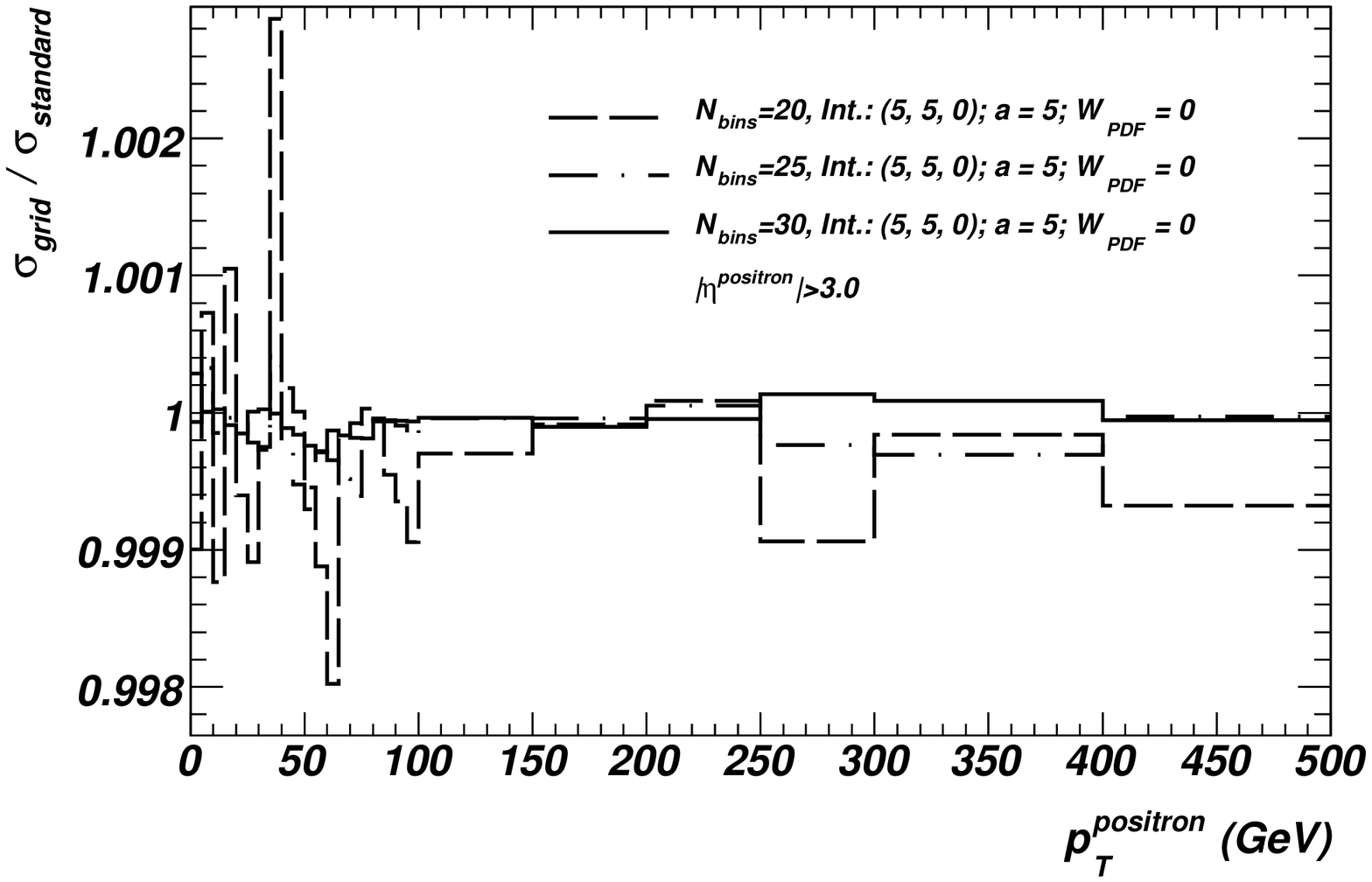}
\begin{picture}(0,0)
\put( -450,0){a)}
\put( -210,0){b)}
\end{picture}
\caption{
  Ratios of grid and standard calculations of the positron $p_T$
  spectrum in $W^+$-boson production, for $ | \eta_{e^+} | < 0.5$ (a)
  and for $| \eta_{e^+} | > 3$ (b).  
  Results are shown for three weights grids with different numbers of
  $x$ bins.  All grids use a coordinate transform parameter $a=5$ and
  third order interpolation.
  The PDF set is CTEQ6mE.
% 
% weight grid
% with a coordinate transform parameter $a=5$ and a third order interpolation
% Ratio of the $W$-boson production cross-section
% calculated using a weight grid to the standard calculation
% as a function of the positron \pt  
% for $ | \eta | < 0.5$ (a) and for $| \eta | > 3$ (b). 
% A weight grid
% with a coordinate transform parameter $a=5$ and a third order interpolation
% is used. 
% The number of $x$-bins is varied. The CTEQ6mE PDF has been used.
\label{zwprodvarbin}
}
\end{figure}
%%%%%%%%%%%%%%%%%%%%%%%%%%%%%%%%%%%%%%%%%%%%%%%%%%%%%%%%%%%%%%%%%%%%%%%%%%%%%%%%%%%%%%%%%%

%%%%%%%%%%%%%%%%%%%%%%%%%%%%%%%%%%%%%%%%%%%%%%%%%%%%%%%%%%%%%%%%%%%%%%%%%%%%%%%%%%%%%%%
\begin{figure}[t]
%\vspace{-2cm}
\centering
\includegraphics[width=0.49\textwidth]{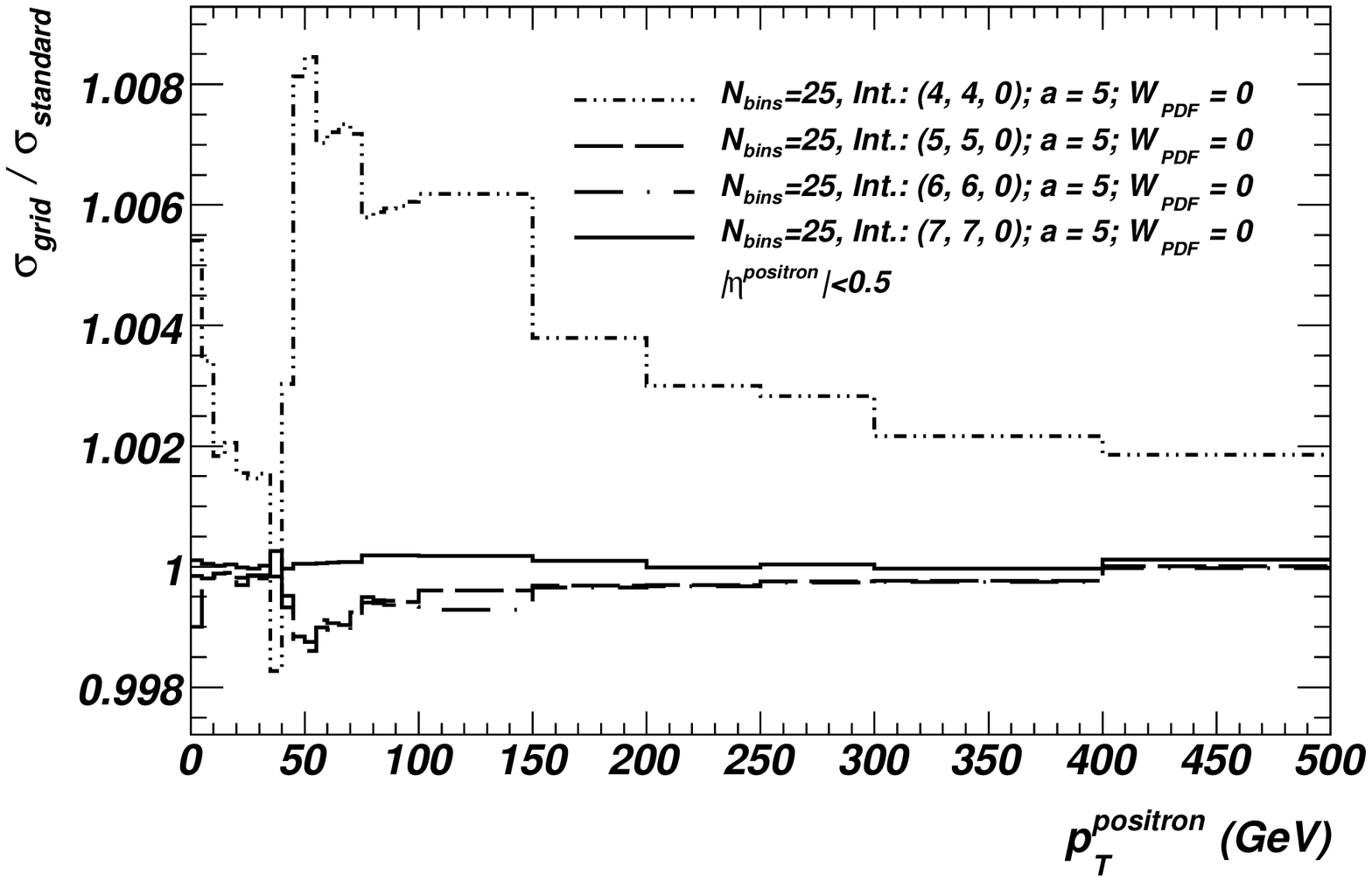}
\includegraphics[width=0.49\textwidth]{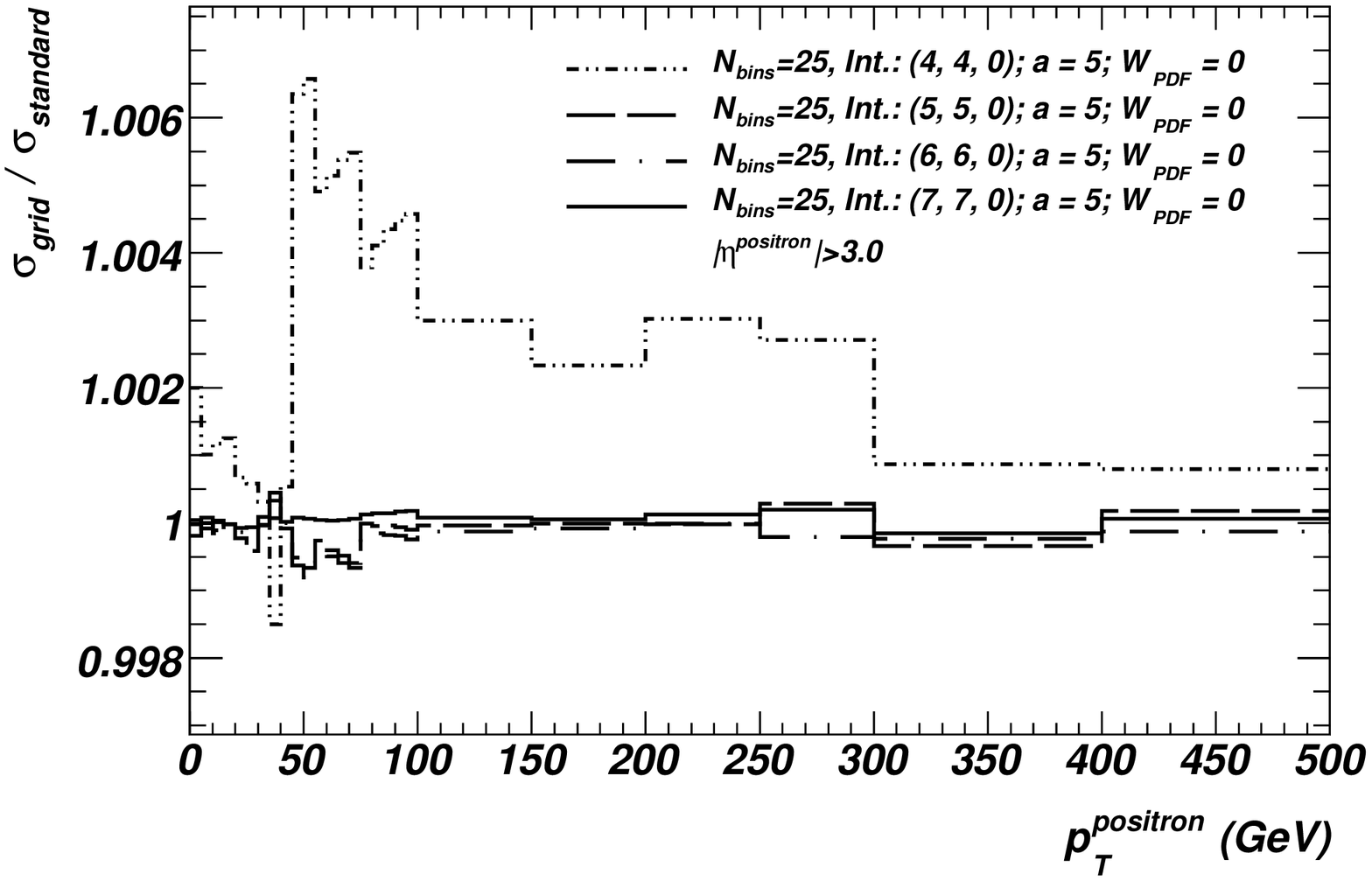}
\begin{picture}(0,0)
%\put( -450,0){c)}
%\put( -210,0){d)}
\put( -450,0){a)}
\put( -210,0){b)}
\end{picture}
\caption{
  Ratios of grid and standard calculations of the positron $p_T$
  spectrum in $W^+$-boson production, for $ | \eta_{e^+} | < 0.5$ (a)
  and for $| \eta_{e^+} | > 3$ (b).  
  Results are shown for four grids, each with a different
  interpolation order.
  All grids have $25$ bins in $x$ and a coordinate transform parameter
  $a=5$. 
  The PDF set is CTEQ6mE.
  %
  % Ratio of the $W$-boson production cross-section
  % calculated using a weight grid to the standard calculation
  % as a function of the positron \pt  
  % for $ | \eta | < 0.5$ (a) and for $| \eta | > 3$ (b).
  % A weight grid with $25$ bins in $x$  
  % and a coordinate transform parameter $a=5$ is used. 
  % The interpolation order is varied.
  % The CTEQ6mE PDF has been used.
\label{zwprodvarint}
}
\end{figure}
%%%%%%%%%%%%%%%%%%%%%%%%%%%%%%%%%%%%%%%%%%%%%%%%%%%%%%%%%%%%%%%%%%%%%%%%%%%%%%%%%%%%%%%%%%

As in the previous section, a default weight grid is defined and variations in a few
parameters are studied.
The default weight grid consists of $25$ bins in $x$.
The points are distributed according to eq.~\ref{eq:ytau} with $a=5$ and a fifth order
interpolation is used. No PDF weight (see eq.~\ref{eq:pdfweight}) is used.
The cross-sections are calculated with the factorisation and renormalisation scale
fixed to the mass of the $W$-boson. Therefore, the weight grid need only be two dimensional.

The influence of the number of bins in $x$ is shown in Fig.~\ref{zwprodvarbin}.
If the number of bins in $x$ is too small ($N_{bins}=20$) the cross-section 
is reproduced to about $0.5\%$ in the central region and $0.2\%$ in the forward
region.

For the default weight grid, lowering the interpolation order from $n=5$ to $n=4$
results in an accuracy loss of about $0.2\%$ over much of the $p_T$ range, as
 shown in Fig.~\ref{zwprodvarint}. The accuracy for positrons
with low transverse momenta degrades to $0.8\%$. The good precision for $n=5$ can
only be improved using $n=7$.

\begin{figure}[tp!]
\centering
\includegraphics[width=0.49\textwidth]{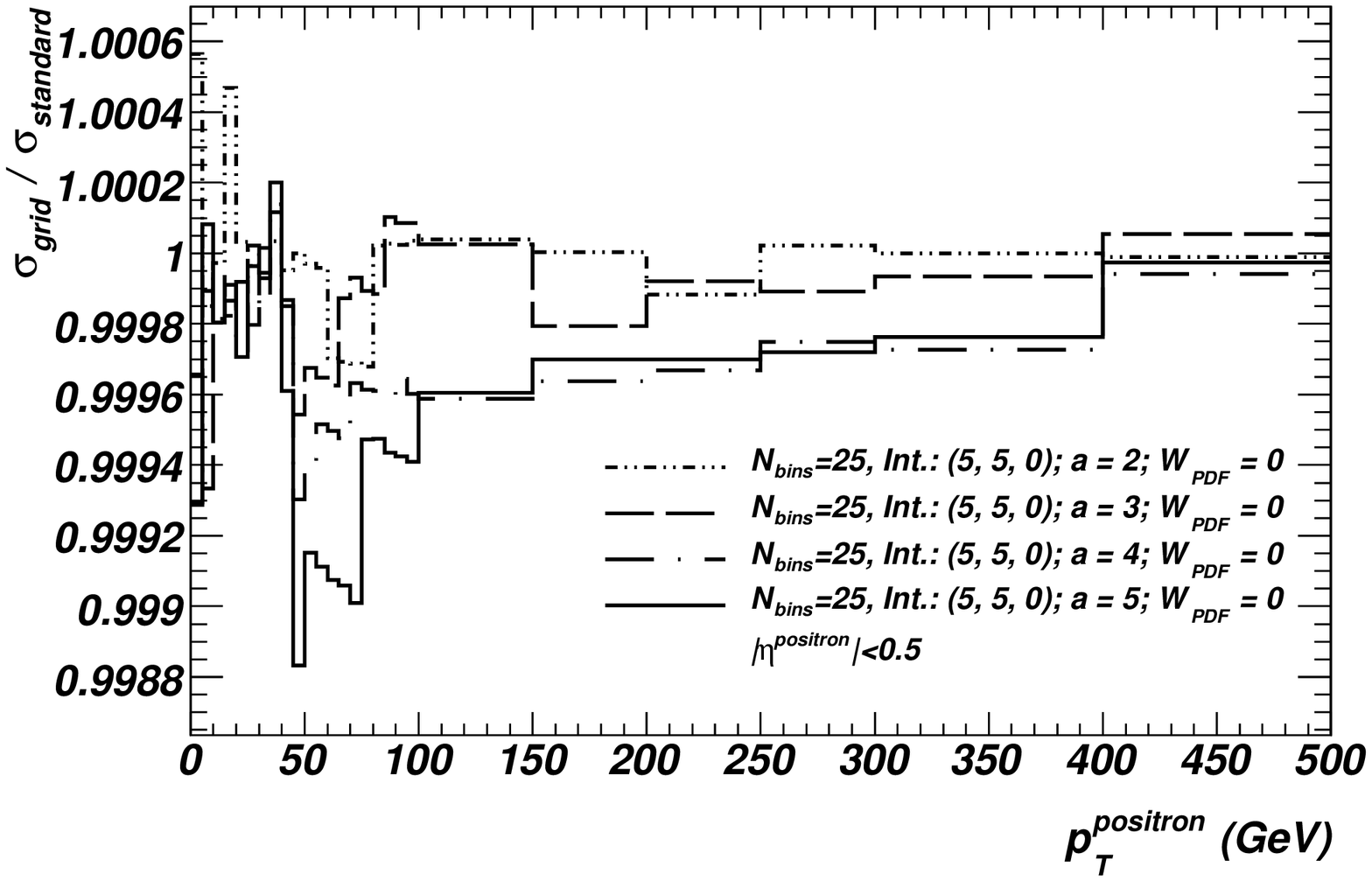}
\includegraphics[width=0.49\textwidth]{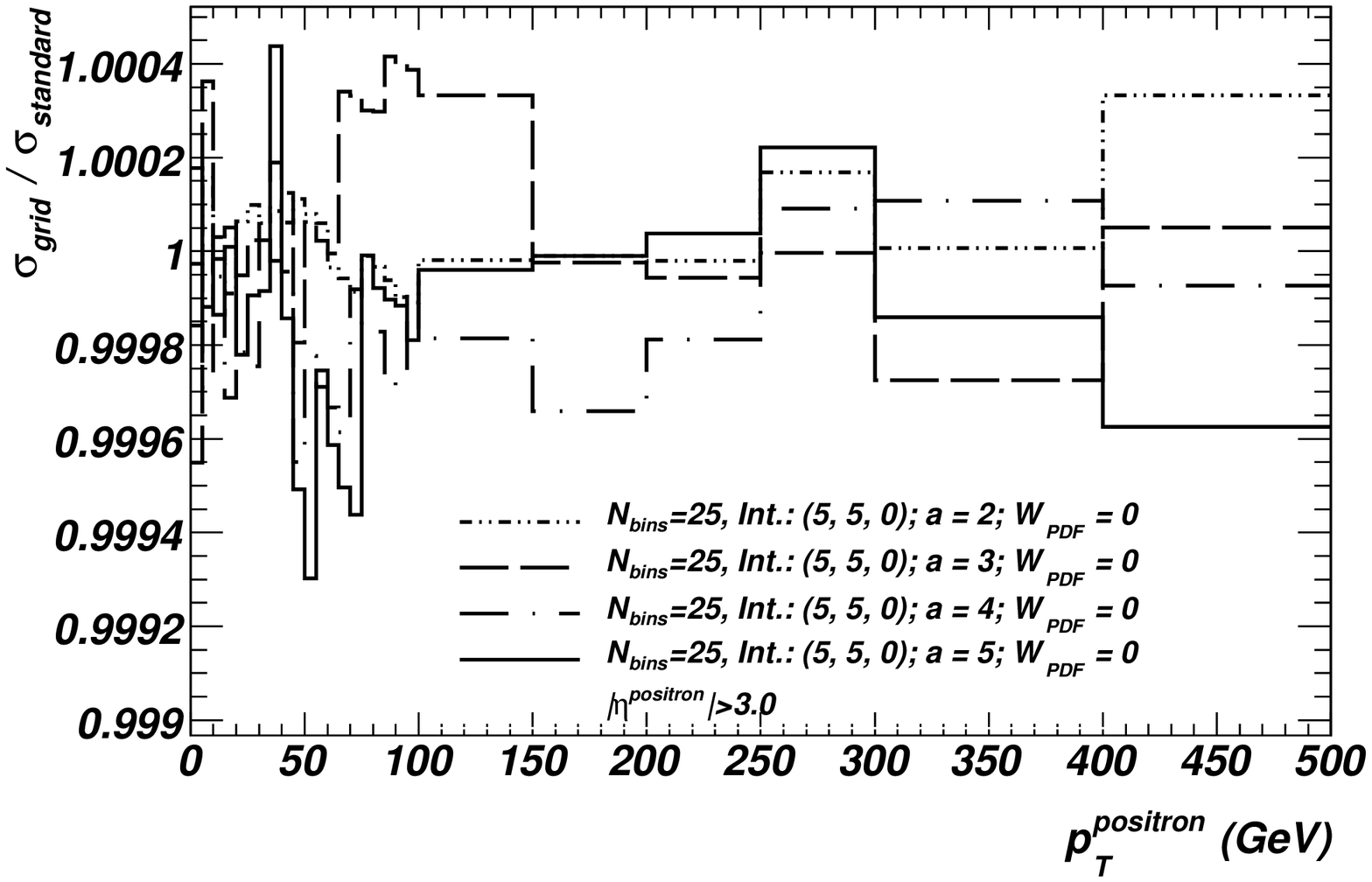}
\begin{picture}(0,0)
\put( -450,0){a)}
\put( -210,0){b)}
\end{picture}
\caption{
  Ratios of grid and standard calculations of the positron $p_T$
  spectrum in $W^+$-boson production, for $ | \eta_{e^+} | < 0.5$ (a)
  and for $| \eta_{e^+} | > 3$ (b).  
  Results are shown for four grids, each with a different
  coordinate transform parameter, $a$.
  All grids have $25$ bins in $x$ and fifth order interpolation.
  The PDF set is CTEQ6mE.
  % Ratio of the $W$-boson production cross-section
  % calculated using a weight grid to the standard calculation
  % as a function of the positron \pt  
  % for $ | \eta | < 0.5$ (a) and for $| \eta | > 3$ (b). 
  % A weight grid with $25$ bins in $x$ 
  % and with a fifth order interpolation is used. 
  % The coordinate transform parameter is varied.
  % The CTEQ6mE PDF has been used.
\label{zwprodvara}
}
\end{figure}

Fig.~\ref{zwprodvara} shows the dependence on the grid spacing parameter corresponding
to the parameter $a$ in eq.~\ref{eq:ytau}.
The $x$-values in the cross-section calculation are not large
 and consequently a fine spacing at
large $x$ (corresponding to a large $a$ parameter) is not needed 
and the result improves for low $a$ values. An accuracy of better than $0.1\%$
is achieved for all variations. 

%%%%%%%%%%%%%%%%%%%%%%%%%%%%%%%%%%%%%%%%%%%%%%%%%%%%%%%%%%%%%%%%%%%%%%%%%%%%%%%%%%%%%%%
%%%%%%%%%%%%%%%%%%%%%%%%%%%%%%%%%%%%%%%%%%%%%%%%%%%%%%%%%%%%%%%%%%%%%%%%%%%%%%%%%%%%%%%%%%

%%%%%%%%%%%%%%%%%%%%%%%%%%%%%%%%%%%%%%%%%%%%%%%%%%%%%%%%%%%%%%%%%%%%%%%%%%%%%%%%%%%%%%%
\begin{figure}[t]
%\vspace{-2cm}
\centering
\includegraphics[width=0.49\textwidth]{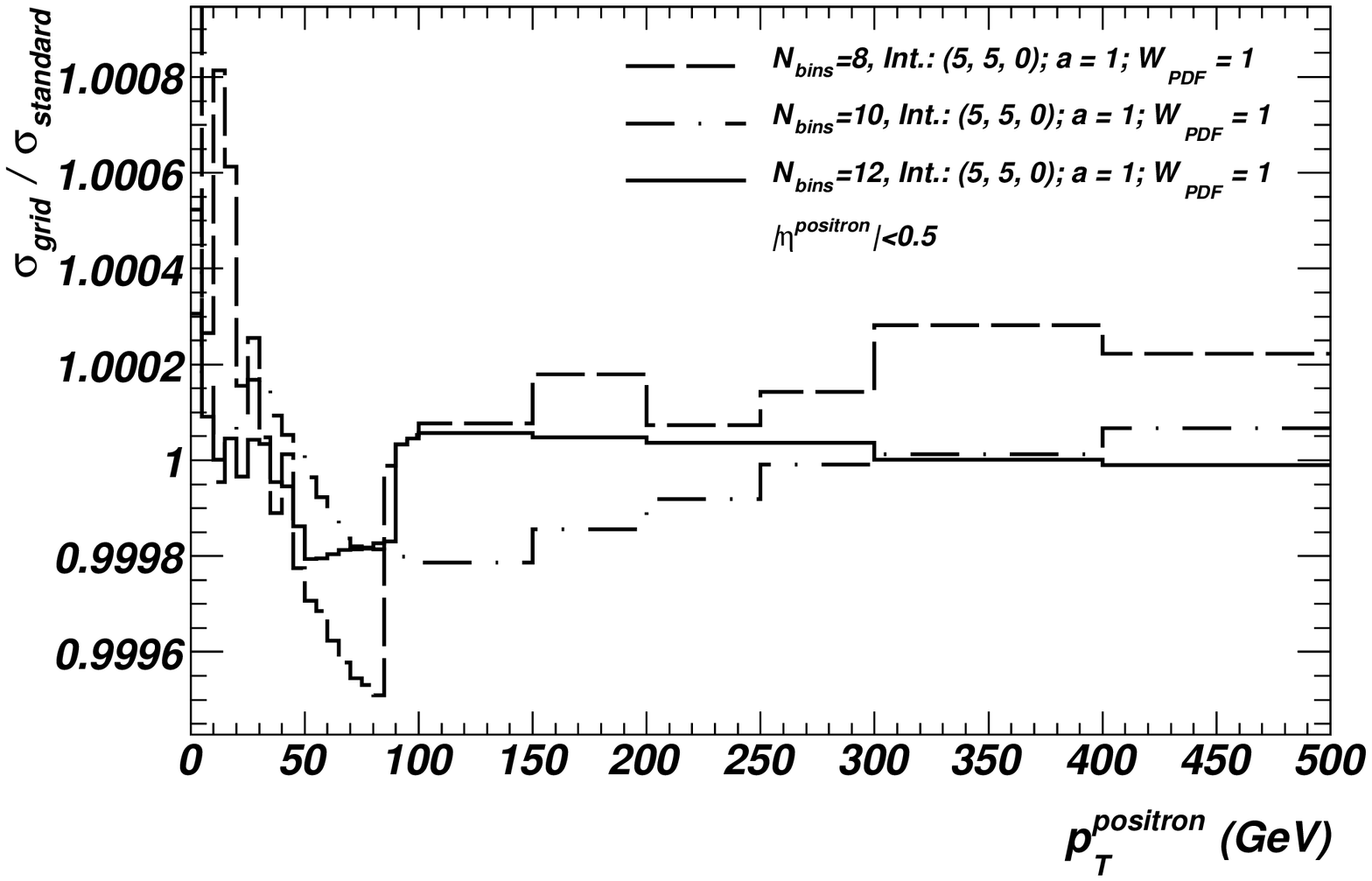}
\includegraphics[width=0.49\textwidth]{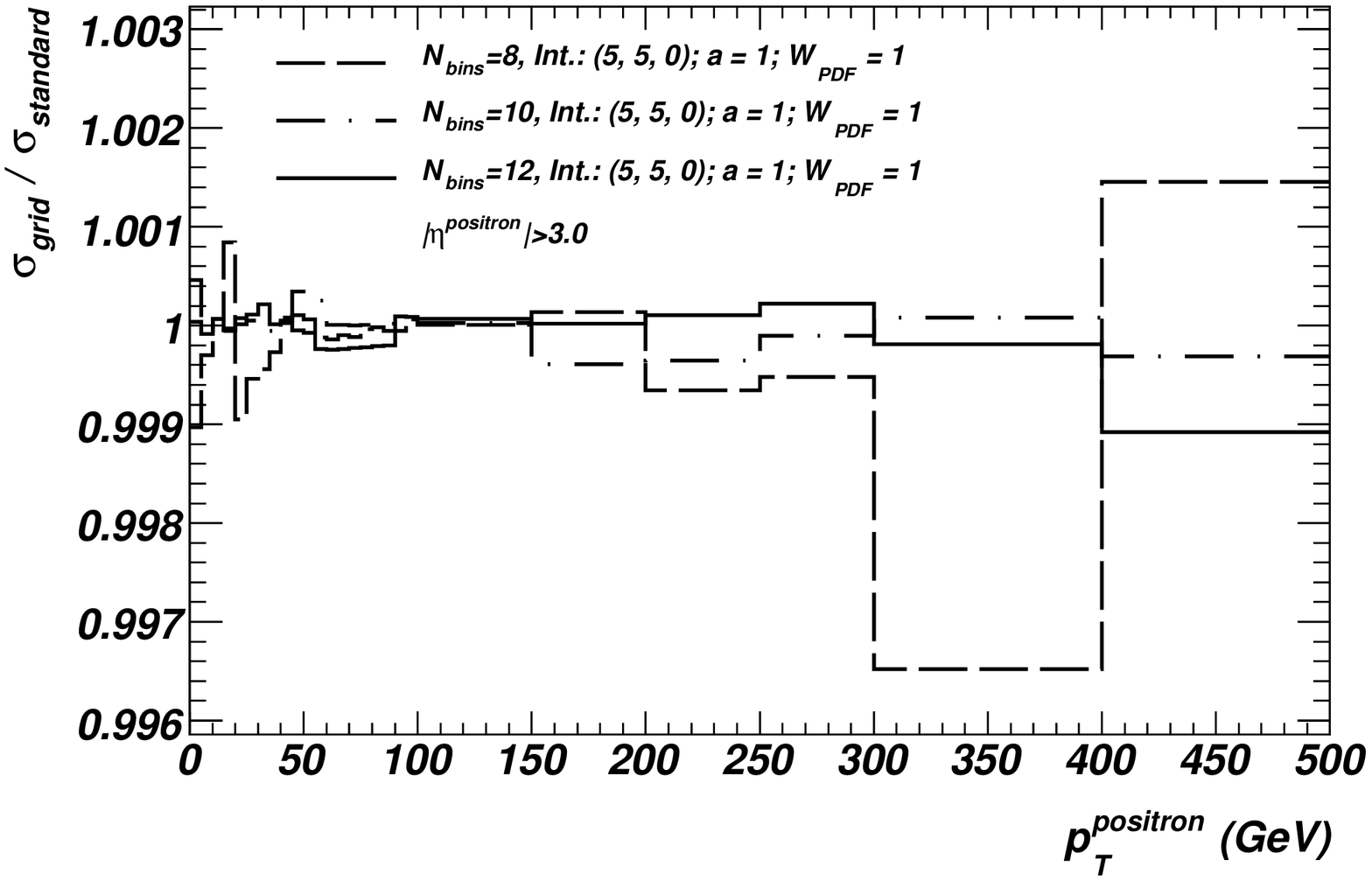}
\begin{picture}(0,0)
%\put( -450,0){c)}
%\put( -210,0){d)}
\put( -450,0){a)}
\put( -210,0){b)}
\end{picture}\caption{
  Ratios of grid and standard calculations of the positron $p_T$
  spectrum in $W^+$-boson production, for $ | \eta_{e^+} | < 0.5$ (a)
  and for $| \eta_{e^+} | > 3$ (b).  
  Results are shown for grids with a reduced number of $x$ bins and
  PDF reweighting. 
  All the grids use second order interpolation and a coordinate
  transform parameter $a=1$. 
  The PDF set is CTEQ6mE.
  %
  % Ratio of the $W$-boson production cross-section
  % calculated using a weight grid to the standard calculation
  % as a function of the positron \pt  
  % for $ | \eta | < 0.5$ (a) and for $| \eta | > 3$ (b). 
  % A minimal weight grid with varying $x$ bins,  
  % with a second order interpolation and a coordinate transform parameter $a=1$
  % is used.
  % The PDF weighting and the CTEQ6mE PDF is used.
\label{zwprodvarPdf}
}
\end{figure}
%%%%%%%%%%%%%%%%%%%%%%%%%%%%%%%%%%%%%%%%%%%%%%%%%%%%%%%%%%%%%%%%%%%%%%%%%%%%%%%%%%%%%%%%%%

Finally, Fig.~\ref{zwprodvarPdf} shows that, if a PDF weighting is used, 
it is possible to use very small grid sizes.
For a weight grid with only eight $x$-bins an accuracy of $0.1 \%$ can be achieved. 
In this case the gain in accuracy is small when increasing the
number of $x$-bins.
Only in the forward region and for high transverse energies
the increase in the number of $x$-bins is beneficial.

In summary, a sufficient accuracy is achieved with about $25$ $x$-bins and a fifth
order interpolation. An equidistant grid spacing ($a=0$) is sufficient.

%%
%\newpage
\subsection{CPU and computer memory performance}
\label{sec:cpuperformance}
\begin{figure}[htp]
\includegraphics[width=0.49\textwidth]{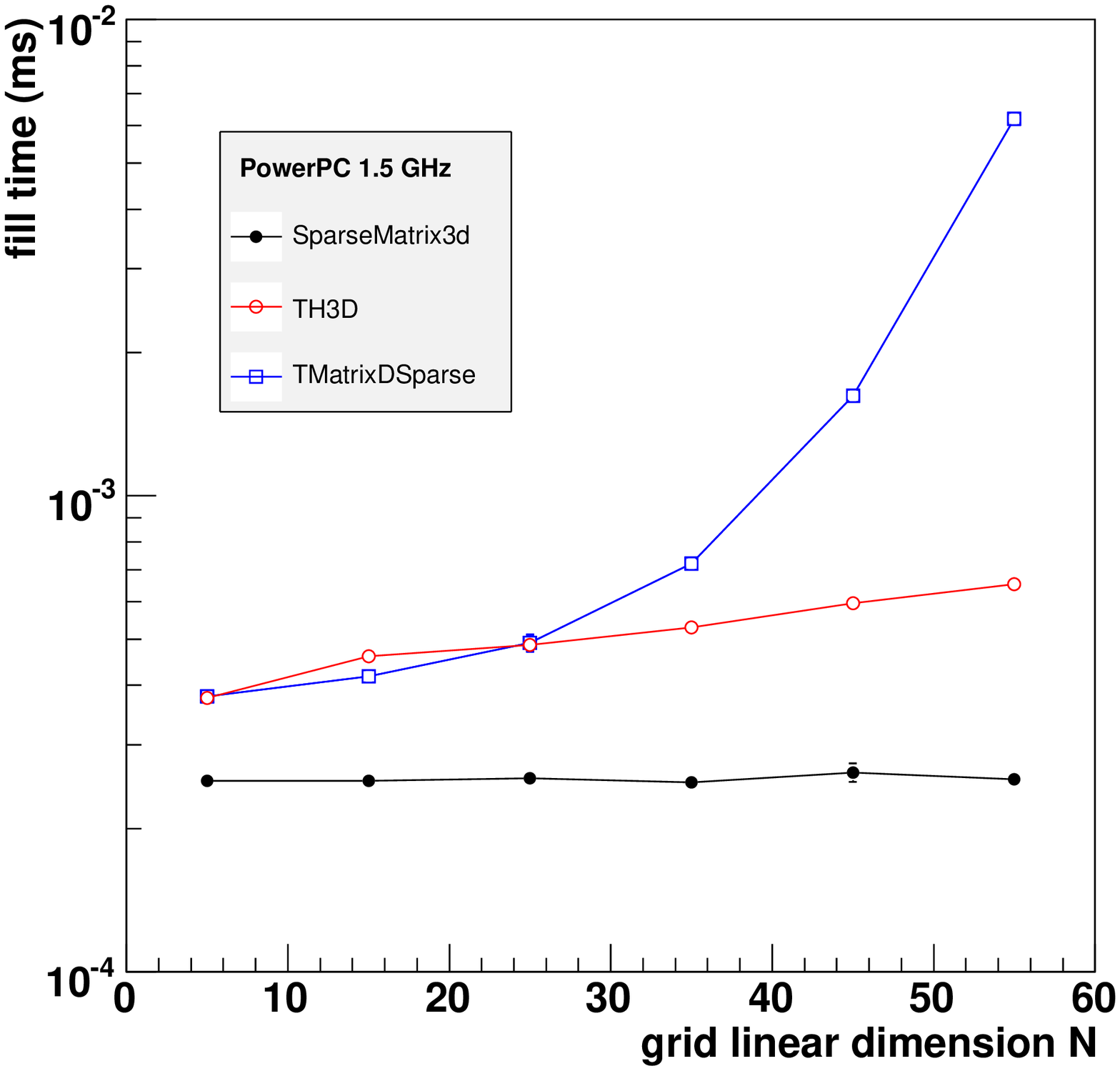}
\includegraphics[width=0.49\textwidth]{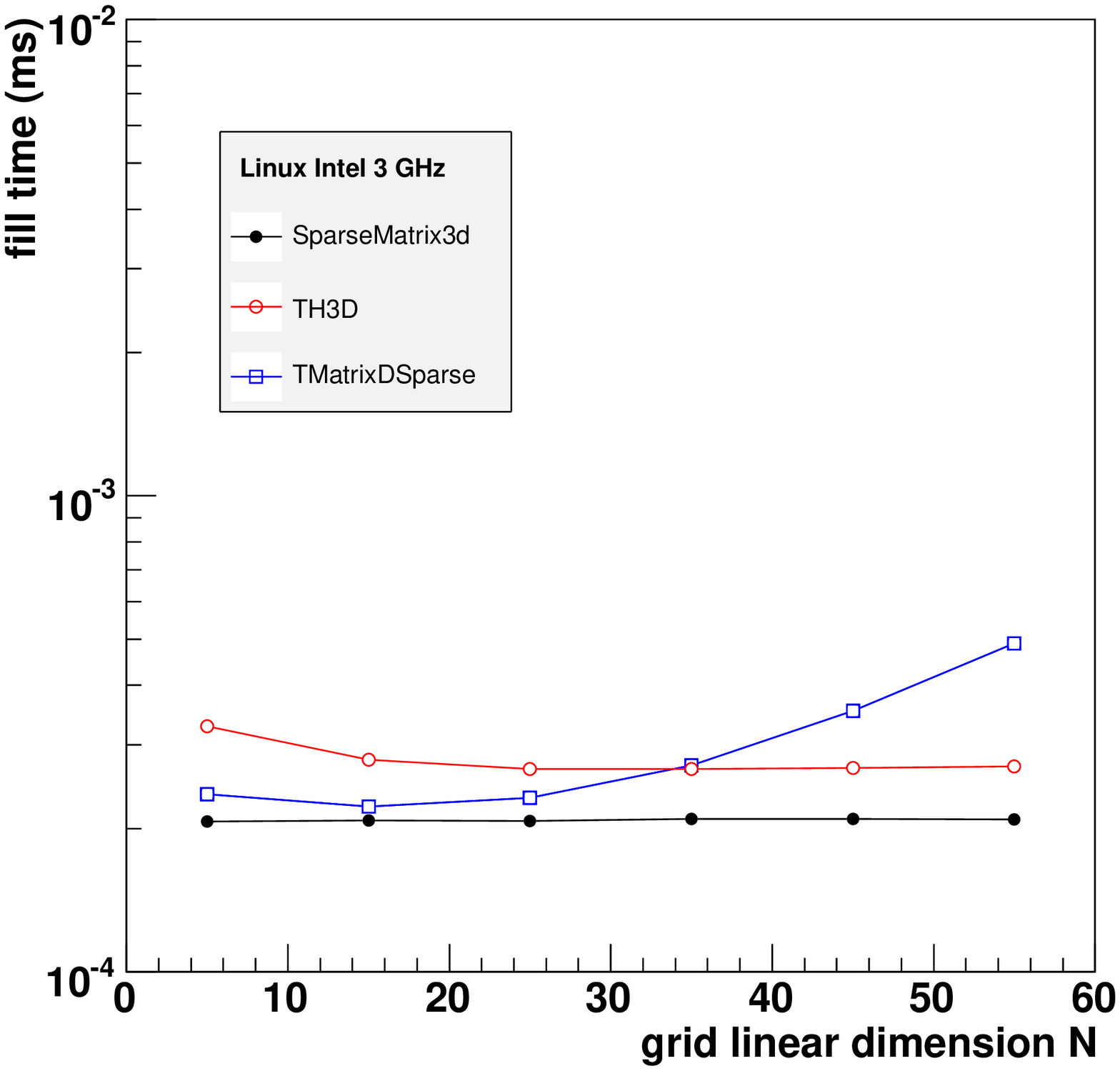}
\caption{The time per call for filling grid classes based on various grid architectures
on a 1.5~GHz PowerPC (left) and 3~GHz Linux PC (right).}
\label{filltimes}
\end{figure}

The execution time for each call to the filling routine for the grid 
has been studied on
a $1.5$~GHz PowerPC and a $3$~GHz Intel Xeon running Linux, using a dummy structure 
with $N$ points in each dimension. 
Fig.~\ref{filltimes} shows the performance for various grid architectures.
The grids are based on either the \ROOT \texttt{TH3D} class, 
the custom sparse class (\texttt{SparseMatrix3d}) described in the
section~\ref{sec:technical}, or 
%are shown together with a grid based on %$60$
% of the phase space is occupied. Also shown is the filling time per fill call for a class based on the \ROOT
the \texttt{TMatrixDSparse} class which implements the 2-dimensional Harwell-Boeing matrix representation.
In the latter case, a sparse 1-dimensional structure of \texttt{TMatrixDSparse} matrices using the classes of
the \texttt{SparseMatrix3d} has been used to create a sparse 3-dimensional structure.
%%%%%%%%%%%%%%%%%%%%%%%%%%%%%%%%%%%%%%%%%%%%%%%%%%%%%%%%%%%%%%%%%%
%%%%%%%%%%%%%%%%%%%%%%%%%%%%%%%%%%%%%%%%%%%%%%%%%%%%%%%%%%%%%%%%%%
As expected the Harwell-Boeing based class is very quick for filling when the grid is small, but as the
grid size becomes larger, since the occupation is reasonably large, the number of entries
that must be examined becomes large and the filling time increases rapidly. For the \texttt{TH3D} and custom
sparse structures, the filling time is largely independent of the grid size.

\begin{figure}[htp]
\includegraphics[width=7cm]{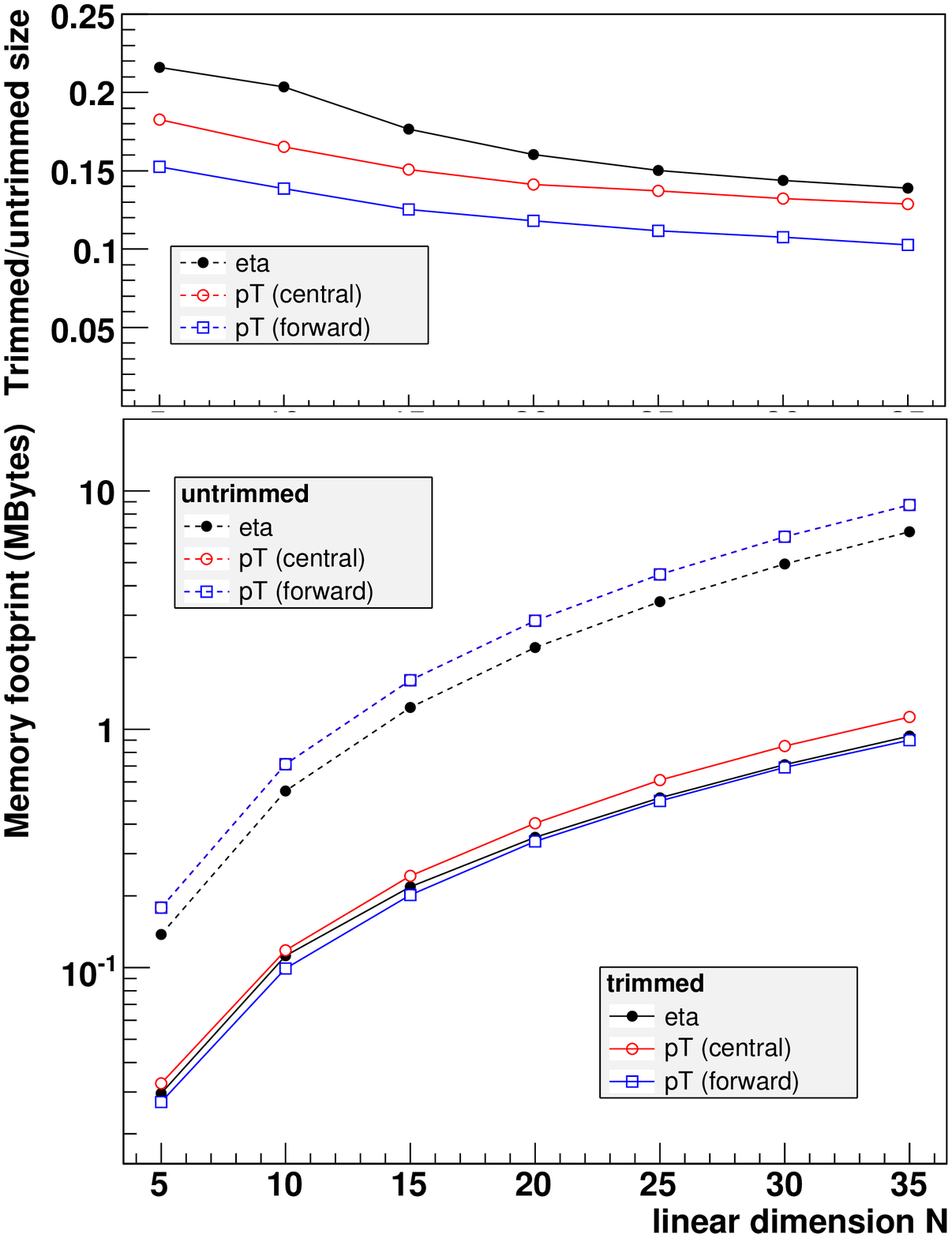}
\includegraphics[width=7.5cm]{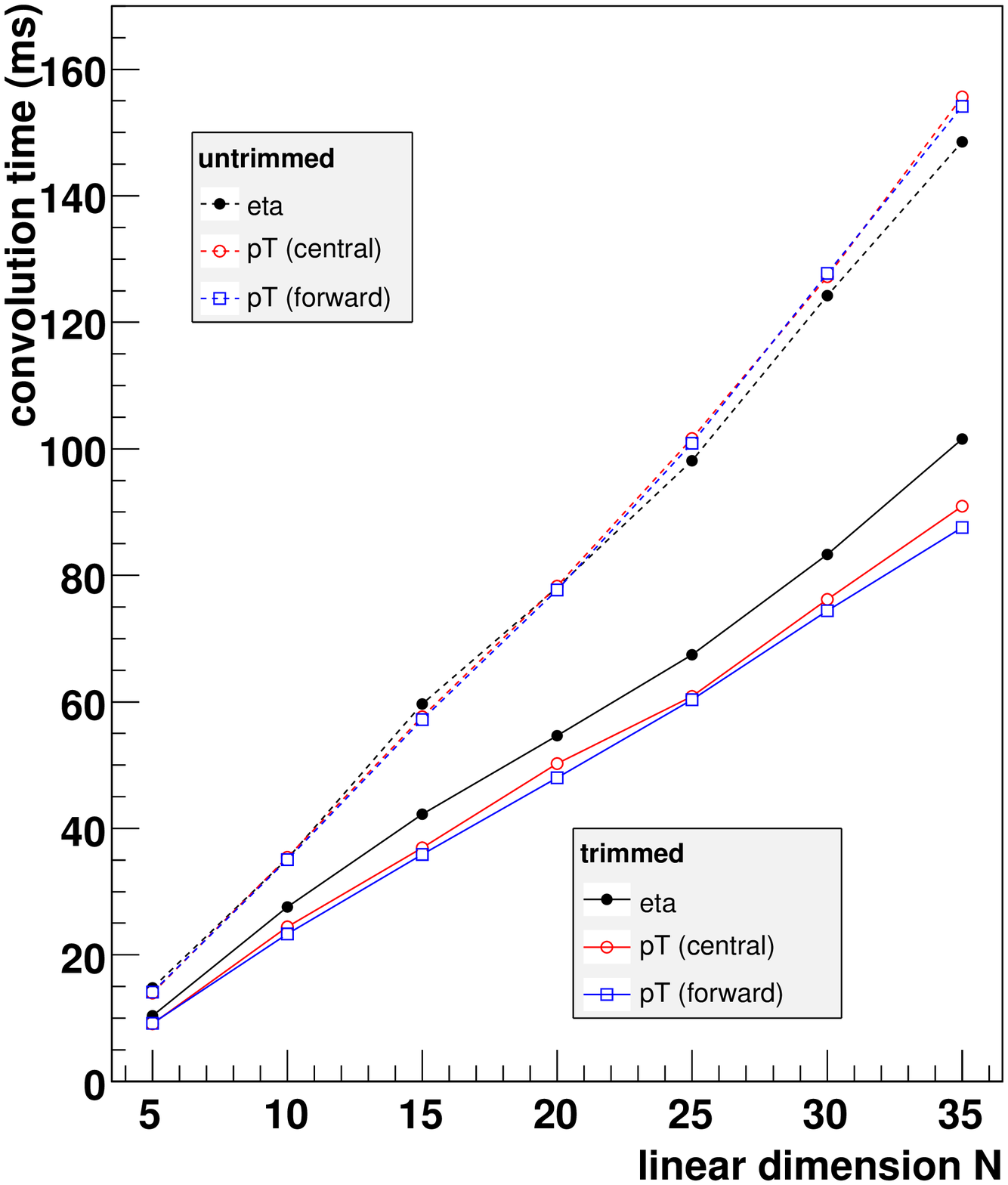}
\begin{picture}(0,0) \put( -390,0){a)} \put( -200,0){b)} \end{picture}
\caption{
a) Memory used for the default grid architecture using a custom sparse grid (untrimmed)
and after removing the unoccupied elements. The top figure shows the ratio
of the reduced to the full case.
b) Time needed to calculate the cross-section by convoluting the coefficients on the
grid with PDFs and $\alpha_s$. The
convolution times are measured on a 1.5~GHz PowerPC 
for a default grid.
The memory and the CPU time performance is evaluated for the $W$-boson cross-sections
as a function of the electron rapidity and transverse momentum.
}
\label{sizestimes}
\end{figure}

The reduction in memory occupied by the custom sparse grid structure after trimming away unoccupied
elements is illustrated in Fig.~\ref{sizestimes}. The bottom-left plot shows the absolute size of the stored
elements in MBytes, both before, and after trimming away unfilled elements. The top-left plot shows the
fraction of the total, untrimmed grid size, occupied by the filled elements. As the grid spacing
decreases, the overall grid size naturally increases.
%%%%%%%%%%%%%%%%%%%%%%%%%%%%%%%%%%%%%%%%%%%%%%%%%%%%%%%%%%%%%%%%%%
%%%%%%%%%%%%%%%%%%%%%%%%%%%%%%%%%%%%%%%%%%%%%%%%%%%%%%%%%%%%%%%%%%

%
The execution time using 
the grid to perform the final cross-section calculation including the PDF convolution
has also been studied using a 1.5~GHz PowerPC. The results are based
on calculations of
differential cross-sections with respect to the positron pseudo-rapidity and transverse momentum
in $W$-boson production using \MCFM\cite{MCFM1, MCFM2}, as presented in 
section \ref{sec:wzresults}. The cross sections involve  20 and 24  bins for the lepton 
pseudo-rapidity and transverse energy distributions respectively.  
Fig.~\ref{sizestimes}b shows the convolution time for grids with $N$ bins in dimensions 
$x_1$ and $x_2$ for the sparse structure. Results are given for the
trimmed and untrimmed structures
% where 
% unfilled data elements have been trimmed and the dimension ranges set appropriately, 
%and the case where the grid has not been trimmed are shown. 
In the case of the untrimmed grid, 
all data elements are retained in the convolution, even those with no entries.
%
%

%It is clear that 
Excluding the unfilled data elements in the convolution improves the
convolution time by a factor approaching two. In addition, we see that the convolution time
varies approximately linearly with the grid linear dimension. This is because the most costly
part of the convolution is the calculation of the PDF at the grid nodes.
With independent grid nodes for $x_1$ and $x_2$, there are $2N$ evaluations of the PDF
for each observable bin, and so the convolution scales linearly with $N$.

In conclusion, the custom sparse structure using trimmed blocks gives the best
performance.

\section{Application example: Calculation of NLO QCD uncertainty for inclusive
jet cross-sections for proton proton collisions at various centre-of-mass energies}
\label{sec:jetuncertainties}
As an example in this section the uncertainties of the inclusive jet cross-section in the central region ($0<y<1$) are evaluated
from the default grid obtained at a centre-of-mass energy of $\sqrt{s} = 14000$~GeV. The jet cross-sections are calculated at various centre-of-mass
energies. 
The most recent PDF parameterisations along with their associated uncertainties
are used, i.e.  CTEQ6.6 \cite{CTEQ66}, MSTW2008 \cite{Martin:2009iq}, HERAPDF01 \cite{herapdf01}
and  NNPDF  \cite{Ball:2008by}. 

\begin{figure}[htp]
%\vspace{-2cm}
\centering
\includegraphics[width=0.49\textwidth]{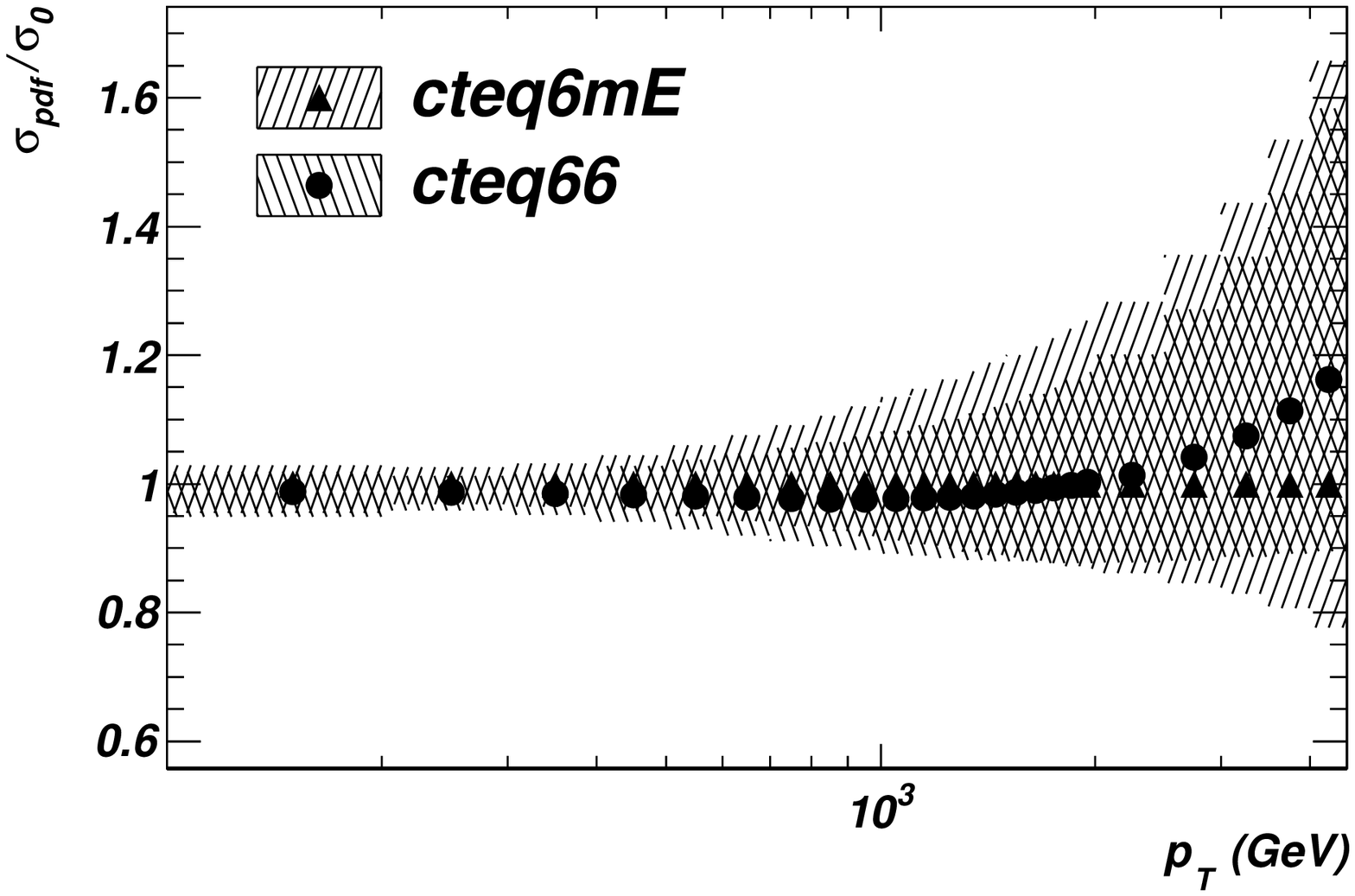}
\includegraphics[width=0.49\textwidth]{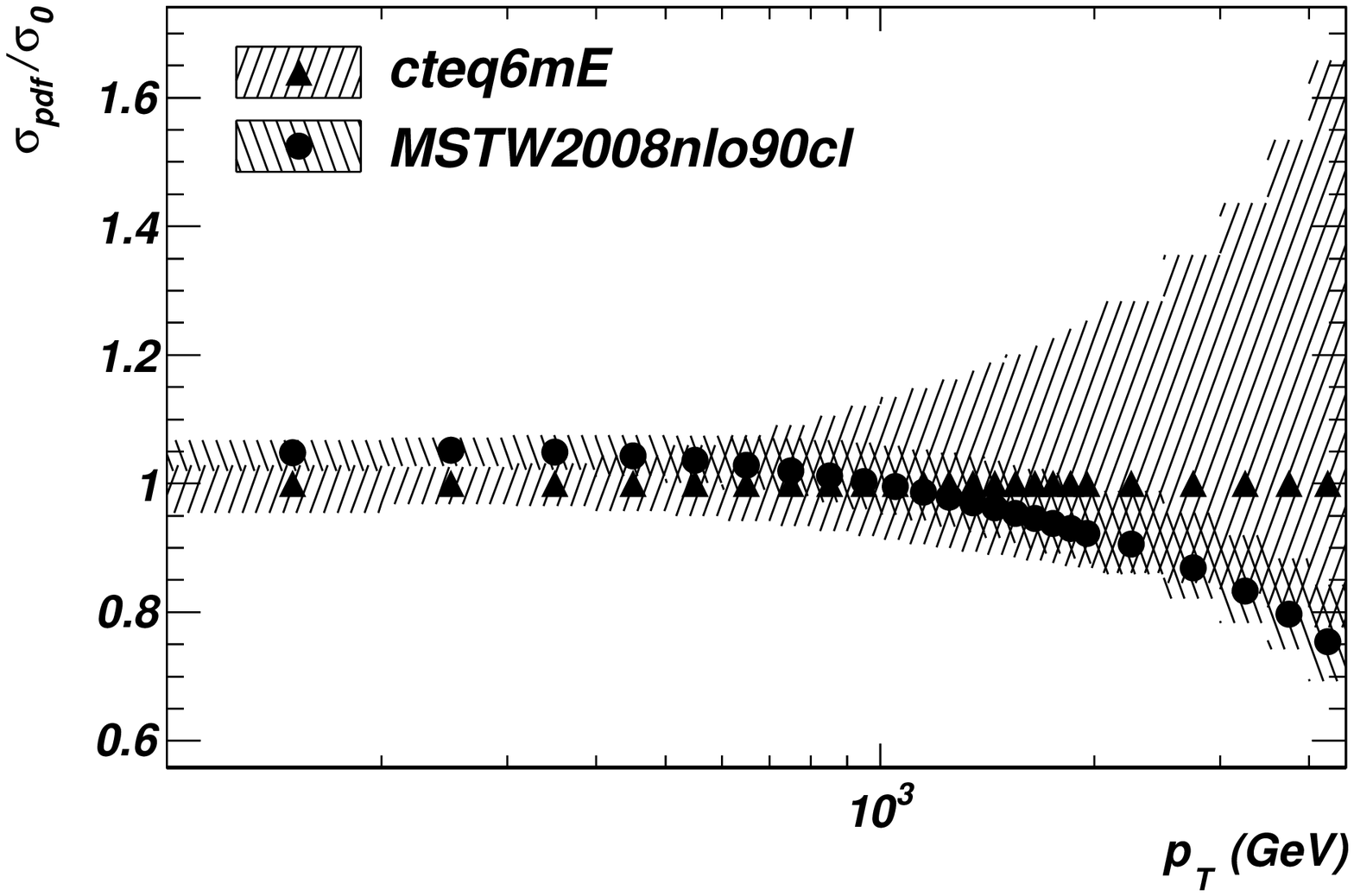}
\includegraphics[width=0.49\textwidth]{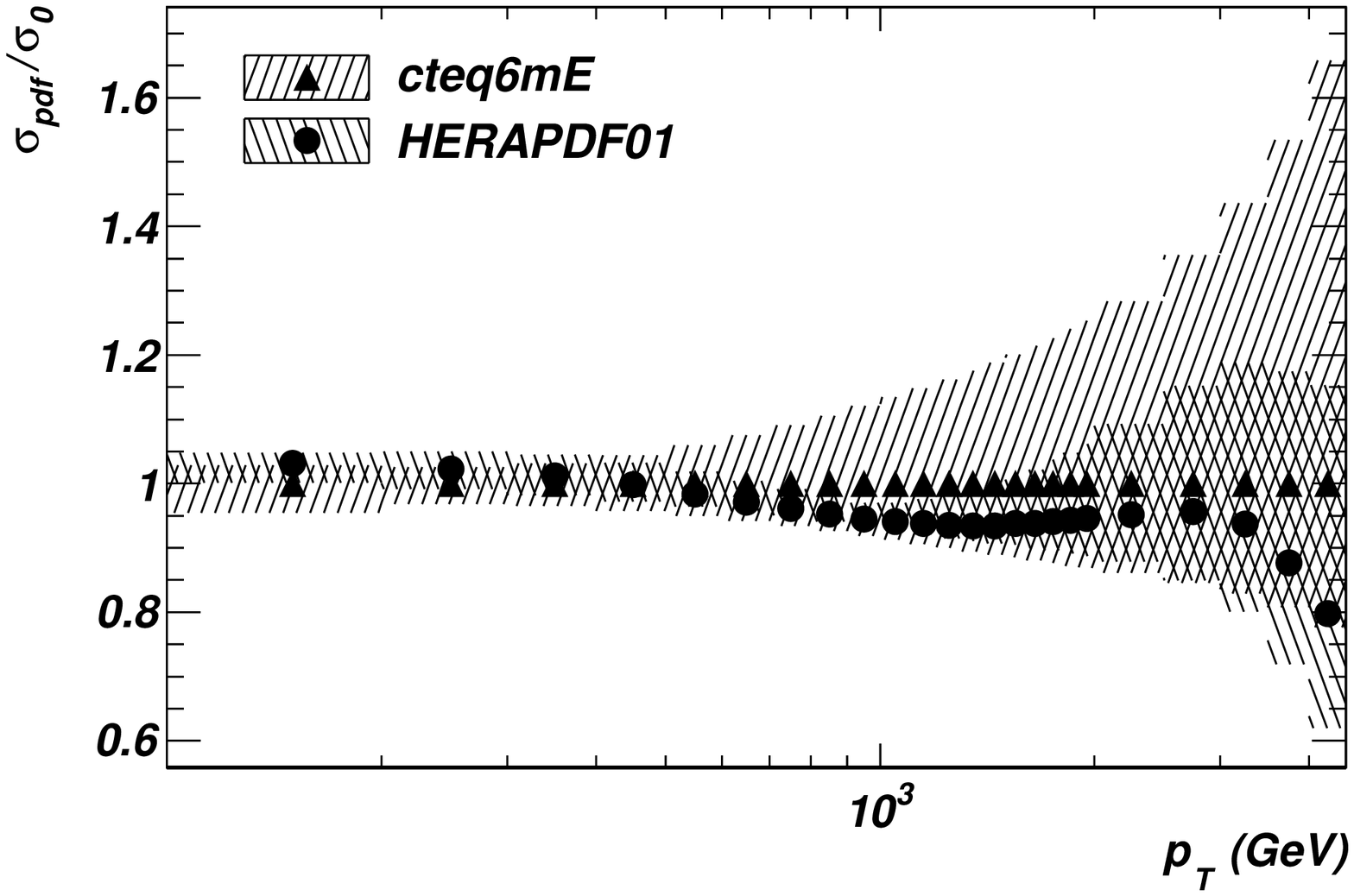}
\includegraphics[width=0.49\textwidth]{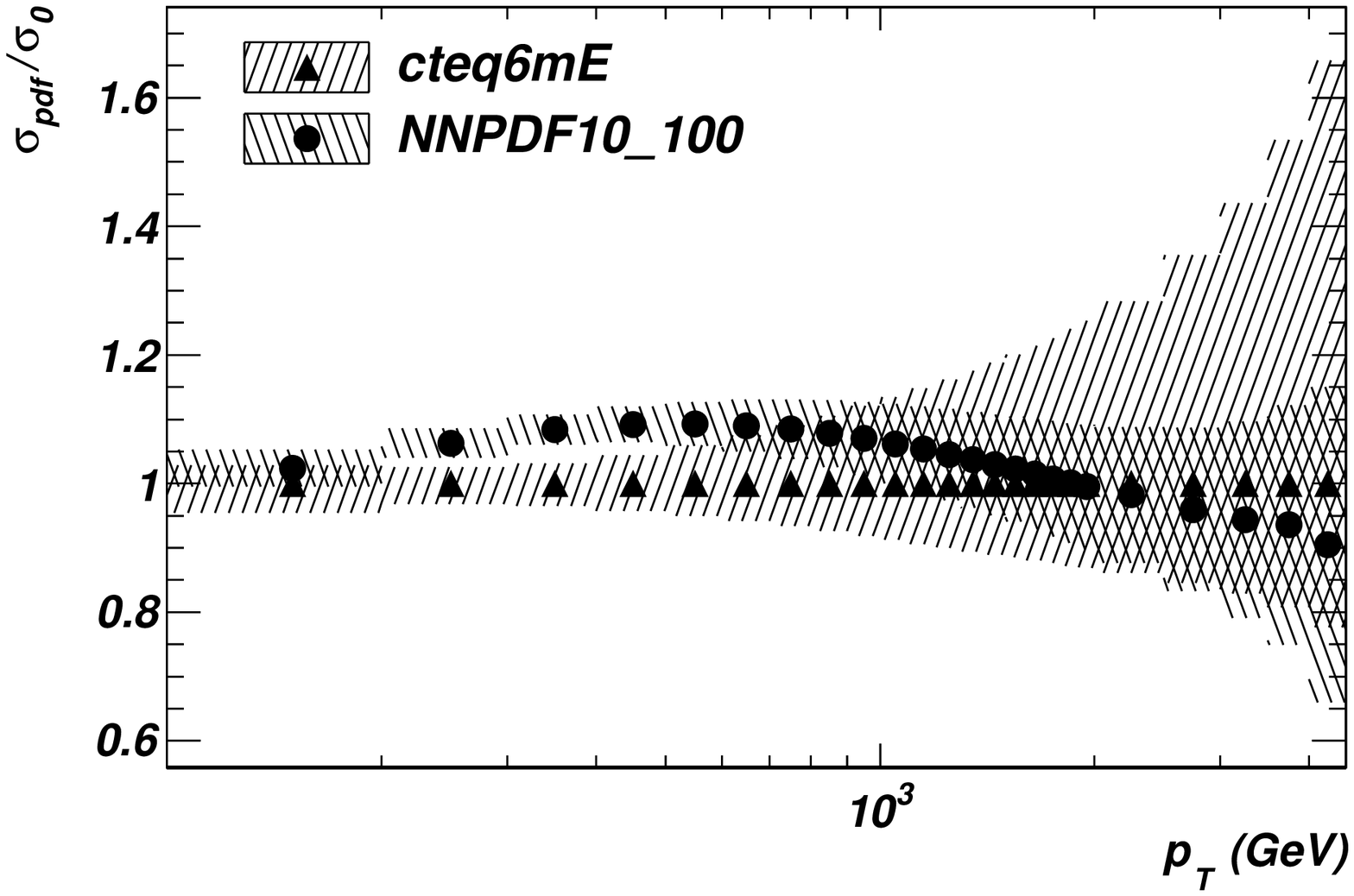}
%\vspace{0.5cm}
\begin{picture}(0,0)
\put( -450,0){c)}
\put( -210,0){d)}
\put( -450,155){a)}
\put( -210,155){b)}
\end{picture}
\caption{PDF uncertainty of the inclusive jet cross-section for jets 
within $0< y < 1$ as a function of the transverse jet momentum \pt
for a centre-of-mass energy of $\sqrt{s} = 14000$~GeV. 
Shown is the jet cross-section uncertainty induced by
the CTEQ6mE PDF and 
the CTEQ6.6 (a) the MSTW2008 (b), the HERAPDF (c) and the NNPDF PDF (d).
The reference cross-section $\sigma_0$ is the one obtained by the central value of the CTEQ6mE PDF.
The default PDF is indicated by a marker.
\label{fig:pdfuncertaintyvspt}
}
\end{figure}

\begin{figure}[htp]
%\vspace{-2cm}
\centering
\includegraphics[width=0.49\textwidth]{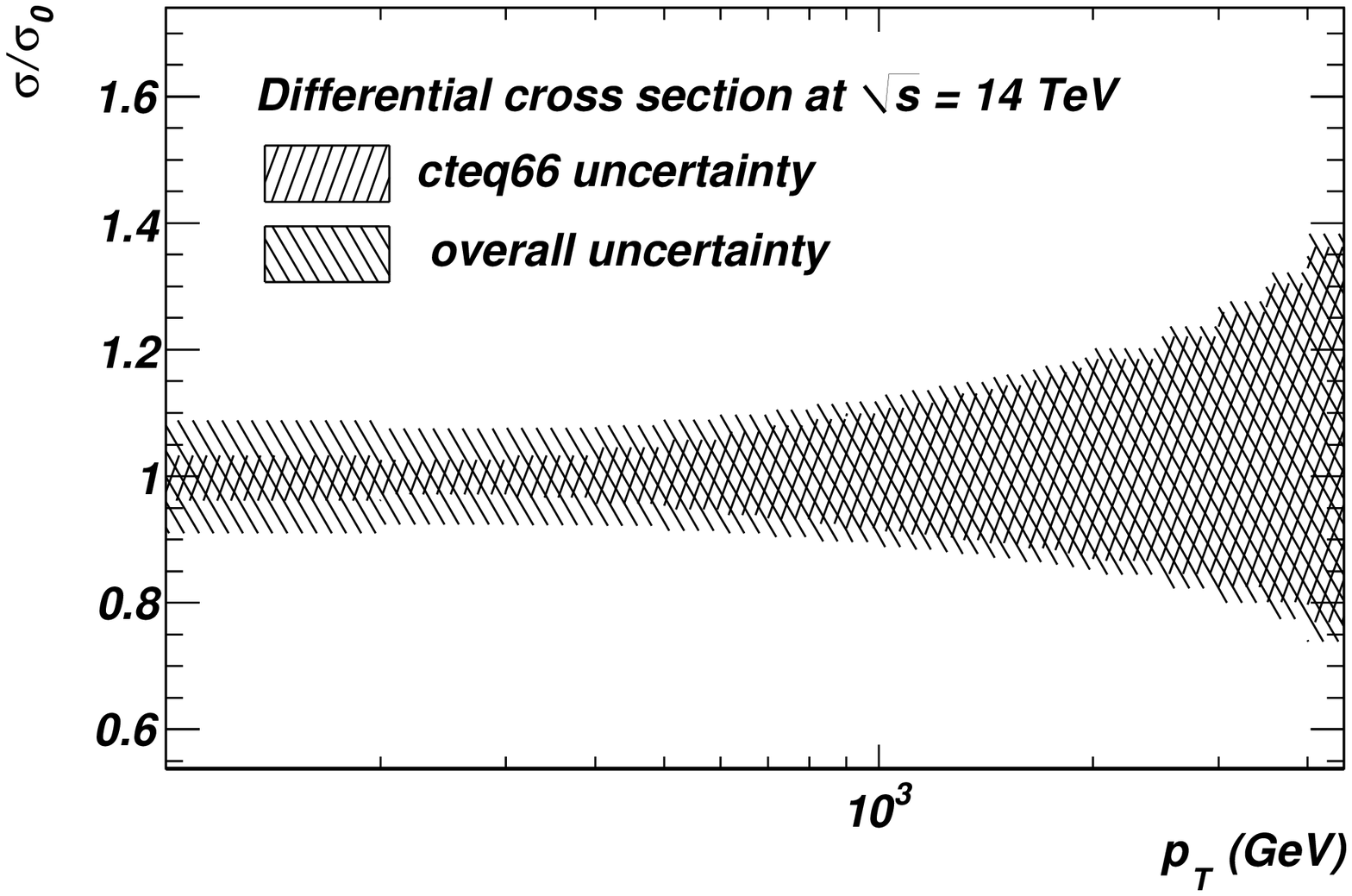}
\includegraphics[width=0.49\textwidth]{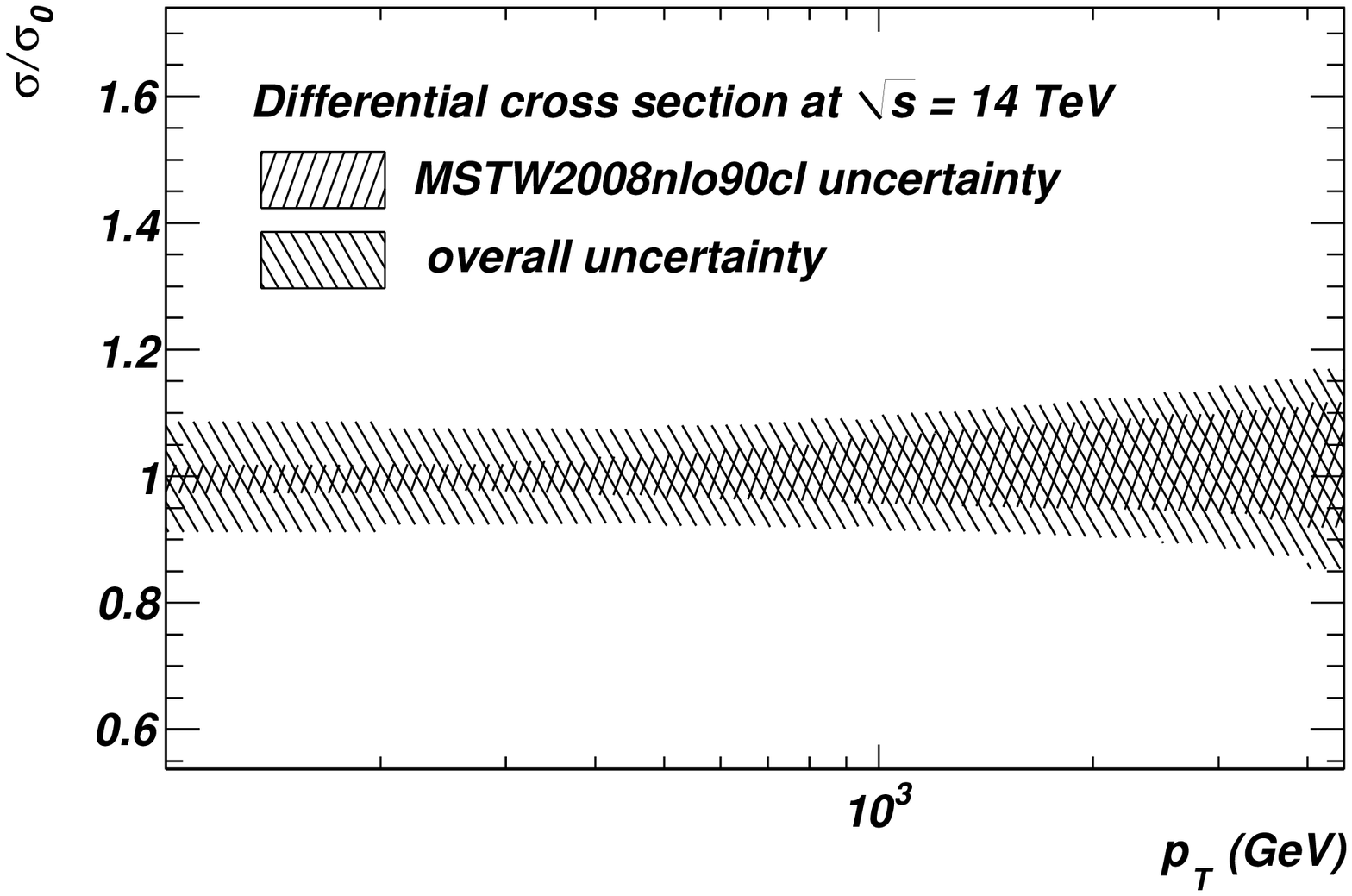}
\includegraphics[width=0.49\textwidth]{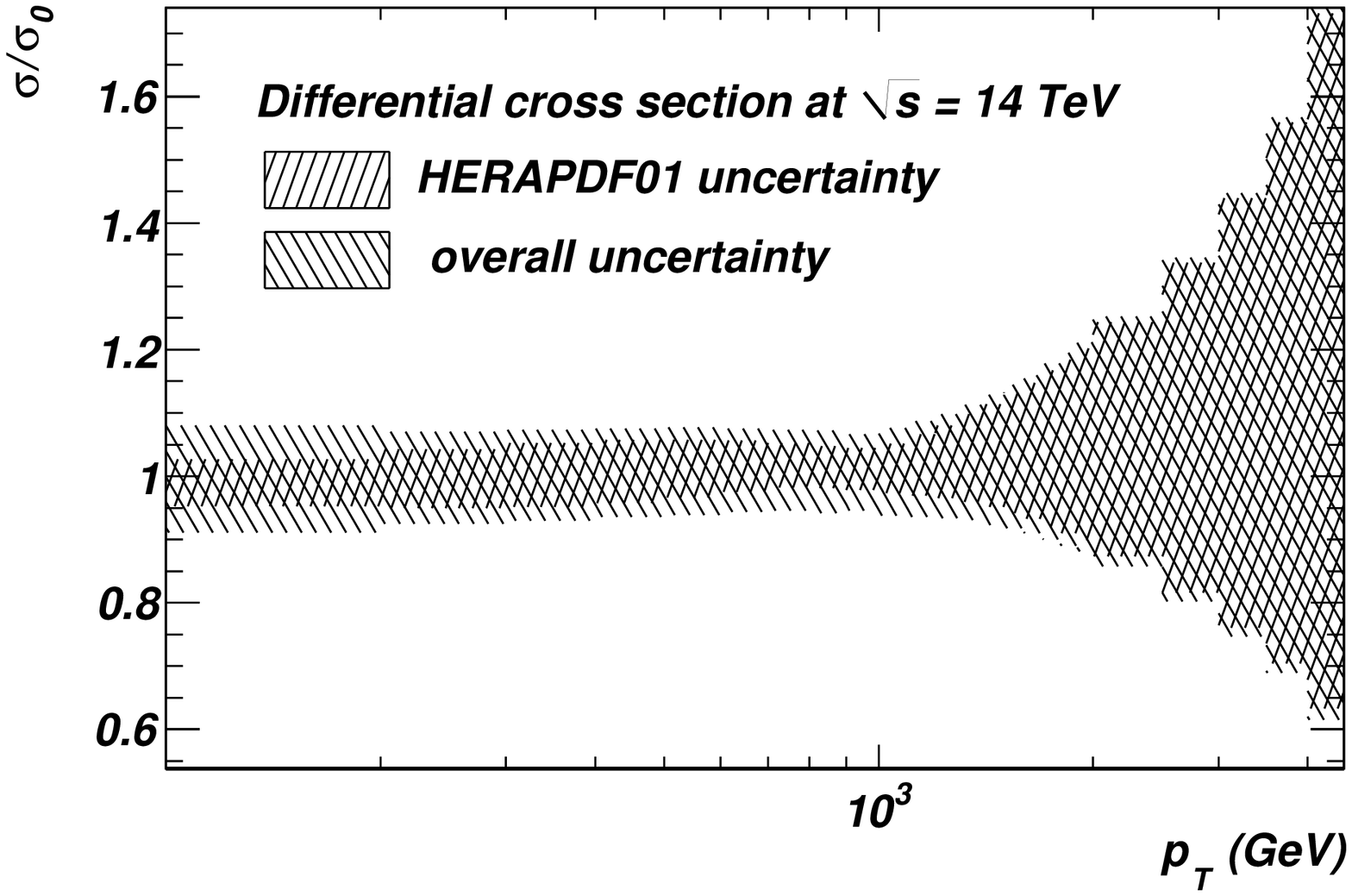}
\includegraphics[width=0.49\textwidth]{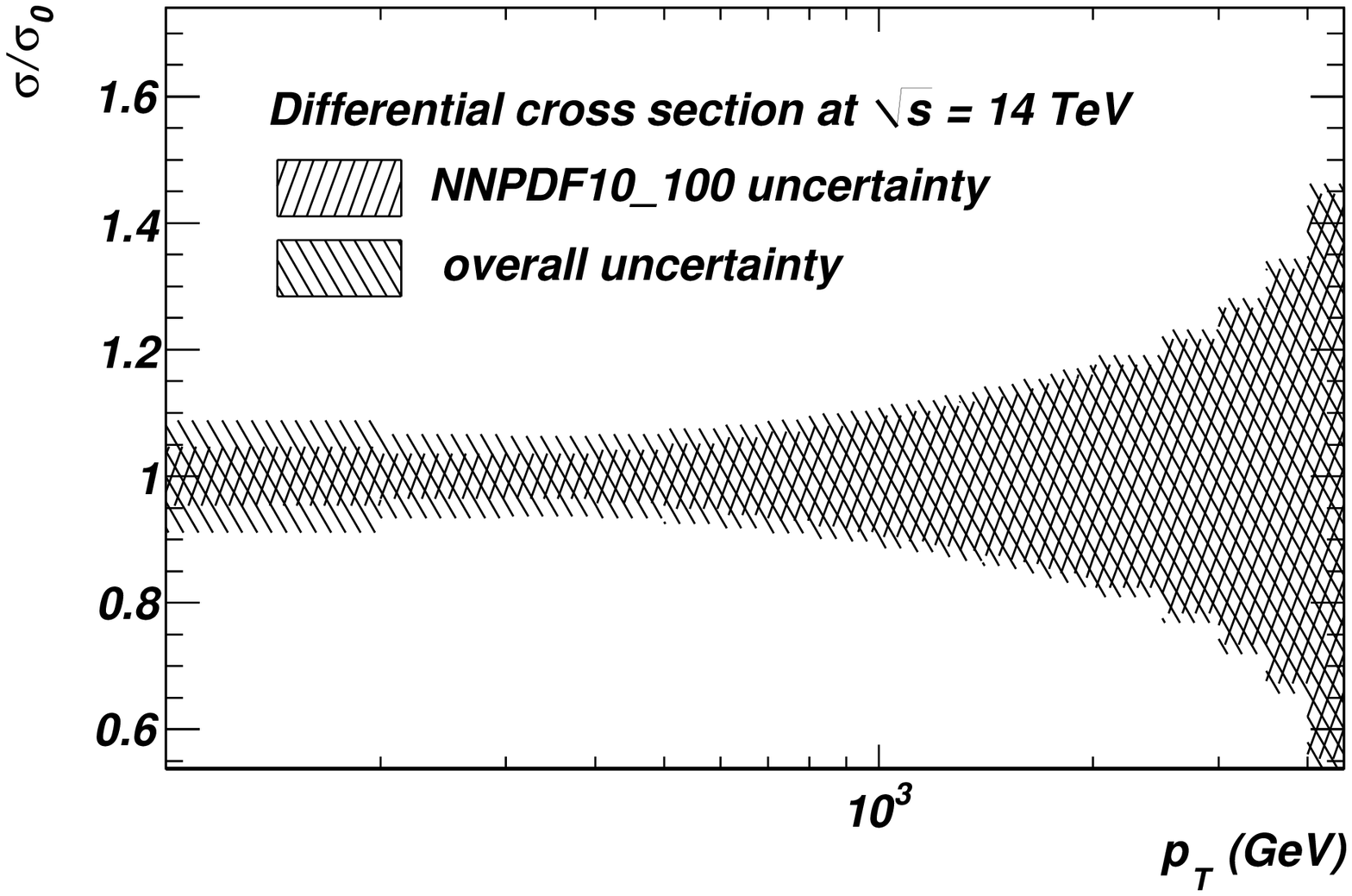}
%\vspace{0.5cm}
\begin{picture}(0,0)
\put( -450,0){c)}
\put( -210,0){d)}
\put( -450,155){a)}
\put( -210,155){b)}
\end{picture}
\caption{Uncertainty of the inclusive jet cross-section for jets 
within $0< y < 1$ as a function of the transverse jet momentum \pt
at fixed  centre-of-mass energy $\sqrt{s} = 14000$~GeV. 
Shown is the ratio of the cross-section with varied PDFs and
renormalisation and factorisation scales ($\sigma$) 
to the cross-section calculated
with the central value of each PDF set and no scale variation,
i.e. $\mu_r = \mu_f = 1$ ($\sigma_0$).
The inner uncertainty band shows only the PDF uncertainty. 
The outer band shows the PDF and the scale uncertainty added
in quadrature. The uncertainty of  CTEQ6.6 is shown in a), 
of MSTW2008 in b), of HERAPDF in c) and of NNPDF in d).
%The reference cross-section $\sigma_0$ is the one obtained by the central PDF
%in each of the cases.
\label{fig:uncertaintyvspt}
}
\end{figure}

Fig.~\ref{fig:pdfuncertaintyvspt} shows the effect of the 
PDF uncertainty from 
CTEQ6.6 (a), MSTW2008 (b), HERAPDF01 (c) and  NNPDF (d) 
on the inclusive jet cross-section with respect to the central value of 
the somewhat older PDF, CTEQ6mE \cite{Pumplin:2002vw}.
The uncertainty from the CTEQ6mE PDF is also overlayed.
The band illustrates the result of adding the jet cross-sections
obtained for each of the PDF variations\footnote{
The uncertainty band is obtained
using eq.~51 and eq.~52 in ref.~\cite{Martin:2009iq}
for the HERA, MSTW and the CTEQ PDFs. This formula
has also been suggested earlier in ref.~\cite{Campbell:2006wx}.
For the NNPDF eq.~164 in ref.~\cite{Ball:2008by} is used.
The uncertainty in the NNPDF corresponds to the standard
deviation of all variations, while
in the case of the other PDFs it corresponds to the 90\%
confidence limit. For better comparison, the uncertainties of the NNPDF and the HERAPDF01
have been scaled up using eq.~$165$ in ref.~\cite{Ball:2008by}.
}.
The marker indicates the central value. 
Fig.~\ref{fig:uncertaintyvspt} shows the PDF uncertainty together with the
renormalisation and factorisation scale uncertainty added in quadrature
with respect to the central value for each of the PDFs.

%%%%%%%%%%%%%%%%%%%%%%%%%%%%%%%%%%%%%%%%%%%%%%%%%%%%%%%%%%%%%%%%%%%%%%%%%%%%%%%%%%%%%%%
%%%%%%%%%%%%%%%%%%%%%%%%%%%%%%%%%%%%%%%%%%%%%%%%%%%%%%%%%%%%%%%%%%%%%%%%%%%%%%%%%%%%%%%%%%

%%%%%%%%%%%%%%%%%%%%%%%%%%%%%%%%%%%%%%%%%%%%%%%%%%%%%%%%%%%%%%%%%%%%%%%%%%%%%%%%%%%%%%%
%%%%%%%%%%%%%%%%%%%%%%%%%%%%%%%%%%%%%%%%%%%%%%%%%%%%%%%%%%%%%%%%%%%%%%%%%%%%%%%%%%%%%%%%%%

For $\ptx < 1000$ GeV the jet cross-section obtained with CTEQ6.6
is about $2\%$ smaller than from CTEQ6mE. 
Above this value
the CTEQ6.6 cross-section increases with respect to the one from  CTEQ6mE
as the jet \pt increases. At $\pt = 2000$ GeV it is about equal
and at $\pt = 4000$ GeV it is about $10\%$ larger.
The uncertainty is reduced for the CTEQ6.6 PDF.
The uncertainty is about $3$\% for $\ptx < 500$ GeV, 
about $8$\% at $\pt = 1000$ GeV
and about $20$\% at $\pt = 3000$ GeV.

The MSTW2008 PDF gives a jet cross-section that is $5\%$ larger than the one
obtained with CTEQ6mE at $\ptx < 500$ GeV and 
is about the same at $\pt = 1000$ GeV and then further decreases.
%$10\%$ 
%lower at very large $\pt = 1000$ GeV. 
The uncertainty is only about $2$ \% for $\ptx < 500$~GeV
and then increases to about $6\%$ at $\pt = 1000$~GeV. 
The MSTW2008 PDF gives a smaller uncertainty than the CTEQ6.6 PDF.
It seems that the differences between the jet cross-section calculated
with  CTEQ6.6 and MSTW2008 are a bit larger than the individual uncertainties.

The result obtained with the HERAPDF01 is more similar to the one obtained
from\linebreak[3] MSTW2008 than the one from CTEQ6.6. At low \pt the central value is about $2\%$
higher than the one from CTEQ6mE. In the region $500 < \ptx < 1500$ GeV
the HERAPDF01 predicts a lower jet cross-section than the other PDFs.
The uncertainty is about $5 \%$ for $\ptx <1000$ GeV and then increases
to about $20 - 40\%$ at $\pt = 3000$~GeV.
The small uncertainty of the jet cross-section calculated with the
HERAPDF01 is remarkable, since only DIS data are used. 
However, model and parametrization
uncertainties are not included in this cross-section calculation.
The MSTW and CTEQ sets do not yet include the most recent HERA data. 
%PDFs might achieve similar accuracy on the jet cross-section
%prediction at lower \ptx, when the most recent HERA data will be included. 
%
The NNPDF predicts jet cross-sections that are $5-10\%$ higher
than the one from the other PDFs; in particular in the region
$300 < \ptx < 1000$~GeV. The uncertainty is about $5\%$ at low \ptx,
$10\%$ at $1000$ GeV and $20-30\%$ at $3000$~GeV.

The overall uncertainty, i.e. including the PDF and the scale
variation added in quadrature, is shown in 
Fig.~\ref{fig:uncertaintyvspt}.
It is about $8\%$  up to a \pt of about $1000$~GeV and 
then increases towards higher \ptx. It is about $20 -30$\% at $\pt = 3000$~GeV.
For very high \pt the PDF uncertainty dominates.
%The scale uncertainty dominates over the PDF uncertainty
%for $\pt < 1000$~GeV.

%%%%%%%%%%%%%%%%%%%%%%%%%%%%%%%%%%%%%%%%%%%%%%%%%%%%%%%%%%%%%%%%%%%%%%%%%%%%%%%%%%%%%%%
\begin{figure}[htp]
%\vspace{-2cm}
\centering
\includegraphics[width=0.49\textwidth]{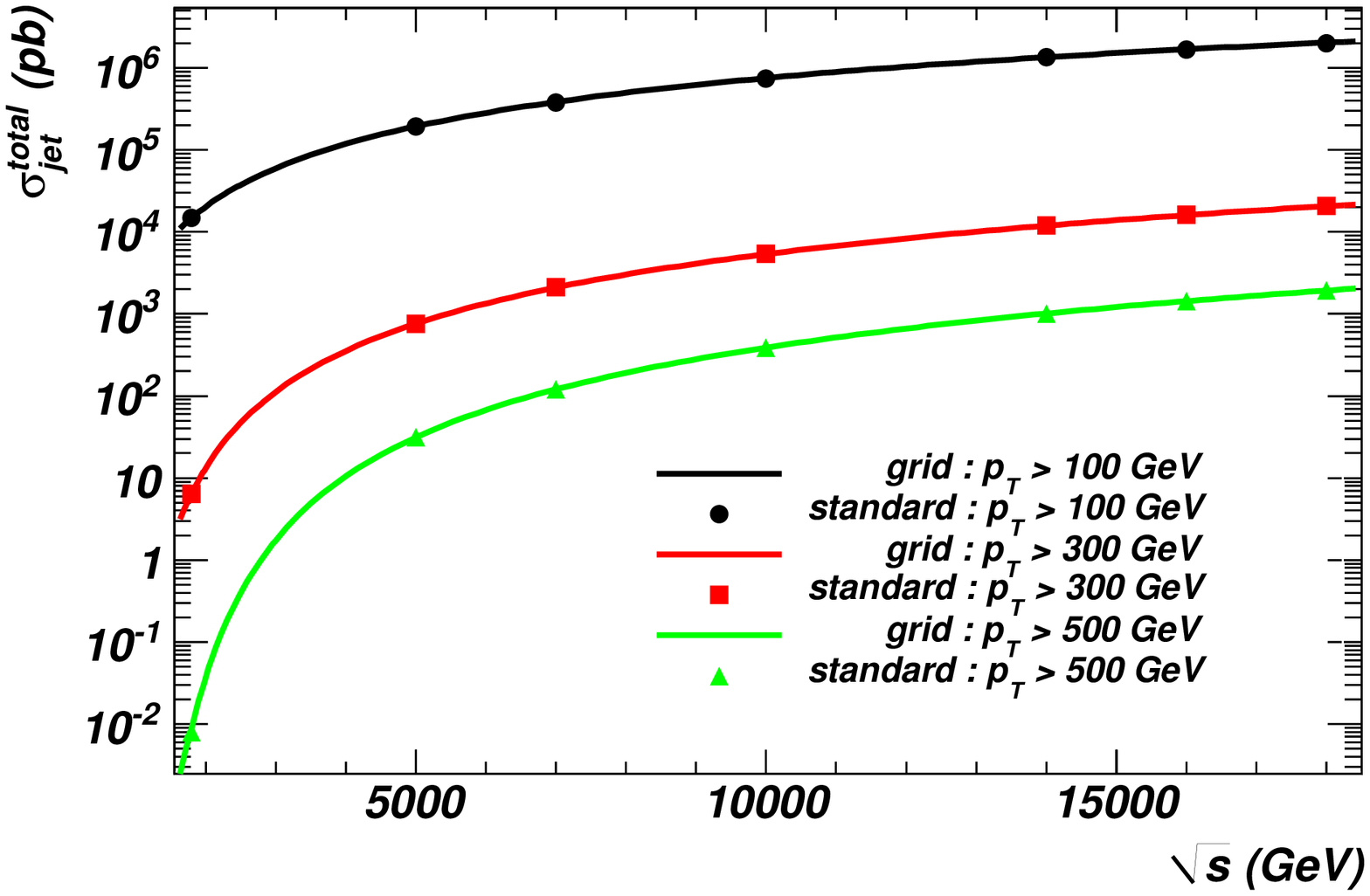}
\includegraphics[width=0.49\textwidth]{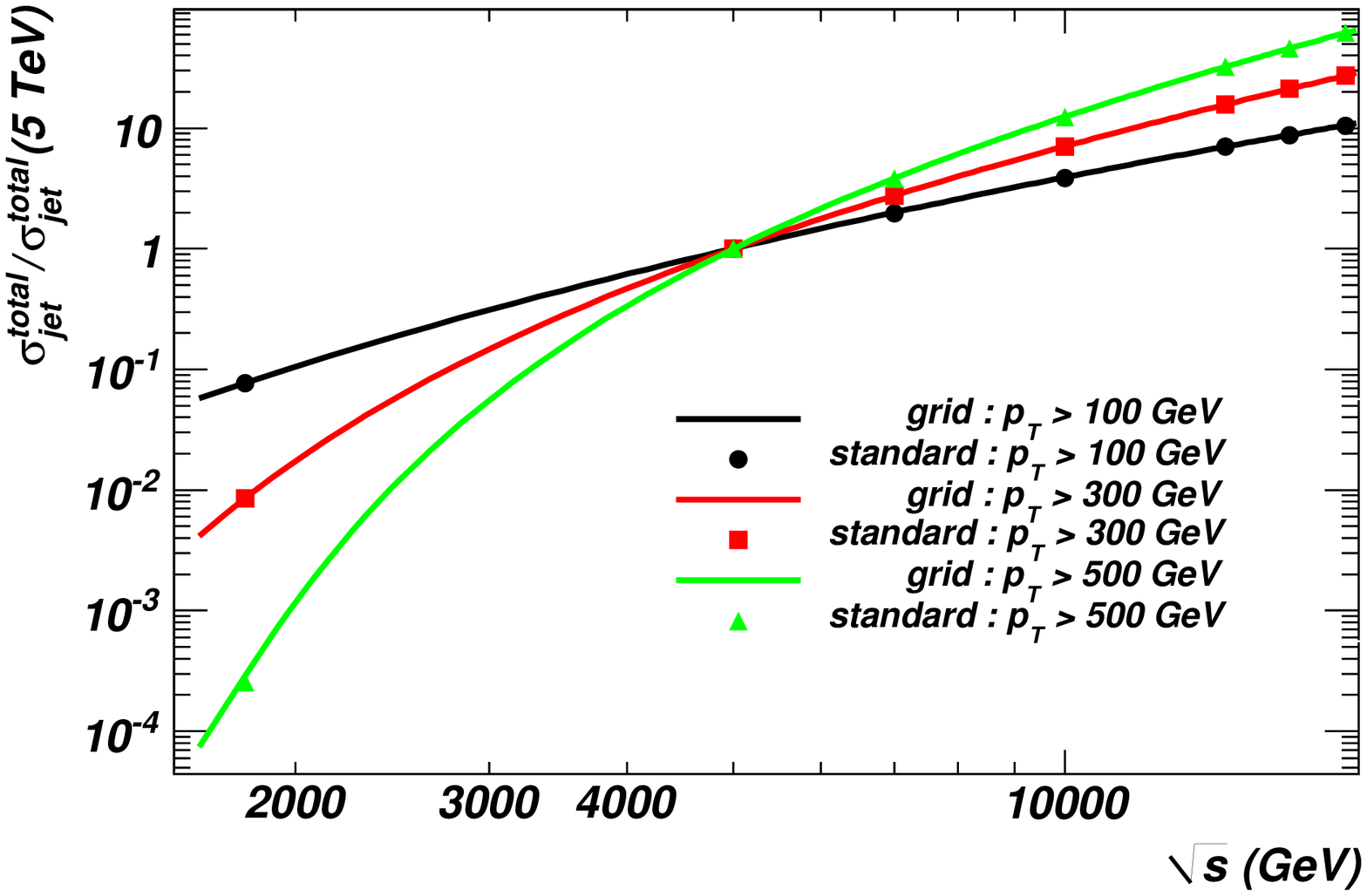}
%\vspace{0.5cm}
\begin{picture}(0,0)
\put( -450,0){a)}
\put( -210,0){b)}
\end{picture}
\caption{a) Inclusive jet cross-section for jets within $0< y < 1$ and with
 transverse jet momenta $\pt > 100$ GeV, $\pt > 300$ GeV and $\pt > 500$ GeV
as a function of the centre-of-mass energy $\sqrt{s}$.
b) Shows the same as a), but all results are normalised to $\sqrt{s} = 5000$ GeV.
The markers indicate the results calculated at each centre-of-mass
energy in the standard way. The lines indicate the
results deduced from the default weight grid produced for a centre-of-mass energy
at $\sqrt{s} = 14000$~GeV. 
\label{fig:totalxs}
}
\end{figure}
%%%%%%%%%%%%%%%%%%%%%%%%%%%%%%%%%%%%%%%%%%%%%%%%%%%%%%%%%%%%%%%%%%%%%%%%%%%%%%%%%%%%%%%%%%

%%%%%%%%%%%%%%%%%%%%%%%%%%%%%%%%%%%%%%%%%%%%%%%%%%%%%%%%%%%%%%%%%%%%%%%%%%%%%%%%%%%%%%%
\begin{figure}[htp]
%\vspace{-2cm}
\centering
\includegraphics[width=0.49\textwidth]{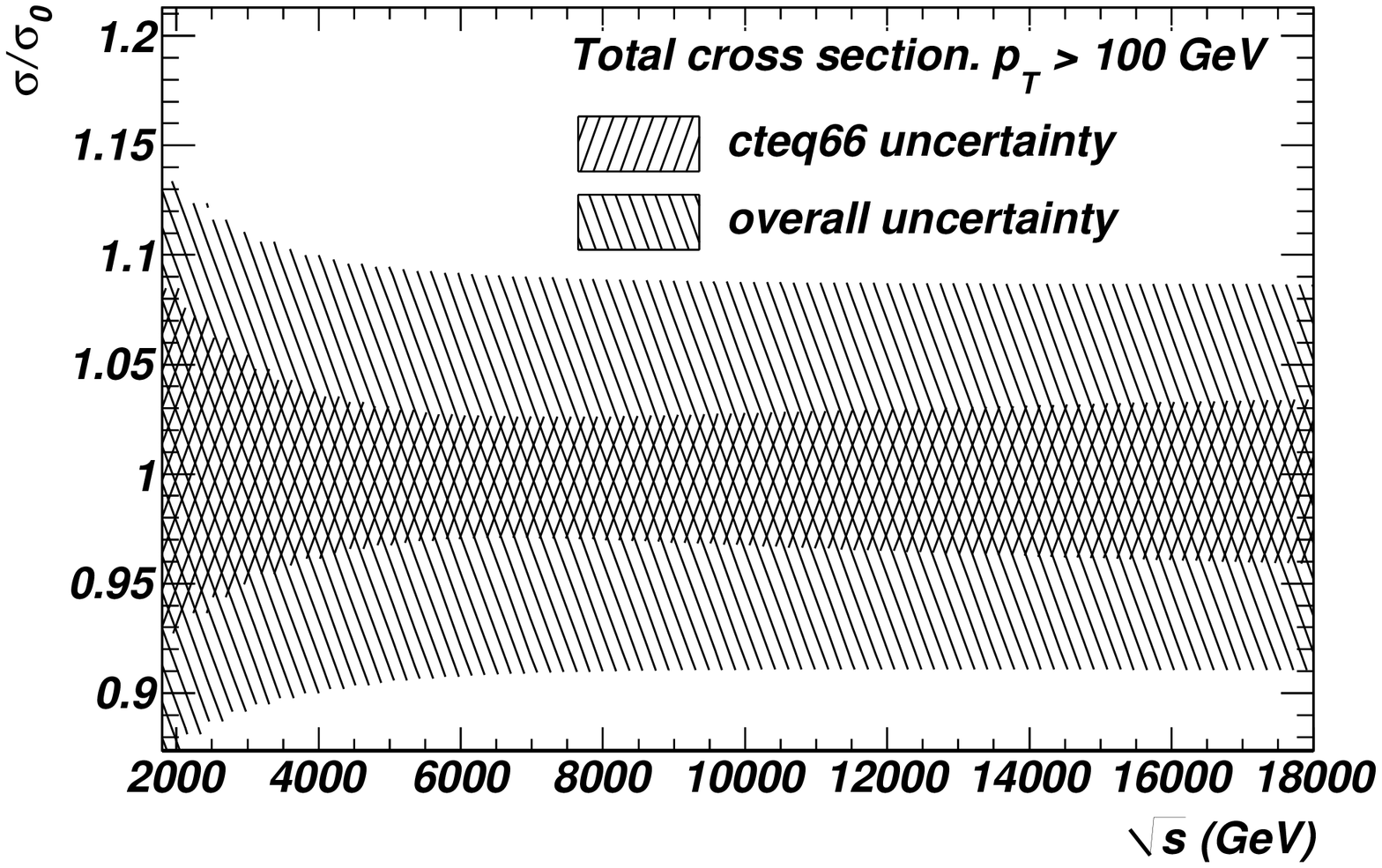}
\includegraphics[width=0.49\textwidth]{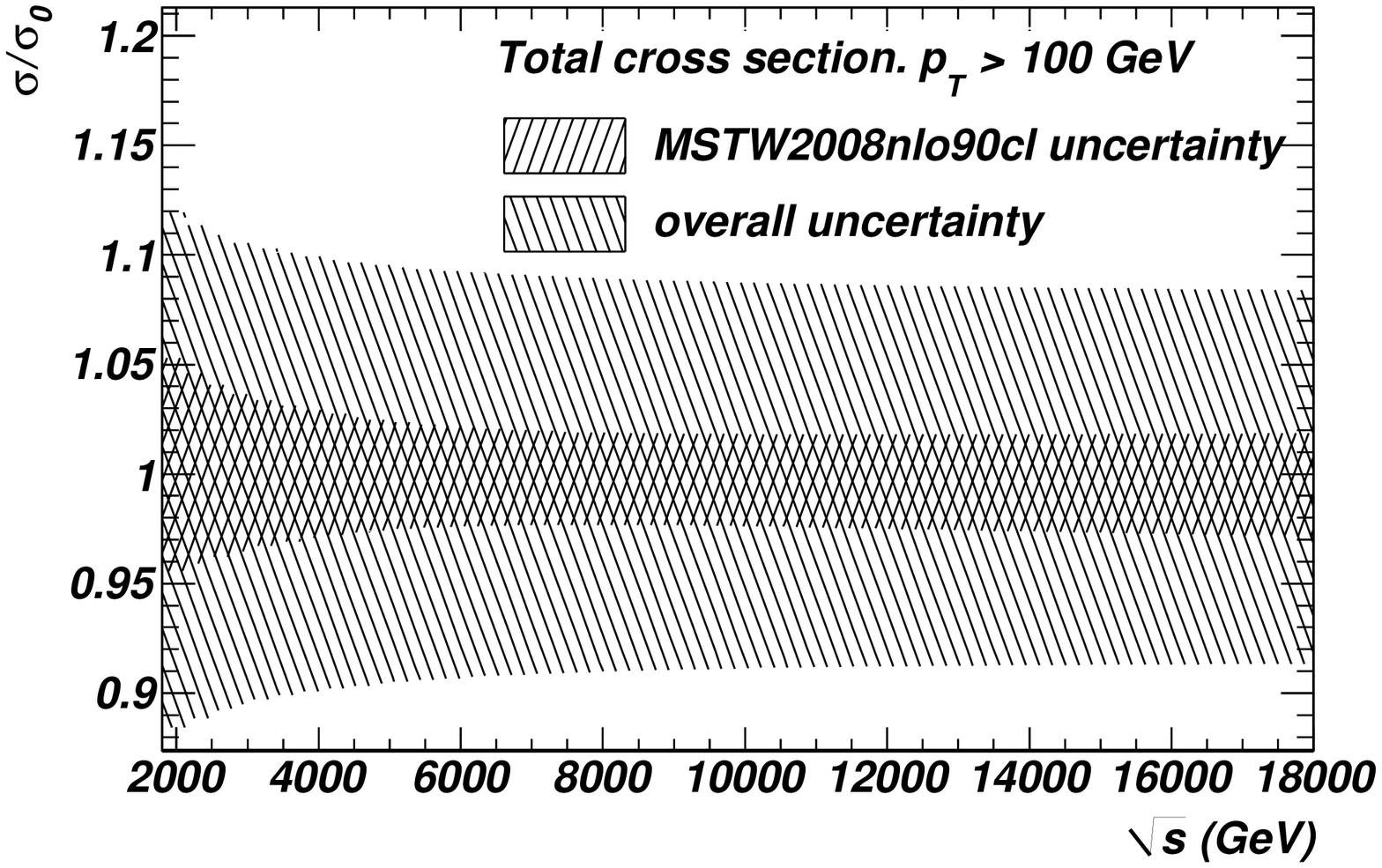}
\includegraphics[width=0.49\textwidth]{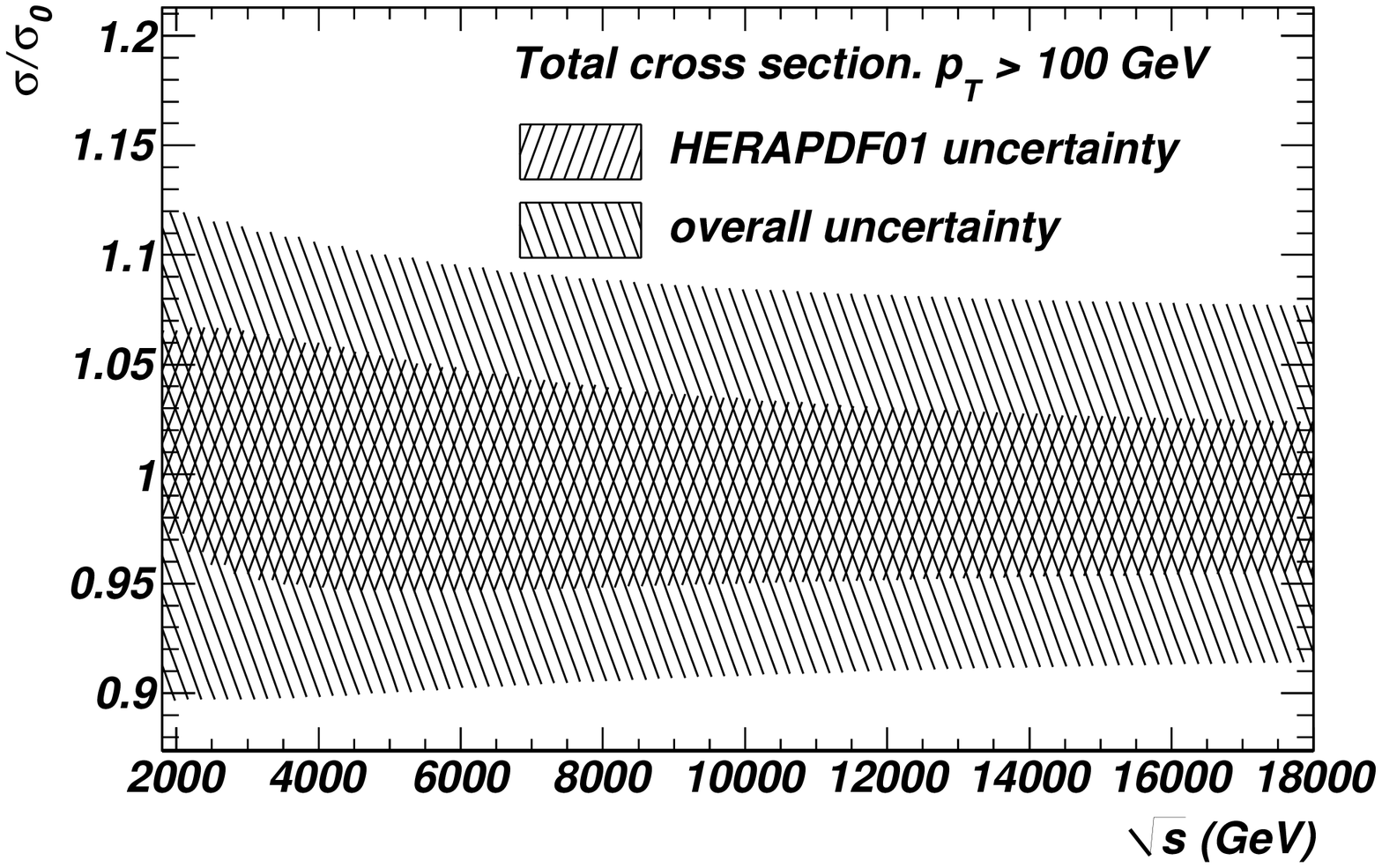}
\includegraphics[width=0.49\textwidth]{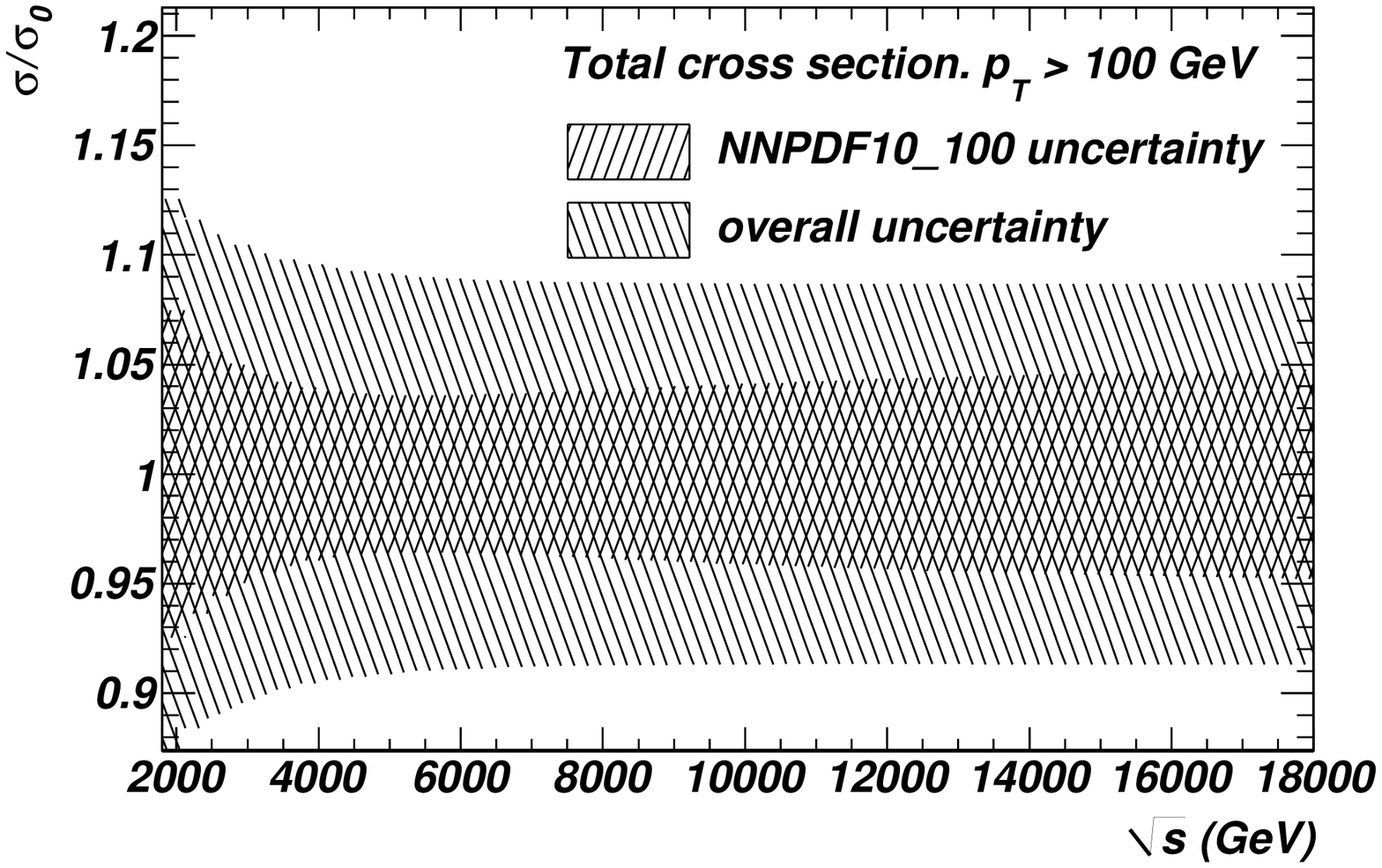}
%\vspace{0.5cm}
\begin{picture}(0,0)
\put( -450,0){c)}
\put( -210,0){d)}
\put( -450,155){a)}
\put( -210,155){b)}
\end{picture}
\caption{Uncertainty of the inclusive jet cross-section for jets with transverse
momenta $\pt > 100$~GeV and within $0< y < 1$ as a function of the centre-of-mass
energy $\sqrt{s}$. Shown is the ratio of the cross-section with varied PDFs and
renormalisation and factorisation scale ($\sigma$)
to the cross-section calculated with the central value of each PDF set
and no scale variation ($\sigma_0$).
The inner uncertainty band shows only the PDF uncertainty. 
The outer band shows the PDF and the scale uncertainty added
in quadrature. The uncertainty from CTEQ6.6 is shown in a), 
from MSTW2008 in b), from HERAPDF in c) and from NNPDF in d).
\label{fig:uncertaintyvscms}
}
\end{figure}
%%%%%%%%%%%%%%%%%%%%%%%%%%%%%%%%%%%%%%%%%%%%%%%%%%%%%%%%%%%%%%%%%%%%%%%%%%%%%%%%%%%%%%%%%%
Fig.~\ref{fig:totalxs}a) shows the total inclusive cross-section for central jets
($0 < y < 1$) integrated for $\ptx > 100$~GeV, $\ptx > 300$~GeV and $\ptx 
> 500$~GeV
as a function of the centre-of-mass energy. The markers denote the reference
cross-section calculated in the standard way. The lines are obtained from a weight
grid produced at $\sqrt{s} = 14000$~GeV. 
The cross-section calculation from the default weight grid reproduces 
the reference cross-sections within $1-2\%$
(see also section \ref{sec:cmsreweightingperformance}).
For each jet transverse momentum threshold
the total jet cross-section rises with increasing centre-of-mass energy.
Fig.~\ref{fig:totalxs}b) shows the centre-of-mass energy dependence
of the jet cross-section normalised to the jet cross-section
at $5000$ GeV for each jet transverse momentum threshold.
As expected the centre-of-mass energy dependence is strongest
for high transverse jet momenta.

Fig.~\ref{fig:uncertaintyvscms} shows for each of the considered PDF sets
the PDF uncertainty along with the renormalisation and factorisation scale
uncertainty added in quadrature for jets with $\ptx > 100$ GeV as a 
function
of the centre-of-mass energy $\sqrt{s}$. Both the PDF and the scale uncertainties
only depend weakly on the centre-of-mass energy.  For high centre-of-mass energies
the uncertainties are a bit smaller.

%\newpage
\section{Application example: PDF fit including DIS data and jet production data at hadron colliders}
\label{sec:pdffit}
An important application of the method outlined above, is the 
consistent inclusion of final state 
measurements from hadronic colliders into the final extraction of PDFs by NLO QCD fits. 
Measurements of final states -- such as jet production 
or the production of lepton pairs via the Drell-Yan process -- 
can provide important additional constraints 
on the proton PDFs, complementary to those from inclusive DIS data.

As a simple ``proof-of-principle'' example, the grid technique 
outlined in this paper has been used to include simulated 
LHC jet data into a NLO QCD fit. % for proton PDFs. 
%High-transverse-momentum jet data is directly sensitive to the high-$x$ gluon PDF.   
The fit framework used here is based on the recent ZEUS-JETS PDF,  
derived from a fit to inclusive DIS and jet data from HERA.
Jet cross-sections from the TEVATRON or any other data than that from HERA
are not used. 
Full details of the data-sets, PDF parameterisation and other assumptions 
are given elsewhere\cite{zeusjets}.

To represent the LHC data for inclusion in the fit, jet production 
from proton-proton collisions at a centre-of-mass energy of $14000$ GeV
was simulated using the \texttt{JETRAD}\cite{jetrad} program, 
using the CTEQ6.1 PDF\cite{cteq61m}. 
Single inclusive jet cross sections, differential in $\ptx$, 
were obtained in three regions of rapidity: $0<|y|<1$, $1<|y|<2$ and $2<|y|<3$. 
%Corresponding event weight grids were also produced,  using 
A grid with default parameters, as described in
Sec.~\ref{sec:jetsresults}, was produced 
%
%default weight grid definition, as described in Sec.~\ref{sec:jetsresults}, were produced 
and interfaced to the ZEUS NLO QCD fit program. 
Several fits were performed, using different assumptions on the statistical 
and systematic uncertainties on the simulated data. The PDF uncertainties  
were calculated using the Hessian method\cite{hessian1,hessian2}, with 
$\Delta\chi^2 =1$\footnote{Note that a using $\Delta\chi^2=1$ in the Hessian method 
is generally considered to underestimate the PDF uncertainties. However, the main aim of this study 
is to provide a proof-of-principle example of the use of the 
grids discussed in this paper, and not to provide qualitative estimates of weight
expected PDF uncertainties. Furthermore, all fits shown in this section 
have used the same definition of the PDF uncertainties, such that any 
comparison should still be valid.}.
%Need to say something here about that this does 
%not ncessarily give a good representation of the uncertainties, but is OK for our 
%purposes, since the main aim is to demonstrate method + we are 
%comparing like with like when we do comparisons of fits?}}. 
%Note that using a $\Delta\chi^2$ value of $1$ is 
%not nessecarily thought to provide a reliable estimate of the PDF uncertainties. 
%However, the main aim of the present study is to 
%demonstrate the use of the event weight grids as applied to LHC data.}. 
%
%Other details of the fit were the same as for the ZEUS NLO QCD analysis\cite{zeusjets}. 

\begin{figure}[ht]
%\vspace{-2cm}
\centering
\includegraphics[width=0.49\textwidth]{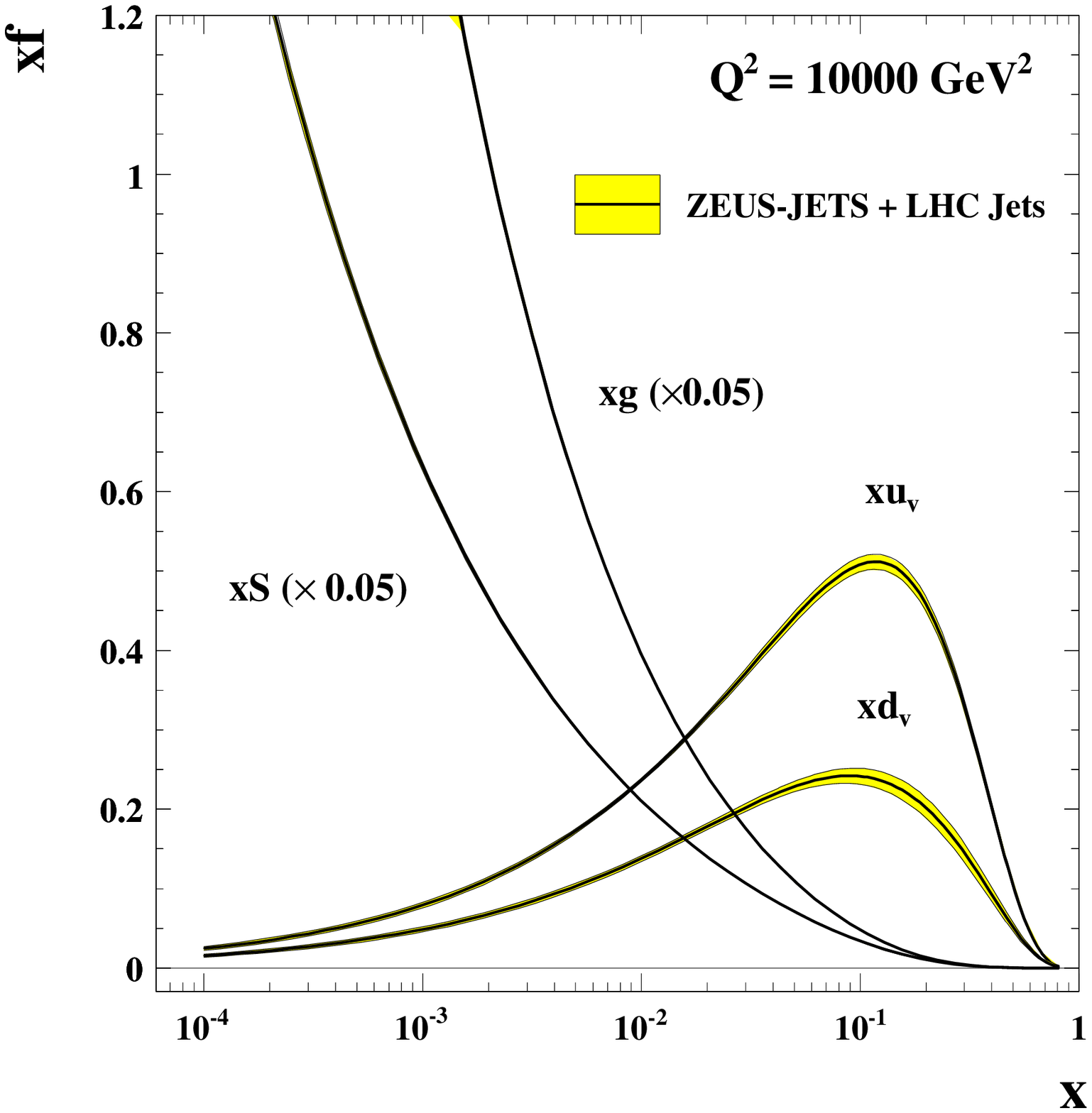}
\vspace{-0.5cm}{\includegraphics[width=0.49\textwidth, height=7.8cm]{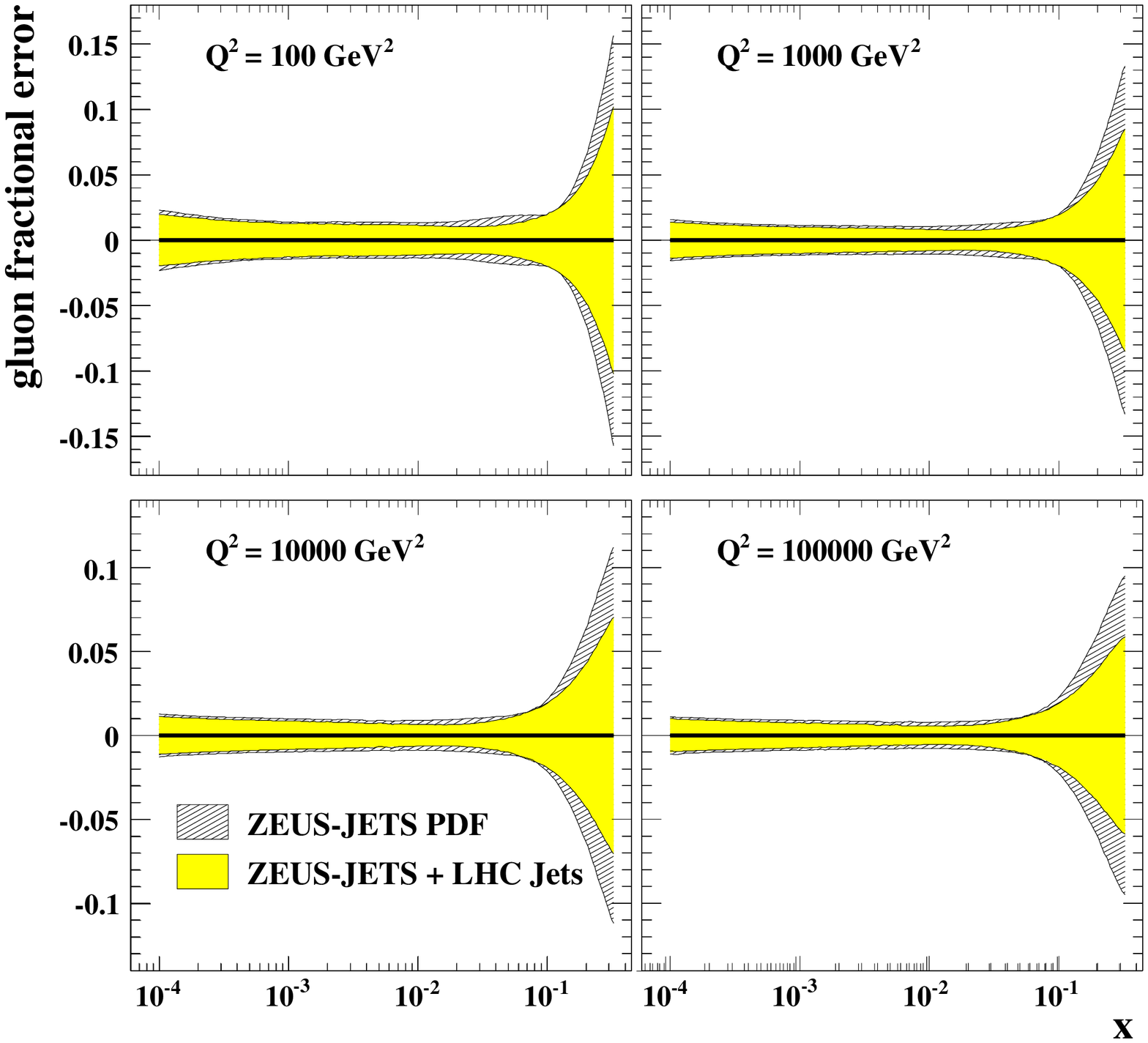}}
\vspace{0.5cm}
\begin{picture}(0,0)
%\put( -450,0){c)}
%\put( -210,0){d)}
\put( -450,0){a)}
\put( -210,0){b)}
\end{picture}\caption{
The distributions of the up-valence, down-valence, 
total sea and gluon PDFs (a),  
%Note that the sea and gluon distributions have been reduced by a factor of $\times20$ in this plot 
and the fractional uncertainty on 
the gluon distribution at a number of $Q^2$ values (b) as a function of the parton
momentum fraction $x$. 
The results from the fit using weight  grids to include 
simulated LHC jet data is shown by the shaded band. For comparison, in (b), 
the results of the ZEUS NLO QCD fit are also shown, indicated by the hatched band. 
The simulated LHC jet data included in the new fit assume a statistical uncertainty corresponding 
to an integrated luminosity of $10$ ${\rm fb}^{-1}$, uncorrelated systematics of $5\%$ and a 
jet energy scale uncertainty of $1\%$.
\label{fig:summarypdf}
}
\end{figure}

A representative result is shown in Fig.~\ref{fig:summarypdf}.  
In this example, the statistical uncertainty on the simulated LHC jet data 
corresponds to an integrated luminosity of
$10$ ${\rm fb}^{-1}$ and uncorrelated systematic uncertainties have been 
assumed to be at a level of $5\%$. A precise jet energy scale uncertainty 
of $1\%$ (corresponding to $\sim 5-15\%$ on the generated cross-sections)  
has also been assumed, and is included as a correlated systematic in the fit.  
Fig.~\ref{fig:summarypdf} a) shows the up-valence, down-valence, total sea 
and gluon PDF distributions as a function of $x$, at $Q^2=10000$ ${\rm GeV}^2$.  
The shaded band shows the results of the fit including the simulated 
LHC jet data. 
In Fig.~\ref{fig:summarypdf} b), the fractional uncertainties on the gluon PDF,   
at a number of $Q^2$ values, are shown. 

Comparison with the results from a fit 
which does not include the simulated LHC jet data indicates that some 
constraint on the high-$x$ gluon could be provided by the LHC single inclusive 
jet data\footnote{Note that the fit without the simulated LHC jet data is not 
identical to the ZEUS-JETS fit since the standard ZEUS fit\cite{zeusjets} 
uses the Offset method to determine the PDF uncertainties. 
The ZEUS fit shown here is a modified 
version of the published analysis, with uncertainties determined using the Hessian method, with $\Delta\chi^2=1$. 
Different treatments of experimental uncertainties in PDF analyses are discussed extensively 
elsewhere\cite{hessian1,hessian2,pdfuncert1,pdfuncert2}.}.
%the same as the published version\cite{zeusjets}, but with uncertainties determined 
%using the Hessian method, with $\Delta\chi^2=1$.}  
%in order to compare directly with the fit including simulated LHC jet data.}.
%, so as to be directly comparable with the fit including simulated jet data.} 
However, this is reliant on a very precise knowledge of the 
jet energy scale. % -- to a level of $\sim 1\%$ -- as assumed here. 

In fact, according to this study, a precise knowledge of the jet energy scale 
is the key factor. Other fits, which assumed a smaller integrated luminosity 
($1$ ${\rm fb^{-1}}$) or larger uncorrelated systematics ($10\%$), still indicated     
an improvement on the gluon uncertainties, 
provided the jet energy scale uncertainty was kept at a level of $\sim 1\%$. 
However, fits in which the latter uncertainty was 
assumed to be larger, indicated little or no improvement in the 
gluon uncertainty compared to the reference. 
More details can be found in ref. \cite{CooperSarkar:2007pj}.

%%%%%%%%%%%%%%%%%%%%%%%%%%%%%%%%%%%%%%%%%%%%%%%%%%%%%%%%%%%%%%%%%%%%%%%%%%%%%%%%%%%%%%%
%%%%%%%%%%%%%%%%%%%%%%%%%%%%%%%%%%%%%%%%%%%%%%%%%%%%%%%%%%%%%%%%%%%%%%%%%%%%%%%%%%%%%%%%%%

Such precision on the jet energy scale 
% will be difficult to achieve.
% A jet-energy-scale of $1\%$ will be rather difficult to achieve. 
%Such precision will be difficult to achieve. 
is achievable, but will require a lot of experimental work on the understanding
of the LHC detectors. The inclusion of TEVATRON jet cross-sections in the
NLO QCD fit might provide further constrains.
% that will make the requirement
%on the LHC jet cross-section uncertainties even more demanding.
%
However, it may be the case that ratios of jet cross sections 
-- for example, in different rapidity regions -- may have 
substantially smaller systematic uncertainties, while retaining 
sensitivity to the gluon density in the proton. %This has not been investigated here. 
Further constraints on the proton PDFs are also expected from Drell-Yan data measured
at LHC or any other data than those from HERA. 
Such data sets can now be consistently included in NLO QCD fits. 

%In this section, an example of the use of the event weight grid technique outlined in 
%this paper has been presented, in order to demonstrate the ability to include 
%hadronic collider final state data.

%\newpage
\section*{Conclusions}
A technique has been developed to store the perturbative coefficients calculated by a NLO
QCD Monte Carlo program in a look-up table (grid) allowing for {\em a posteriori} inclusion of 
an arbitrary parton density function (PDF) set and of alternative values of
 the strong coupling constant
as well as for {\em a posteriori} variation of the renormalisation and 
factorisation scale.
This extends a technique that has already been successfully used to analyse
HERA data to the more demanding case of proton-proton collisions at LHC energies.

The technique can be used to constrain PDF uncertainties by allowing 
the consistent inclusion of final state observables
in global QCD PDF fit analyses. This will help to increase 
the sensitivity of the LHC
to find new physics as deviations from the Standard Model predictions.

An accuracy of better than $0.1$\% can be reached with reasonably
small look-up tables for the single inclusive jet cross-section 
in the central rapidity region $|y|<1$, for jet transverse momenta (\ptx) 
from $100$ to $4500~\mathrm{GeV}$ and about $0.2 \%$ for jets in the
forward rapidity region $2 < y < 3$.
Similar accuracy can be achieved for the differential cross-sections
in rapidity and transverse momentum of electrons produced 
in $Z$ and $W$-boson decays. This was examined in the central
$ y < 0.5$ and very forward $y > 3$ regions for transverse momentum up
to $ \pt < 500\GeVx$.

The look-up tables provide a powerful tool to quickly evaluate
the PDF and scale uncertainties of the cross-section at various centre-of-mass
energies. The most recent PDFs predict jet cross-sections in the
central rapidity region within a few percent accuracy over a large range
of jet transverse momenta.

This technique has been successfully applied to a PDF fit using inclusive
deep-inelastic scattering and jet data measured at the electron-proton collider
HERA and using simulated LHC jet cross-sections.
%An improvement on the uncertainty of the gluon density can be achieved, if the
%jet energy scale uncertainty can be reduced to $1\%$. 
An improvement on the uncertainty of the gluon density can only be achieved
if the jet energy scale is very precisely known.
A more comprehensive
analysis will be possible in the future, since the grid technique 
can be applied
to most of the available NLO QCD calculations.

%Even for the large kinematic range for the parton momentum fractions $x_1$ and $x_2$ and
%of the squared momentum transfer $Q^2$ accessible at LHC, grids of moderate size
%seem to be sufficient. The single inclusive jet cross-section in the central region $|y|<0.5$
%can be calculated with a precision of $0.01\%$ in a realistic example with $100$ bins
%in the transverse jet energy range $100\, \leq \pt \leq 5000\,\mathrm{GeV}$.
%In this example, the grid  occupies about $300$~{\rm Mbyte} computer memory.
%With smaller grids of order $10$~{\rm Mbyte}
%the reachable accuracy is still $0.5\%$. This is probably sufficient for all practical
%applications.

\section*{Acknowledgements}
We would like to thank 
J.~Campbell, 
T.~Kluge, 
Z.~Nagy, 
K.~Rabbertz,  
T.~Sch\"orner-Sadenius, 
M.~H.~Seymour, 
P.~Uwer
and M. Wobisch
for useful discussions on the grid technique and A.~Vogt for
discussions on moment-space techniques. We thank Z. Nagy for help and support with
\NLOJET and J. Campbell for help and support with \MCFMx. 
C. Gwenlan and M. Sutton would like to thank the UK Science and Technology Facilities Council for support. 
G.~P.~Salam would like to thank Rutgers University for hospitality
while this work was being completed.
F.~Siegert would like to thank CERN for the Summer Student Program. 
P. Starovoitov would like to thank for the support given from INTAS. 
We would also like to thank J. Ribassin for the work he did towards the \MCFM interface.

% GPS: removed "Break-up" from title (too colloquial)
% GPS: removed "in proton-proton collisions" (it also holds for ppbar).
\section*{Appendix A: sub-processes for W- and Z-boson production} 
\label{sec:wzsubprocesses}
The production of $W$- and $Z$-bosons in proton-proton collisions involves flavour-dependent electro-weak couplings.
Therefore, the number of sub-processes that need to be defined is larger than in the case
of jet production. To reduce the number of sub-processes as much as possible,
quarks are assumed to be massless and 
the CKM matrix elements \cite{CKM,CKMCabibbo} to describe the contributions of the various quark flavours
are used.

 In the case of $Z$-boson production $12$ combinations of initial state partons need to be distinguished: 
  \begin{align}
    U\bar{U}: \qquad F^{(0)} \left(x_1,x_2,Q^2\right) &= U_{12}(x_1,x_2)\notag\\
    D\bar{D}: \qquad F^{(1)} \left(x_1,x_2,Q^2\right) &= D_{12}(x_1,x_2)\notag\\
    \bar{U}U: \qquad F^{(2)} \left(x_1,x_2,Q^2\right) &= U_{21}(x_1,x_2)\notag\\
    \bar{D}D: \qquad F^{(3)} \left(x_1,x_2,Q^2\right) &= D_{21}(x_1,x_2)\notag\\
    gU      : \qquad F^{(4)} \left(x_1,x_2,Q^2\right) &= G_1(x_1)U_2(x_2)\notag\\
    g\bar{U}: \qquad F^{(5)} \left(x_1,x_2,Q^2\right) &= G_1(x_1)\overline{U}_2(x_2)\notag\\
    gD      : \qquad F^{(6)} \left(x_1,x_2,Q^2\right) &= G_1(x_1)D_2(x_2)\notag\\
    g\bar{D}: \qquad F^{(7)} \left(x_1,x_2,Q^2\right) &= G_1(x_1)\overline{D}_2(x_2)\notag\\
    Ug      : \qquad F^{(8)} \left(x_1,x_2,Q^2\right) &= U_1(x_1)G_2(x_2)\notag\\
    \bar{U}g: \qquad F^{(9)} \left(x_1,x_2,Q^2\right) &= \overline{U}_1(x_1)G_2(x_2)\notag\\
    Dg      : \qquad F^{(10)}\left(x_1,x_2,Q^2\right) &= D_1(x_1)G_2(x_2)\notag\\
    \bar{D}g: \qquad F^{(11)}\left(x_1,x_2,Q^2\right) &= \overline{D}_1(x_1)G_2(x_2),
\label{Zprocesses}
  \end{align}
  where $g$ denotes gluons and $U (D)$ denotes up (down)-type quarks.
Use is made of the generalised PDFs defined as:
\begin{align}
  G_{H}(x) &= f_{0/H}\left(x,Q^{2}\right),\notag\\
  U_{H}(x) &= \sum\limits_{i = 2,4,6} f_{i/H}\left(x,Q^{2}\right),\qquad
  \overline{U}_{H}(x) = \sum\limits_{i = 2,4,6} f_{-i/H}\left(x,Q^{2}\right), \notag \\
  D_{H}(x) &= \sum\limits_{i = 1,3,5} f_{i/H}\left(x,Q^{2}\right),\qquad
  \overline{D}_{H}(x) = \sum\limits_{i = 1,3,5} f_{-i/H}\left(x,Q^{2}\right), \notag \\
 U_{12}(x_{1}, x_{2}) &= \sum_{i = 2,4,6}  f_{i/H_1}\left(x_{1},Q^2\right) f_{-i/H_2}\left(x
_{2},Q^{2}\right),\notag\\
 D_{12}(x_{1}, x_{2}) &= \sum_{i = 1,3,5}  f_{i/H_1}\left(x_{1},Q^2\right) f_{-i/H_2}\left(x
_{2},Q^{2}\right), \notag\\
 U_{21}(x_{1}, x_{2}) &= \sum_{i = 2,4,6} f_{-i/H_1}\left(x_{1},Q^2\right)  f_{i/H_2}\left(x
_{2},Q^{2}\right),\notag\\
 D_{21}(x_{1}, x_{2}) &= \sum_{i = 1,3,5} f_{-i/H_1}\left(x_{1},Q^2\right)  f_{i/H_2}\left(x
_{2},Q^{2}\right),
\label{Zpdf}
\end{align}
where $f_{i/H}$ is the PDF of flavour $i=-6 \dots 6$ for hadron $H$ and
$H_1$ ($H_2$) denotes the first or second hadron.

In the case of $W^+$-boson production\footnote{The case of $W^-$-boson can be treated
in an analogous way.} 
$6$ initial state combinations are needed:
 \begin{eqnarray}
 \label{eq:WZgenpdf}
    \bar{D}U:\qquad F^{(0)}\left(x_1,x_2,Q^2\right) &= S_{12}\left(x_1,x_2\right)\notag\\
    U\bar{D}:\qquad F^{(1)}\left(x_1,x_2,Q^2\right) &= S_{21}\left(x_1,x_2\right)\notag\\
    \bar{D}g:\qquad F^{(2)}\left(x_1,x_2,Q^2\right) &= \overline{D}_1(x_1) G_2(x_2)\notag\\
    Ug:      \qquad F^{(3)}\left(x_1,x_2,Q^2\right) &= U_1(x_1) G_2(x_2)\notag\\
    g\bar{D}:\qquad F^{(4)}\left(x_1,x_2,Q^2\right) &= G_1(x_1)\overline{D}_2(x_2)\notag\\
    gU:      \qquad F^{(5)}\left(x_1,x_2,Q^2\right) &= G_1(x_1)U_2(x_2),
\label{Wprocesses}
 \end{eqnarray}
where the generalised PDFs are used. They are defined as:
  \begin{align}
  G_{H}(x) =& f_{0/H}\left(x,Q^{2}\right),\notag\\
  U_H(x) =& f_{2/H}\left(x,Q^{2}\right)\left(\vud+\vus\right)+ f_{4/H}\left(x,Q^{2}\right)\left(\vcd+\vcs\right),\notag\\
  \overline{D}_H(x) =& f_{-1/H}\left(x,Q^{2}\right)\left(\vud+\vcd\right)+ f_{-3/H}\left(x,Q^{2}\right)\left(\vus+\vcs\right),\notag\\
  S_{12}(x_1,x_2) =&
                     f_{-3/H_1}\left(x_1,Q^{2}\right)f_{2/H_2}\left(x_2,Q^{2}\right)\vus+\notag\\
                    &f_{-3/H_1}\left(x_1,Q^{2}\right)f_{4/H_2}\left(x_2,Q^{2}\right)\vcs+\notag\\
                    &f_{-1/H_1}\left(x_1,Q^{2}\right)f_{2/H_2}\left(x_2,Q^{2}\right)\vud+\notag\\
                    &f_{-1/H_1}\left(x_1,Q^{2}\right)f_{4/H_2}\left(x_2,Q^{2}\right)\vcd,\notag\\
  S_{21}(x_1,x_2) =&
                     f_{2/H_1}\left(x_1,Q^{2}\right)f_{-3/H_2}\left(x_2,Q^{2}\right)\vus+\notag\\
                    &f_{4/H_1}\left(x_1,Q^{2}\right)f_{-3/H_2}\left(x_2,Q^{2}\right)\vcs+\notag\\
                    &f_{2/H_1}\left(x_1,Q^{2}\right)f_{-1/H_2}\left(x_2,Q^{2}\right)\vud+\notag\\
                    &f_{4/H_1}\left(x_1,Q^{2}\right)f_{-1/H_2}\left(x_2,Q^{2}\right)\vcd,
\label{Wpdf}
\end{align}
where $V_{ij}$ are the CKM matrix elements.\footnote{
The CKM matrix elements are stored together with the weight grid in the same file.
This ensures that the same values are used in the NLO calculation and in the PDF combinations. This choice can be changed {\em a posteriori} according to the needs to the user. 
In \MCFMx, only four non-zero CKM matrix elements are used.}

For simplicity in the former equations we omitted the top contribution,
since the parton densities are zero for most practical applications.
% 
%%%%%%%%%
%%
%%%%%%%%
\newpage
\section*{Appendix B: Automated identification of sub-processes}
\label{sec:automatedsubprocesses}

In general there are $169$ ($13\times 13$ flavour) possible PDF
combinations for proton proton collisions. In order to store only the
minimal amount of information, one needs to establish which of those
combinations always come with correlated weights, or equivalently one
should identify the underlying physical sub-processes.
So far, for each process under study, the sub-processes have been
identified manually, on a case-by-case basis.
However, the sub-processes can also be found in an automated way.

To simplify the discussion (without loss of generality) it will be
convenient to assume that the PDFs are always evaluated at fixed
values of $x$.
%
%Define $p_j$ to be however many PDF channels we have (e.g.\ 169 for
%$13\times13$ flavours in $pp$ collisions).
%
For each event $i$ and for each of the $169$ PDF combination $j$ (with PDF
weight $p_j$), the NLO QCD program calculates matrix-element weights
$W_{ij}$. The total weight for the event $i$ is $\sum_j W_{ij} \:
p_j$. 
The PDF combinations are called channels in the following.

To identify the sub-processes,  one determines the $W_{ij}$ weights for
$169$ events, giving a $169\times169$ matrix, whose $i$ (event) index labels the
rows and whose $j$ (PDF channel) index labels the columns.
%
%In $W_{ij}$ $i$ will
%be taken to label rows, $j$ columns of the $W_{ij}$ matrix.
%
One then carries out an eigenvalue decomposition of the $W_{ij}$
matrix.
%
%
%
%The relevant entries in the matrix can be found by generating $169$
%events and carrying out an eigenvalue decomposition of the resulting
%square matrix.  
%
If $v_n$ denotes the $n^{th}$ eigenvalue and $L_n$ and
$R_n$ the left and right eigenvectors (with components $L_{ni}$,
etc.), then as long as the there are no degenerate eigenvalues, an
orthonormality relation can be written:
\begin{equation}
  \label{eq:orthonorm}
  L_n \cdot R_m  = \delta_{mn},
\end{equation}
where the normalisation is our specific choice. 
Then one can rewrite the $W_{ij}$ matrix as
%Furthermore one can write
\begin{equation}
  \label{eq:decomp}
  W_{ij} = \sum_n R_{ni} \, v_n \, L_{nj}\,,
\end{equation}
it being straightforward, for example, to verify that both sides
satisfy $\sum_i L_{ni} W_{ij} = v_n L_{nj}$ and $\sum_j W_{ij} R_{mj}
= v_m R_{mi}$.
Let us now assume that only $N$ of the eigenvalues are
non-zero.\footnote{This contradicts the requirement that the
  eigenvalues be non-degenerate. In practice the rounding errors in
  the original calculation of the $W_{ij}$  cause the
  nominally zero eigenvalues to be slightly non-zero, thus alleviating
  this issue in practice.} %
Then eq.~\ref{eq:decomp} can be interpreted as follows: there are
$N$ relevant sub-processes; each $n\le N$ corresponds to a sub-process
that multiplies a linear combination of PDF channels in which the
contribution of channel $j$ is $L_{nj}$.
In event $i$, sub-process $n$ comes with a weight $R_{ni} v_n$. It will
be convenient to denote this by $w_{in}$. By virtue of the
orthornormality condition eq.~\ref{eq:orthonorm}, we have that
$w_{in} = \sum_j W_{ij} R_{nj}$, i.e.\ to determine the weight of
sub-process $n$ (whose PDF channel combination is given by the left
eigenvector $L_{nj}$) we take the right-eigenvector that corresponds
to this channel and use it to right-multiply the full weight matrix,
so to as to eliminate all but the contribution to the $n^\mathrm{th}$
sub-process.

The next step is to observe that the sub-processes determined for the
first 169 events should hold for all remaining events.\footnote{This
  is guaranteed as long as the NLO Monte Carlo weights include all
  sub-processes for each of the first 169 events.}
So now for \emph{any} event $i$, we can determine the weight for
sub-process $n$, $w_{in} = \sum_j W_{ij} R_{nj}$, using the $R_{nj}$
determined from the initial events.
Having stored the $w_{in}$ one can then subsequently reconstruct the
full $W_{ij}$ as $W_{ij} = \sum_n w_{in} L_{nj}$.

We have verified, in the context of a number of MCFM processes, 
that this approach is viable in practice\footnote{For carrying out this step in practice, we found the
  \texttt{ALGLIB} library to be useful, because it provides the
  necessary software tools written in \texttt{C++} and has the option
  of using the \texttt{mpfr} multiple-precision library (available on
  most distributions), which turned out to be necessary in order to
  get numerically convergent results for the full set of flavours.
  The \texttt{ALGLIB} algorithms are translations of the
  \texttt{LAPACK} ones.  }.
However, it is has yet to be fully integrated with the rest of our grid code 
and the results shown above are based on the manual sub-process
decompositions explicitly spelled out 
in sections~\ref{sec:jetsubprocesses} and in the Appendix A.

One should be aware that while the automated suprocess decomposition
yields a number of subprocesses that is identical to what can be found
with manual decomposition, the specific linear combinations of PDF
channels are usually different. 
To help understand why, one can take the example of jet production
with the 7 subprocesses of eq.~\ref{eq:jetgenpdf}. There, rather
than using $\mathrm{qq}$ and $\mathrm{q\bar q}$ channels, one might
have chosen instead to store weights for the combinations of
$\mathrm{qq + q\bar q}$ and $\mathrm{qq - q\bar q}$ channels.
More generally, one would have been free to base the grid on any 7
linearly independent combinations of the channels of
eq.~\ref{eq:jetgenpdf}.
For the automated channel decomposition process, the particular
independent linear combinations that emerge depend on the random
weights of the events used to identify the channels.

\bibliographystyle{atlasstylem.bst}
{\raggedright
\bibliography{applgridpaper}
}

\end{document}